\documentclass{aa}  

\usepackage{graphicx}
\usepackage{txfonts}

\usepackage{natbib}
\bibpunct{(}{)}{;}{a}{}{,} 
\usepackage{amstext}

\begin{document}

   \title{Laplace-like resonances with tidal effects}

   \author{A. Celletti
          \inst{1}, 
          E. Karampotsiou\inst{1,2}, 
          C. Lhotka\inst{1}, 
          G. Pucacco\inst{3}
          \and
          M. Volpi\inst{1}
          }

   \institute{Department of Mathematics, University of Roma Tor
  Vergata, Via della Ricerca Scientifica 1, 00133 Roma (Italy)\\
              \email{karampot@mat.uniroma2.it, lhotka@mat.uniroma2.it}
         \and
            Department of Physics, Aristotle University
  of Thessaloniki, 54124, Thessaloniki (Greece)
          \and
          Department of Physics, University of Roma Tor Vergata, Via della Ricerca Scientifica 1,
00133 Roma (Italy)\\
             }

 
  \abstract{
The first three Galilean satellites of Jupiter, Io, Europa, and Ganymede, move
in a dynamical configuration known as the \sl \textup{Laplace resonance},
\rm which is characterized by a 2:1 ratio of the rates of
variation in the mean longitudes of Io-Europa and a 2:1 ratio
of Europa-Ganymede. We refer to this configuration as a
2:1\&2:1 resonance. We generalize the Laplace
resonance among three satellites, $S_1$, $S_2$, and $S_3$, by
considering different ratios of the mean-longitude variations. These resonances, which we call \sl \textup{Laplace-like}\rm, are classified as first order in the cases of the 2:1\&2:1, 3:2\&3:2, and 2:1\&3:2 resonances, second order in the case of the 3:1\&3:1 resonance, and mixed order in the case of the 2:1\&3:1 resonance.
We consider a model that includes the gravitational interaction with the central
body together with the effect due to its oblateness, the mutual
gravitational influence of the satellites $S_1$, $S_2$, and $S_3$ and the secular
gravitational effect of a fourth satellite $S_4$, which plays
the role of Callisto in the Galilean system. In addition, we consider the
dissipative effect due to the tidal torque between the inner
satellite and the central body.
We investigate these Laplace-like
resonances by studying different aspects: $(i)$ we study the survival of the resonances
when the dissipation is included, taking two different expressions
for the dissipative effect in the case of a fast- or a slowly
rotating central body, $(ii)$ we investigate the behavior of the
Laplace-like resonances when some parameters are varied, specifically,
the oblateness coefficient, the semimajor axes, and the
eccentricities of the satellites, $(iii)$ we analyze the
linear stability of first-order resonances for different values of
the parameters, and $(iv)$ we also
include the full gravitational interaction with $S_4$ to
analyze its possible capture into resonance.
The results show a marked difference between first-, second-, and mixed-order resonances, which might find applications when the evolutionary history of the satellites in the Solar
System are studied, and also in possible actual configurations of extrasolar
planetary systems.}


   \maketitle

\section{Introduction}

The Laplace resonance is a well-known configuration that was
discovered by P.-S. de Laplace during his observations of the
Galilean satellites of Jupiter: Io, Europa, Ganymede, and Callisto.
The resonance consists of a set of commensurability relations
between the mean
longitudes $\lambda_1$, $\lambda_2$, and $\lambda_3$, and the
longitudes of perijoves $\varpi_1$, $\varpi_2$,
and $\varpi_3$,
\begin{eqnarray} 
\label{laplace}
\lambda_1-2\lambda_2+\varpi_1&=&0\nonumber\\
\lambda_1-2\lambda_2+\varpi_2&=&180^{\circ}\nonumber\\
\lambda_2-2\lambda_3+\varpi_2&=&0\ . 
\end{eqnarray}
We denote by
$\Phi_L$ the Laplace angle defined as
\begin{equation}
\label{Phi}
\Phi_L\equiv \lambda_1-3\lambda_2+2\lambda_3\ ;
\end{equation}
due to \eqref{laplace}, $\Phi_L=180^{\circ}$, which implies that a
triple conjunction between Io, Europa, and Ganymede can never be
applied.  \cite{lie97} observed (see also \cite{pai18}) that the
Galilean satellites are such that $\Phi_L=180^\circ$ up to a libration
with small amplitude and period of about 2071 days.

We consider a generalization of \eqref{laplace} by
assuming that the resonance relation between the mean longitudes
and the longitudes of pericenters involves combinations of the form
$j\lambda_1-k\lambda_2$, $m\lambda_2-n\lambda_3$ that we
denote as a j:k\&m:n resonance. We adopt the following
terminology:

\begin{itemize}
    \item[$(i)$] when both $j-k=1$ and $m-n=1$, we speak of a first-order
Laplace-like resonance;
    \item[$(ii)$] if $j-k=2$ and $m-n=2$, we speak of a
second-order Laplace-like resonance;
    \item[$(iii)$]when $j-k=2$ and $m-n=1$, or
$j-k=1$ and $m-n=2$, we speak of a mixed-order Laplace-like resonance.
\end{itemize}
We consider the set of resonances including the 2:1\&2:1,
3:2\&3:2, 3:1\&3:1, 2:1\&3:1, and 2:1\&3:2 resonances.

Although we worked with data and initial conditions corresponding to
the Galilean satellites, many examples of first. and
second-order resonances exist. They are typically found in satellite systems as
well as in extrasolar planetary systems. To mention some
examples, multiple mean-motion commensurabilities of three of the
largest Uranian satellites were studied in \cite{TW1990}: Miranda and Umbriel are in a 3:1 mean motion resonance,
Miranda and Ariel are in a 5:3 resonance, and Ariel and Umbriel are in a
2:1 resonance. Another example of multi-body resonance is
observed in the satellite system of Pluto because its satellites
Styx, Nix, Kerberos, and Hydra are close to 3:1, 4:1, 5:1, and 6:1
resonances, respectively, with Charon, the largest moon of Pluto.
Furthermore, according to \cite{sho15}, the satellites Styx, Nix,
and Hydra are in a 3:2\&3:2 resonance. In addition to the
Solar System, a few examples of three-body Laplace-like resonances have been observed in extrasolar planetary systems.
According to \cite{chr18}, the K2-138 extrasolar planetary system
contains five sub-Neptune planets that are close to a first-order
3:2 resonant chain. Similar to this system, HD 158259 is the
host star of five planets that are close to a 3:2 resonance chain (see
\cite{har20}). Another well-known example of resonance chains of
planets is TRAPPIST-1 (\cite{lug17}), where the seven known
planets are in a chain of Laplace resonances.
An analytical model of multiplanetary resonant chains has 
been developed in \cite{del17} and was applied to
the four planets orbiting Kepler-223. Finally,
the planetary system around the young star V1298 Tauri consists of
four planets, two of which are in a 3:2 resonance (see \cite{dav16}). Other systems with chains of first-order resonances
were presented in \cite{Pic19}.

\vskip.1in

The evolutionary history of the Galilean satellites has been
studied by many authors. It is commonly accepted that
dissipative tidal effects play a dominant role (\cite{lai09},
\cite{mal91}, \cite{sho97}, \cite{tit90}, \cite{yod79},
\cite{yod81}). In particular, as mentioned in \cite{tit90}, while
Io and Europa are assumed to be captured in a 2:1
resonance very soon after formation, Europa and Ganymede might have been in a 3:1
resonance in their early evolutionary history. We try to explore
this scenario by exploring similarities and differences between
the different sets of resonances mentioned before. We
remark that we considered all mutual interactions between the
satellites, while \cite{tit90} considered only the interaction
between Europa and Ganymede. 

We performed a thorough analysis that led to a clear
distinction between first-, second-, and mixed-order resonances. Our results
are based upon a model that includes the gravitational effect
of the central planet, the mutual gravitational interactions of
the satellites, the effects due to the oblateness of the planet, and the
secular gravitational interaction with $S_4$. We also
considered the tidal interaction between $S_1$ and the central body, which is described by a set of equations affecting the evolution of the
semimajor axis, eccentricity, and inclination
(\cite{fer07}). In the remarkable paper \cite{desitter1928}, W. de
Sitter investigated the question of whether the Laplace resonance
is maintained under dissipation. This means that the three
satellites should increase their semimajor axes in such a way that
the ratios of the mean motions are kept fixed. It is then
conjectured that the angular momentum acquired by Io from Jupiter
is transferred to Europa, and then from Europa to Ganymede so that
the Laplace resonant configuration survives. The conclusion drawn
in \cite{desitter1928} is that a condition for the survival of the
Laplace resonance is that the effect of the dissipation is small;
when it is large, the semimajor axes do not adjust, and the consequence
is that the commensurability is destroyed in the course of time.
We attach this question in the more general context of
Laplace-like resonances because they are interesting in solar and
extrasolar multibody systems other than the Galilean satellites.

We approached the problem by starting to investigate the sensitivity to 
the initial conditions and parameters involved, as well as by examining the 
changes in the dynamical evolution of a system of three satellites 
in different first-, second-, and mixed-order resonances. Next, we studied
the linear stability of the first-order resonances, which
are found to preserve the resonant dynamics as some parameters 
are varied.

Finally, we investigated the possibility that a fourth satellite that revolves around the planet is captured into resonance when it is studied
within the context of the different Laplace-like resonances. In
the case of the classical Laplace resonance (i.e., the actual
resonance between the Galilean satellites), Callisto is captured
into resonance, as has been remarked in \cite{lar20} and \cite{cklpv}. We extend the
study to the other Laplace-like resonances and study theoretically
the possibility of the trapping into resonance. Our analysis leads
us to conclude that $S_4$ is captured into resonance when first-order resonances are considered, while this is not the case for
second- and mixed-order resonances.

\vskip.1in

This work is organized as follows. In Section~\ref{sec:model} we present the
averaged and resonant Hamiltonians and the corresponding
equations of motion. In Section~\ref{sec:tidalsec} we provide the
equations for the tidal interactions between the central body and
the first orbiting body. We study by numerical
experiments the conservation of the Laplace argument considering the cases of a fast-rotating and a slowly rotating central body.
The focus of the study is on the evolution in time of the 
resonant arguments, that is, the orbital elements of the satellites,
close to resonant initial conditions, and on varying system 
parameters.
In Section~\ref{sec:linear} we study the linear stability of the first-order 
Laplace-like resonances. Section~\ref{sec:callisto} describes
the study of the capture into resonance of the fourth satellite.
Finally, we add some conclusions in Section~\ref{sec:conclusions}.


\section{Hamiltonian model}\label{sec:model}

%
To study the dynamical evolution of the three inner satellites 
around the central body, the following
contributions need to be considered: the gravitational interaction due to the central
body, the mutual gravitational interactions of the satellites, the
gravitational effects due to the oblateness of the planet, and the
secular gravitational interaction with satellite $S_4$ and a distant star, such as the Sun.

In our model, we make a simplification by retaining only the resonant
and secular parts, so that the Hamiltonian we consider
consists of the following contributions: (a) the Keplerian part
representing the interaction between the satellites and the
central body, (b) the secular part of the mutual gravitational
interaction of the three satellites, (c) the resonant part of the
mutual gravitational interaction of the three satellites, (d) the
secular gravitational effects of the oblateness of the planet, and
(e) the secular gravitational interaction with $S_4$ and a distant star. We mainly concentrate on the study of resonant
configurations among the three inner satellites, $S_1$, $S_2$ , and $S_3$; 
only in Section~\ref{sec:callisto} does the model
include the full (not only secular) gravitational interaction with
$S_4$. When different resonances are studied, only the resonant
part of the Hamiltonian changes, and the other contributions
are the same for all resonances.

\vskip.1in

To compute the Hamiltonian function that describes our model, we adopted the following definitions: $m_0$ is the mass of the
central planet, and $m_j$ is the mass of the $j$-th satellite. The
orbital elements of the $j$-th satellite are the semi-major axis
$a_j$, the eccentricity $e_j$, the inclination $I_j$ with respect
to the equatorial reference frame, the mean longitude $\lambda_j$,
the longitude of the pericenter $\varpi_j$, and the longitude of the
ascending node $\Omega_j$. In addition, we introduce the auxiliary
variable $s_j = \sin(I_j/2)$. We briefly describe the
different contributions to the Hamiltonian below and refer to
\cite{cel19} for full details.

The Keplerian part can be written as
\begin{equation*}
H_{Kep}=-{{\mathcal{G}M_1\mu_1}\over {2a_1}}-{{\mathcal{G}M_2\mu_2}\over {2a_2}}-{{\mathcal{G}M_3\mu_3}\over {2a_3}},
\label{kepHam}
\end{equation*}
where $\mathcal{G}$ is the gravitational constant and the
following auxiliary variables are introduced:
\begin{alignat*}{4}
 M_1 &=& m_0 + m_1, \qquad M_2 &=& M_1 + m_2, \qquad M_3 &=& M_2 + m_3 \nonumber\\
\mu_1 &=& \frac{m_0 m_1}{M_1}, \qquad \mu_2 &=& \frac{M_1 m_2}{M_2},
\qquad \mu_3 &=& \frac{M_2 m_3}{M_3}.
\end{alignat*}

The secular and resonant Hamiltonians describing the mutual interaction of the satellites depend on the specific resonance that is studied. When we consider a specific resonance, we can write the Hamiltonian for the interaction between satellites $S_U$ and $S_V$ in the form
\begin{equation*}
\begin{aligned}
&{H_P}^{U, V} = - \mathcal{G} \frac{m_U m_V}{a_V} \Big\{ F_{sec}
(a_U, a_V, e_U, e_V, s_U, s_V) \\
&+ F_{res} (a_U, a_V, e_U, e_V, s_U, s_V, \lambda_U,
\lambda_V, \varpi_U, \varpi_V, \Omega_U, \Omega_V) \Big\},
\end{aligned}
\end{equation*}
where $F_{sec}$ and $F_{res}$ denote the secular
part and the resonant term, respectively, the latter being a trigonometric
function that we truncate to second order in $e_U, e_V, s_U, \text{and }
s_V$. We provide in Appendix A the explicit expressions of the
Hamiltonian ${H_P}^{U, V}$ for each resonant case.

The contribution due to the oblateness of the planet, limited to the secular terms, is given by
\begin{equation*}
  \label{eq:Hj2}
  \begin{aligned}
    H_{obl}=&-\sum_{i=1}^{3} \frac{\mathcal{G} M_i\mu_i}{2a_i}\bigg[ J_2\left(\frac{R_J}{a_i}\right)^2\left(1+\frac{3}{2}e_i^2-\frac{3}{2}s_i^2\right) \\
&-\frac{3}{4} J_4\left(\frac{R_J}{a_i}\right)^4\left(1+\frac{5}{2}e_i^2-\frac{5}{2}s_i^2\right)\bigg],
\end{aligned}
\end{equation*}
where $R_J$ is the radius of the planet, and $J_2$ is the spherical harmonic coefficient of degree two.

The part of the Hamiltonian that represents the secular gravitational attraction of the fourth satellite or a distant star is given by
\begin{equation*}
  \label{ham:suncal}
  \begin{aligned}
H_\sigma&=-\sum_{i=1}^3 \Bigg [ \frac{\mathcal{G} m_im_\sigma}{a_\sigma}\left\{\frac{1}{2}b_{1/2}^{(0)}\left(\frac{a_i}{a_\sigma}\right)-1+\frac{1}{8}\frac{a_i}{a_\sigma}
b_{3/2}^{(1)}\left(\frac{a_i}{a_\sigma}\right)(e_i^2+e_{\sigma}^2)\right\}  \\
&-\frac{\mathcal{G} m_i m_\sigma}{a_\sigma}\left\{-\frac{1}{2}\frac{a_i}{a_\sigma}
b_{3/2}^{(1)}\left(\frac{a_i}{a_\sigma}\right)(s_i^2+s_{\sigma}^2)\right\} \Bigg],
\end{aligned}
\end{equation*}
where $\sigma = 4$ refers to the secular interaction of the fourth satellite, $\sigma = Sun$ refers to the secular interaction by a distant star, such as the Sun in the case of the Laplace resonance, and ${b_n}^{(s)}$ are the Laplace coefficients (\cite{mur99}).

The final Hamiltonian is given by the sum of the following contributions:
\begin{equation}
\label{final-Ham} H = H_{Kep} + {H_P}^{1,2} + {H_P}^{2,3} +
{H_P}^{1,3} + H_{obl} + H_4 + H_{Sun}.
\end{equation}

\vskip.1in

It is convenient to express the Hamiltonian in modified Delaunay variables, which are given by
\begin{equation*}
\begin{aligned}
& L_i = \mu_i \sqrt{\mathcal{G} M_i a_i} \\
& P_i = L_i (1-\sqrt{1-{e_i}^2}) \qquad for \quad i=1,2,3\\
& \Sigma_i = L_i\sqrt{1-{e_i}^2} (1- \cos I_i)
\end{aligned}
\end{equation*}
with conjugate angles $\lambda_i$, $p_i = -\varpi_i$ , and $\sigma_i = - \Omega_i$.

In the classical Laplace resonance between Io, Europa, and
Ganymede, the longitudes and the arguments of perijove of the
three satellites are found to satisfy relations \eqref{laplace},
and the Laplace angle $\Phi_L$ defined in Eq. \eqref{Phi}
is shown to librate around the value $\Phi_L = 180^o$. In
this case, we speak of a 2:1\&2:1 resonance because Eq.
\eqref{laplace} includes the combinations of the longitudes
$\lambda_1-2\lambda_2$, $\lambda_2-2\lambda_3$. We generalize Eq. \eqref{laplace} by considering j:k\&m:n
resonances that involve combinations of the longitudes of the form
$j \lambda_1-k\lambda_2$, $m \lambda_2-n\lambda_3$.


\vskip.1in

\begin{table}
\begin{tabular}{ |c|c|c|c| }
\hline
& 3:2\&3:2 & 2:1\&3:2 & 2:1\&2:1 \\
\hline
$q_1$ & $3\lambda_2- 2 \lambda_1- \varpi_1$ & $2\lambda_2- \lambda_1- \varpi_1$ & $2\lambda_2- \lambda_1- \varpi_1$ \\
$q_2$ & $3\lambda_2- 2 \lambda_1 - \varpi_2$ & $2\lambda_2- \lambda_1 - \varpi_2$ & $2\lambda_2- \lambda_1-\varpi_2$  \\
$q_3$ & $3\lambda_3- 2 \lambda_2-\varpi_3$ & $3\lambda_3- 2 \lambda_2-\varpi_3$ & $2\lambda_2-\lambda_1-\varpi_3$ \\
$q_4$ & $5\lambda_2-2 \lambda_1-3\lambda_3$ & $3\lambda_3-4 \lambda_2+\lambda_1$ & $3\lambda_2-\lambda_1-2\lambda_3$ \\
$q_5$ & $\lambda_1 - \lambda_3$ & $\lambda_1 - \lambda_3$ & $\lambda_3 - 3 \lambda_2 + 2 \lambda_1$  \\
$q_6$ & $\lambda_3$ & $\lambda_3$ & $\lambda_2 + \frac{1}{3}(\lambda_3 - \lambda_1)$ \\
$q_7$ & $3\lambda_2- 2 \lambda_1- \Omega_1$ & $2\lambda_2- \lambda_1- \Omega_1$ & $2\lambda_2-\lambda_1- \Omega_1$ \\
$q_8$ & $3\lambda_2- 2 \lambda_1 - \Omega_2$ & $2\lambda_2- \lambda_1 - \Omega_2$ & $2\lambda_2-\lambda_1 - \Omega_2$ \\
$q_9$ & $3\lambda_3- 2 \lambda_2-\Omega_3$ & $3\lambda_3- 2 \lambda_2-\Omega_3$ & $2\lambda_3-\lambda_2-\Omega_3$ \\
\hline
\end{tabular}
\begin{tabular}{|c|c|c|}
\hline
& 2:1\&3:1 & 3:1\&3:1 \\
\hline
$q_1$ & $2\lambda_2- \lambda_1- \varpi_1$ & $3\lambda_2- \lambda_1-2 \varpi_1$ \\
$q_2$ & $2\lambda_2- \lambda_1 - \varpi_2$ & $3\lambda_2- \lambda_1-\varpi_1 - \varpi_2$ \\
$q_3$ & $3\lambda_3- \lambda_2-\varpi_3 -\varpi_2$ & $3\lambda_3- \lambda_2-\varpi_3 -\varpi_2$ \\
$q_4$ & $3\lambda_2-3 \lambda_3-\lambda_1 +\varpi_2$ & $4\lambda_2-3 \lambda_3-\lambda_1 -\varpi_1 +\varpi_2$ \\
$q_5$ & $\lambda_1 - \lambda_3$ & $\lambda_1 - \lambda_3$ \\
$q_6$ & $\lambda_3$ & $\lambda_3$ \\
$q_7$ & $2\lambda_2- \lambda_1- \Omega_1$ & $3\lambda_2- \lambda_1-2 \Omega_1$ \\
$q_8$ & $2\lambda_2- \lambda_1 - \Omega_2$ & $3\lambda_2- \lambda_1-\Omega_1 - \Omega_2$ \\
$q_9$ & $3\lambda_3- \lambda_2-\Omega_3 -\Omega_2$ & $3\lambda_3- \lambda_2-\Omega_3 -\Omega_2$ \\
\hline
\end{tabular}
\vskip.1in
\caption{Angular coordinates for the different resonances} \label{tab:q}
\end{table}

In the remaining paper we concentrate on the resonances
2:1\&2:1, 3:2\&3:2, 3:1\&3:1, 2:1\&3:2, and 2:1\&3:1. We
find it convenient to introduce new angular variables for
each specific resonance that replace the angles $\lambda_1$,
$\lambda_2$, $\lambda_3$, $\varpi_1$, $\varpi_2$, $\varpi_3$,
$\Omega_1$, $\Omega_2$, and $\Omega_3$; to this end, we introduce the variables
$q_1, ..., q_9$, as defined in Table 1. Especially in the case of the $2:1 \& 2:1$ resonance, the new set of angular variables was chosen by following \cite{puc21}.
Table \ref{parameters} presents the values of the masses, eccentricities, and inclinations of the Galilean satellites. These
values are used throughout this paper in the numerical
integrations of the equations of motion.
In addition to the parameters, we used the initial conditions for the semimajor axes
of the satellites that are shown in Table~\ref{initial_conditions}.

\begin{table}
\begin{tabular}{|c|c|c|c|}
\hline
& $m_j$ & $e_j$ & $I_j$ \\
\hline
Io & $8.933 \times 10^{22}$ & $4.721 \times 10^{-3}$ & $3.758 \times 10^{-2}$\\
Europa & $4.797 \times 10^{22}$ & $9.819 \times 10^{-3}$ & $4.622 \times 10^{-1}$ \\
Ganymede & $1.482 \times 10^{23}$ & $1.458 \times 10^{-3}$ & $2.069 \times 10^{-1}$\\
Callisto & $1.076 \times 10^{23}$ & $7.44 \times 10^{-3}$ & $1.996 \times 10^{-1}$ \\
\hline
\end{tabular}
\vskip.1in
\caption{Values of the masses $m_j$, the eccentricities $e_j$ and the inclinations $I_j$ of the Galilean satellites.} \label{parameters}
\end{table}

\begin{table}
\begin{tabular}{|c|c|c|c|}
\hline
& 2:1\&2:1 & 3:2\&3:2 & 2:1\&3:2 \\
\hline
Io & $4.22 \times 10^{5}$ & $4.22 \times 10^{5}$ & $4.22 \times 10^{5}$ \\
Europa & $6.713 \times 10^{5}$ & $5.53 \times 10^{5}$ & $6.713 \times 10^{5}$ \\
Ganymede & $10.705 \times 10^{5}$ & $7.25 \times 10^{5}$ & $8.78 \times 10^{5}$ \\
Callisto & $18.828 \times 10^{5}$ & $12.751 \times 10^{5}$ & $15.442 \times 10^{5}$ \\
\hline
\end{tabular}
\begin{tabular}{|c|c|c|}
\hline
& 3:1\&3:1 & 2:1\&3:1 \\
\hline
Io & $4.22 \times 10^{5}$ & $4.22 \times 10^{5}$ \\
Europa & $8.78 \times 10^{5}$ & $6.713 \times 10^{5}$ \\
Ganymede & $18.26 \times 10^{5}$ & $13.94 \times 10^{5}$ \\
Callisto & $32.115 \times 10^{5}$ & $24.517 \times 10^{5}$ \\
\hline
\end{tabular}
\vskip.1in
\caption{Initial values of the semimajor axes of the four satellites, expressed in kilometers.} \label{initial_conditions}
\end{table}

\section{Effect of tides on the dynamics}
\label{sec:tidalsec}

In this section we investigate the role of tides on the dynamics, that is, 
on the persistence of the Laplace-like resonances in the dissipative
framework, and their effect on the orbital parameters in dependence
of the system parameters (e.g., the dependence on the $J_2$ value
and the initial orbital configuration in terms of semimajor axis $a$ 
and orbital eccentricity $e$).

\subsection{Tidal models}\label{sec:tidal}

The tidal interaction between the central body and the closest
satellite affects the evolution of the semimajor axis,
eccentricity, and inclination (\cite{fer07}). The equations
describing the rates of variation in $a, e,\text{and } I$ due to the
dissipation part that we used are
\begin{equation}
  \label{fast-rotating}
  \begin{aligned}
    \frac{\dot{a}}{a} &=\frac{2}{3}c\left(1+\frac{51}{4}e^2 - D(7 e^2 + {S_B}^2)\right)\\
    \frac{\dot{e}}{e} &= -\frac{1}{3}c\left(7D-\frac{19}{4}\right)\\
    \dot{I} & = - \frac{3}{4} S_B c (1 + 2D)\ .
  \end{aligned}
  \quad .
  \end{equation}
The parameters $c$ and $D$ are defined as
\begin{equation}\label{eq:c}
  \begin{aligned}
    c &= \frac{9}{2}\frac{k_0}{Q_0}\frac{m_{1}}{m_0}\left(\frac{R_0}{a_{1}}\right)^5n_1\\
    D &= \frac{Q_0}{Q_{1}}\frac{k_{1}}{k_0}\left(\frac{R_{1}}{R_0}\right)^5\left(\frac{m_0}{m_{1}}\right)^2,
    \end{aligned}
\end{equation}
where for j = 0,1 $k_j$ is the Love number, $Q_j$ is the quality factor, $k_j/Q_j$ is the tidal ratio, and $R_j$ is the radius of the central body and the innermost satellite, respectively. Moreover, $n_1$ is the mean motion of the first satellite,
and $S_B = \sin(I_1)$.

Equations \eqref{fast-rotating}, translated into Delaunay variables, was added to the equations for the variables 
$L, P,\text{and } \Sigma$ that can be derived from the Hamiltonian \eqref{final-Ham}. Equations \eqref{fast-rotating} hold when the central body rotates fast and under the assumption that the phase lags remain constant and are all equal. This tidal model is valid in the case of planetary satellites. However, when an extrasolar planetary system is studied, the central body rotates slowly. In this case, equations \eqref{fast-rotating} can be modified as follows (\cite{fer07}):
\begin{equation}
  \label{slow-rotating}
  \begin{aligned}
    \frac{\dot{a}}{a} &=-\frac{2}{3}c\left(1+\frac{57}{4}e^2 +7 D e^2 \right)\\
    \frac{\dot{e}}{e} &= -\frac{1}{3}c\left(7D+\frac{25}{4}\right)\\
    \dot{I} & = - \frac{3}{4} S_B c (1 + 2D).
  \end{aligned}
  \end{equation}
In the next sections, we investigate the dynamics of first- and second-order Laplace-like resonances
with tidal torque for fast-rotating planets using Eq. \eqref{fast-rotating} and slow rotating
central bodies using Eq. \eqref{slow-rotating}. In our model the tidal effects are only considered between the central body and the innermost satellite. The interaction between the
central body and the remaining moons can usually be neglected in this type of study.

\subsection{Survival of the Laplace-like resonances under the tidal dissipation}\label{sec:survival}

In this section, we present the evolution of the Laplace angle
that results from the numerical integrations in all five different
first-, second-, and mixed-order resonances. The equations of motion were integrated numerically using a Runge-Kutta eighth-order scheme with a variable step size performed using a Wolfram Mathematica program. The resonant arguments that
involve the longitudes of the three satellites are given in
Table~\ref{resangle}.

\begin{figure}[h]
\centering
\includegraphics[width=.99\linewidth]{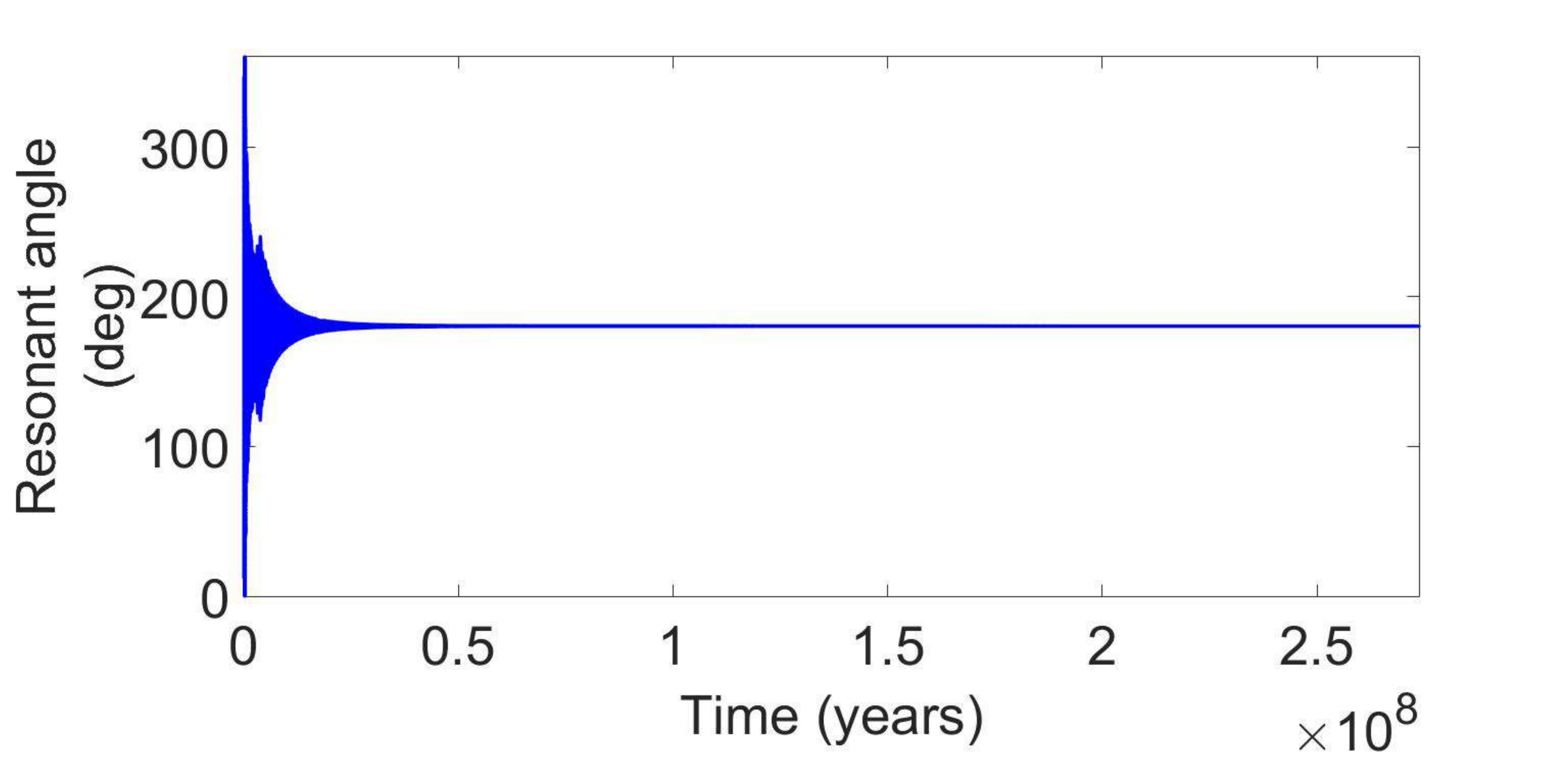}
\includegraphics[width=.99\linewidth]{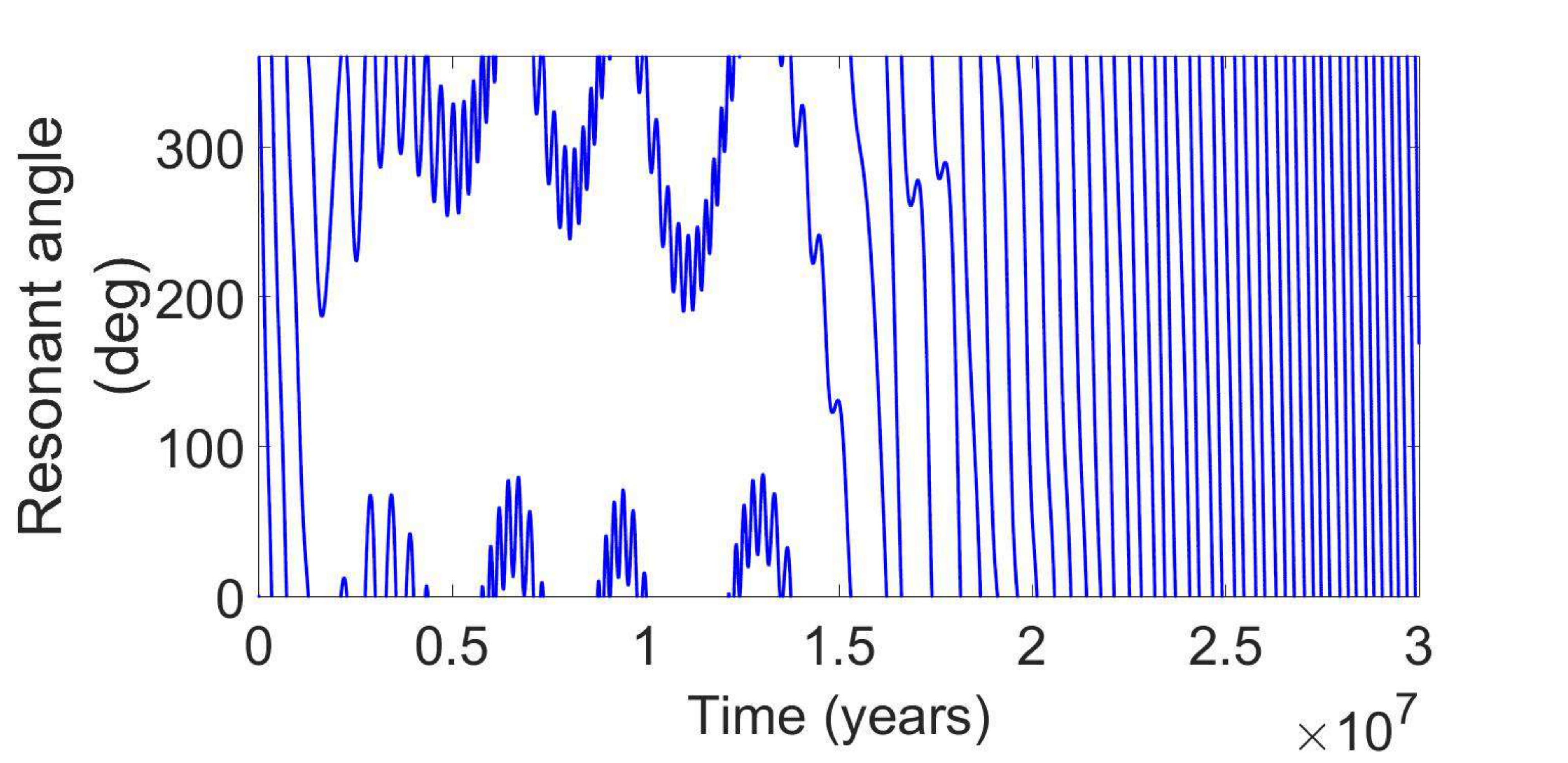}
\includegraphics[width=.99\linewidth]{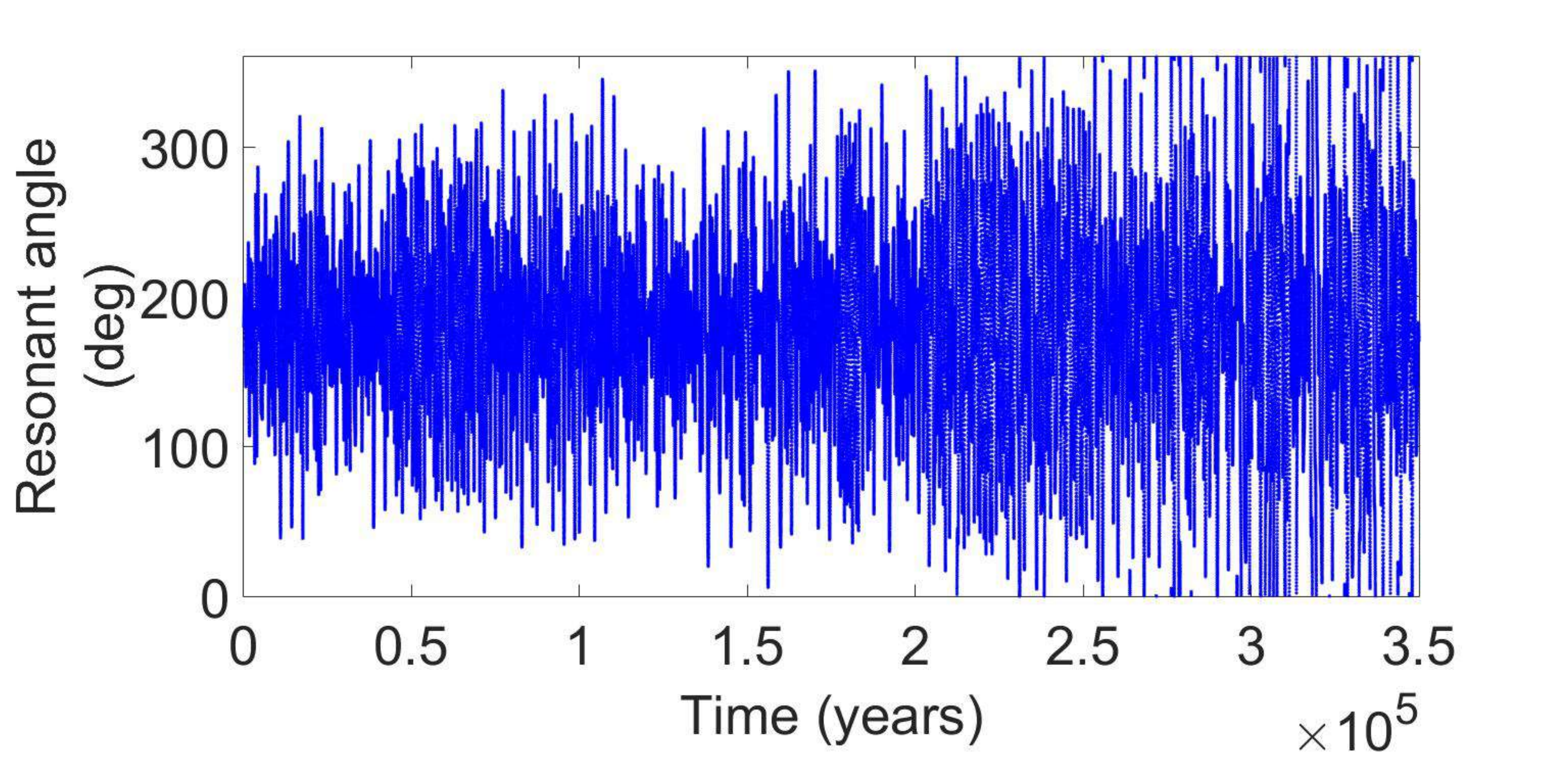}
    \caption{Evolution of the resonant argument in three cases: the 2:1\&3:2 resonance
    considering a fast-rotating central body (top panel), the 2:1\&3:1 resonance
    considering a fast-rotating central body (middle panel), and the 3:2\&3:2 resonance
    considering a slowly rotating central body (bottom panel).}\label{res_argument}
\end{figure}

A major point in the study of the Laplace-like resonances is the
survival of the resonance when the tidal effects between the
central body and one or more satellites are considered. According
to \cite{fer81}, the Laplace resonance is kept under a
dissipative force, leading to quadratic inequalities in
the mean longitudes. This question was already raised in \cite{desitter1928}.
\cite{lai09} used numerical integrations and
stated that the Laplace resonance is destroyed because of the
inward migration of Io and the outward migration of Europa and
Ganymede. Recently, \cite{lar20} showed that the Laplace
resonance is kept for about 1 Gyr, and then Callisto is captured
into resonance with Ganymede. \cite{lar20} also studied the stability of the resonance after 1 Gyr using
different initial conditions of Callisto and testing different
scenarios. In most of the simulations, the Laplace resonance
survives. In this section, we study the survival of the
Laplace-like resonances taking the tidal interaction
between the central planet and $S_1$ into account and considering the cases of a
fast-rotating and slowly rotating central body, as described in
Section 3.

\subsubsection{Fast-rotating central body}

In this section, we analyze the case of a fast-rotating central
planet for which we consider the dissipation given by Eq.
\eqref{fast-rotating}. For each resonance, we started by examining
the behavior of the resonant argument, which corresponds to the
angles listed in Table~\ref{resangle}.

\begin{table}
\centering
\begin{tabular}{ |c|c| }
 \hline
 Resonance & Resonant argument \\
 \hline
 2:1\&2:1 & $2 \lambda_3 - 3 \lambda_2 + \lambda_1$ \\
 \hline
 3:2\&3:2 & $3 \lambda_3 - 5 \lambda_2 + 2 \lambda_1$ \\
 \hline
 2:1\&3:2 & $3 \lambda_3 - 4 \lambda_2 + \lambda_1$ \\
 \hline
 3:1\&3:1 & $3 \lambda_3 - 4 \lambda_2 + \lambda_1$ \\
 \hline
 2:1\&3:1 & $3 \lambda_3 - 5 \lambda_2 + 2 \lambda_1$ \\
 \hline
\end{tabular}\caption{Resonant arguments in the different resonances that involve the mean longitudes of the three satellites.}
\label{resangle}
\end{table}

During the evolution, we multiplied the tidal effect by a pumping factor
$\alpha = 10^5$ to increase the strength of the tides and speed up numerical 
integrations (in absence of the Hamiltonian contributions, the presence of 
$\alpha$ can be seen as a rescaling of time; see \cite{sho97} and \cite{lar20}).
Using as parameters the masses of the Galilean
satellites and the initial conditions given in Tables \ref{parameters}
and \ref{initial_conditions}, we find that the 2:1\&2:1,
3:2\&3:2, and 2:1\&3:2 resonances survive under the dissipation,
while the 3:1\&3:1 and 2:1\&3:1 resonances are destroyed by the
dissipation in the sense that the resonant argument circulates
instead of librating around a fixed value. This makes a
marked difference between first-, second-, and mixed-order Laplace-like
resonances evident.
Figure~\ref{res_argument} shows the sample of a first-order resonance with the Laplace argument
trapped into libration (top panel) and a mixed-order resonance with an oscillation of
the resonant angle (bottom panel). 

\subsubsection{Slowly rotating central-body}

When the dissipation associated with a slowly rotating central
body as in Eq. \eqref{slow-rotating} is used, the situation is different from
the fast-rotating planet case. In the classical
2:1\&2:1 resonance, a migration of the closest
satellite to the planet and a circulation of the resonant
argument are observed. The same situation occurs for the 3:2\&3:2 resonance,
in which the decrease in semimajor axis of the inner satellite
provokes a close encounter with the planet. Here as well as in
the 2:1\&3:2, 3:1\&3:1, and 2:1\&3:1 cases, the resonance is not
preserved, and the resonant argument circulates after a short time
interval. An example is given in Figure~\ref{res_argument} (bottom
panel), showing a short period of libration followed by a
circulation of the Laplace angle.

As a conclusion, the choice of the tidal model strongly affects the presence of the resonant evolution of the Laplace angles because the variation rate of the semi-major axis has a different sign for fast and slow rotation. This is true for the actual Galilean system, but also for the other resonances.

\subsection{Resilience of the Laplace-like resonances}\label{sec:resilience}

\begin{figure}[h]
\centering
\includegraphics[width=.99\linewidth]{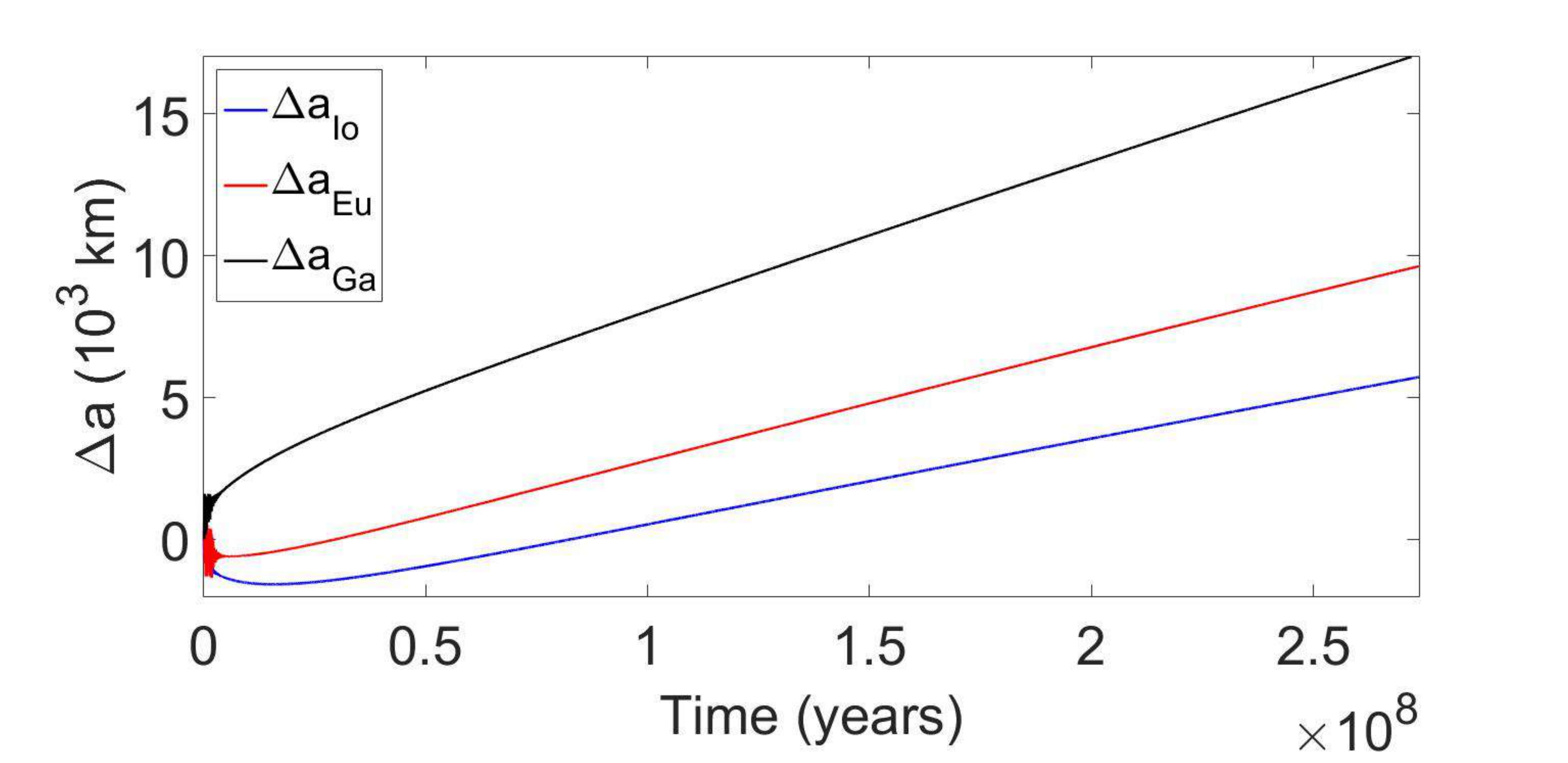}
\includegraphics[width=.99\linewidth]{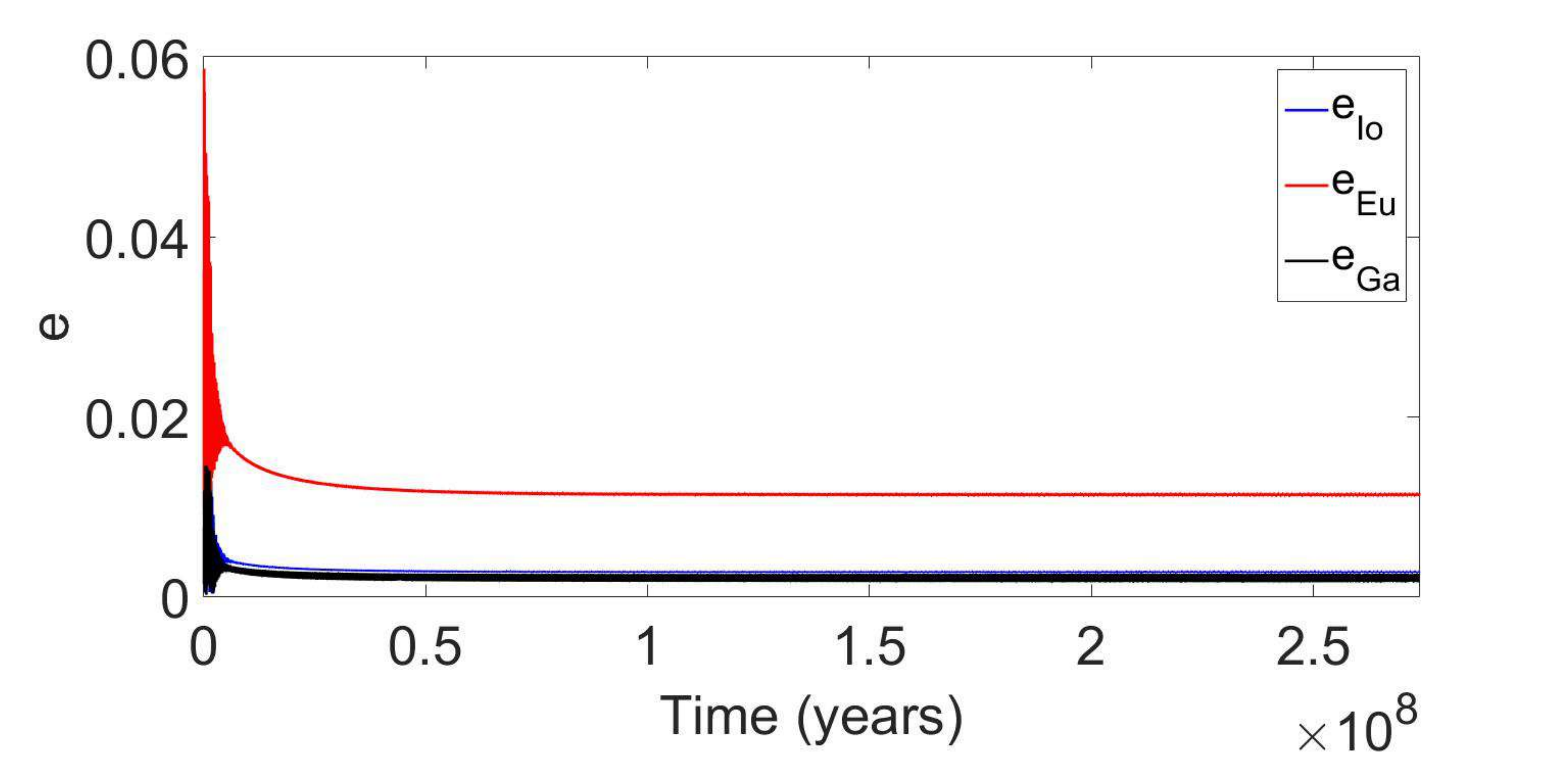}
    \caption{Variation in semimajor axes (top panel) and
    eccentricities (bottom panel) of the three satellites due to the dissipative
    effects between the planet and the inner satellite considering a fast-rotating central body and assuming a 3:2\&3:2 resonance The tidal effects are multiplied by a factor of $\alpha = 10^5$. }\label{evolution1}
\end{figure}

\begin{figure}[h]
\centering
\includegraphics[width=.99\linewidth]{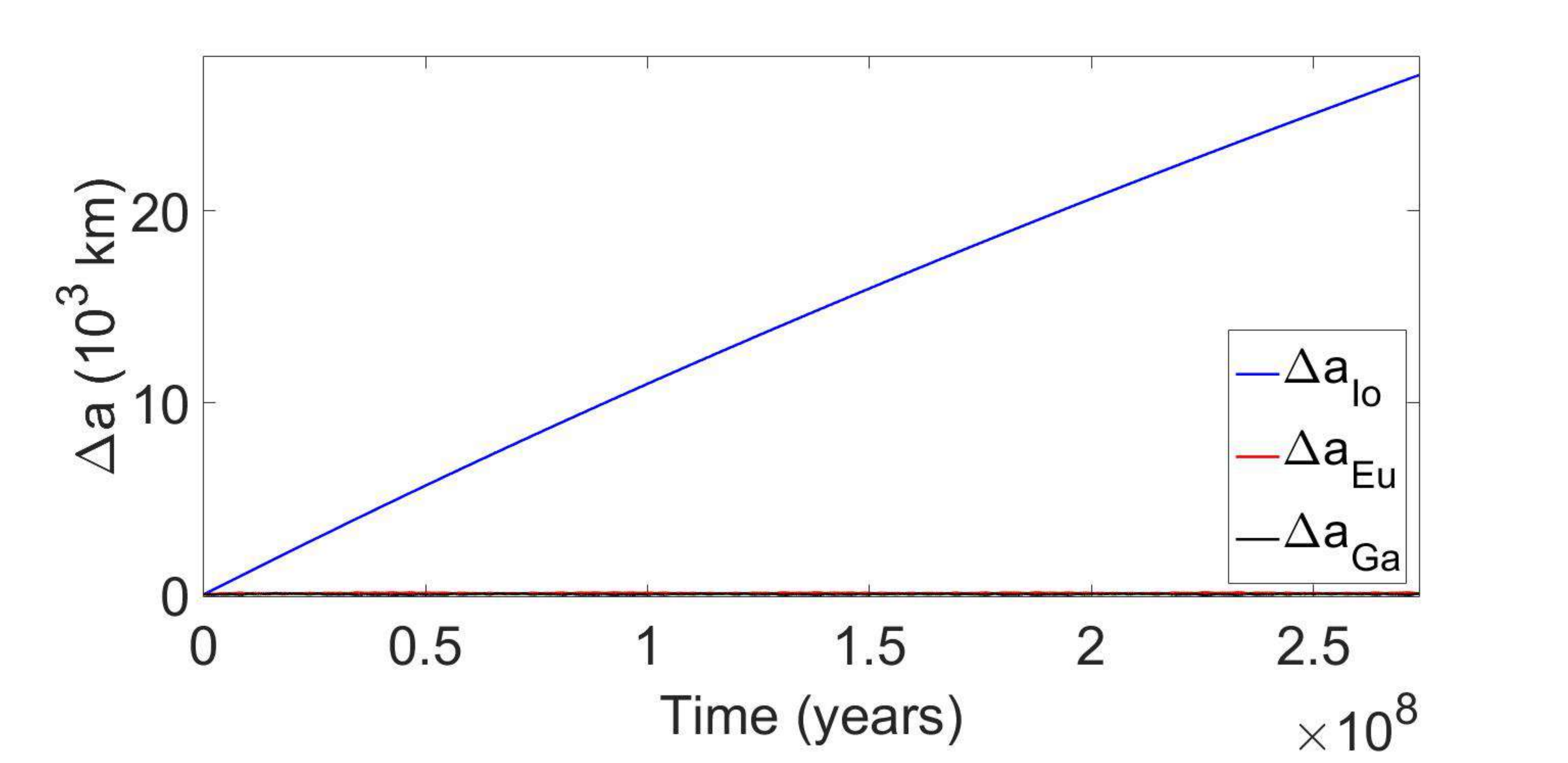}
\includegraphics[width=.99\linewidth]{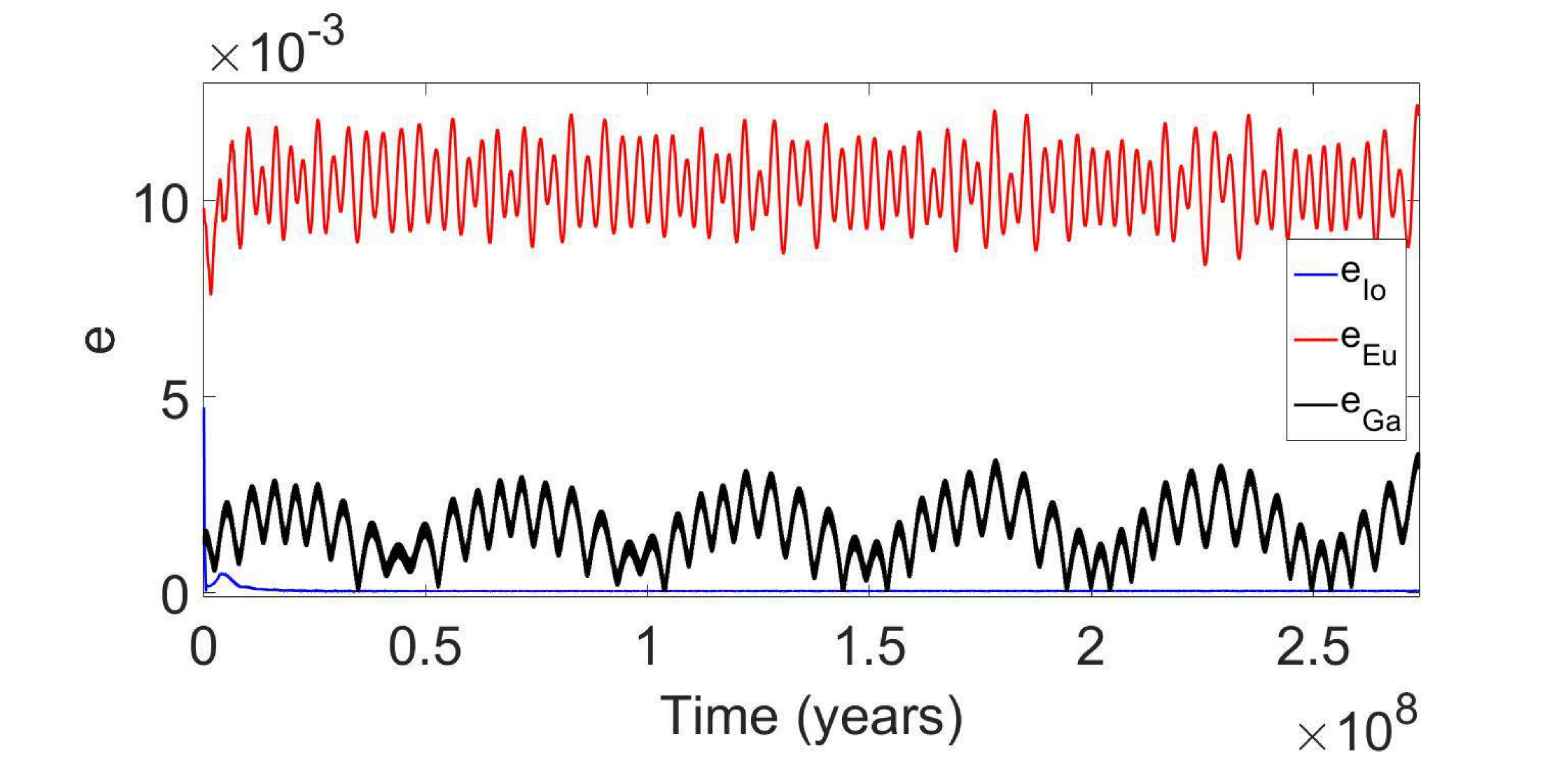}
    \caption{Variation in semimajor axes (top panel) and
    eccentricities (bottom panel) of the three satellites due to the dissipative
    effects between the planet and the inner satellite considering a fast-rotating central body and assuming a 3:1\&3:1 resonance. The tidal effects are multiplied by a factor of $\alpha = 10^5$. }\label{evolution2}
\end{figure}

In this section, we investigate the evolution of the orbital elements of the
three satellites under the variation of the initial conditions and parameters
that are involved in the model. As a first step, we study the evolution of the semimajor axes, eccentricities, and inclinations of the three satellites. We remark that they include the tidal effects
between a fast-rotating central body and the inner satellite only. In some cases,
the simultaneous migration of all three satellites is observed, as
in the 3:2\&3:2 and 2:1\&3:2 resonances, while in the
3:1\&3:1 resonance, only the inner satellite moves outward and the semimajor axes of the other two
satellites remain constant. In the 2:1\&3:1 resonance, 
the two inner satellites migrate and the semimajor axis of the 
third satellite remains constant. In addition, the eccentricity of the inner
satellite converges to zero, while the eccentricity of the other
two satellites oscillates around an average value. The results are
presented in Figures~\ref{evolution1} and \ref{evolution2} for the 3:2\&3:2 and 3:1\&3:1 
resonances, respectively. Moreover, we analyze the behavior of the different
resonances when varying $J_2$ (Section 3.3.1), the initial value of the semimajor
axis (Section 3.3.2), and the initial value of the eccentricity (Section 3.3.3) in the case of the Laplace-like resonances

The increase in semimajor axis of the inner satellite is
expected theoretically due to the tidal torque exerted by the
fast-rotating central body. In the cases of first-order
resonances, the migration of the inner satellite is transferred
through the gravitational attractions to all the bodies, and as a
result, they all migrate outward, maintaining the resonance. This can be seen, for example, in the case of the 3:2\&3:2 resonance, where all three
satellites move outward, while the three eccentricities converge
to limit values. These results indicate a marked difference between
first- and second-order resonances. The linear stability of first order
is addressed in a semi-analytical way in Section~\ref{sec:linear}.

\begin{figure}[h]
\centering
\includegraphics[width=.99\linewidth]{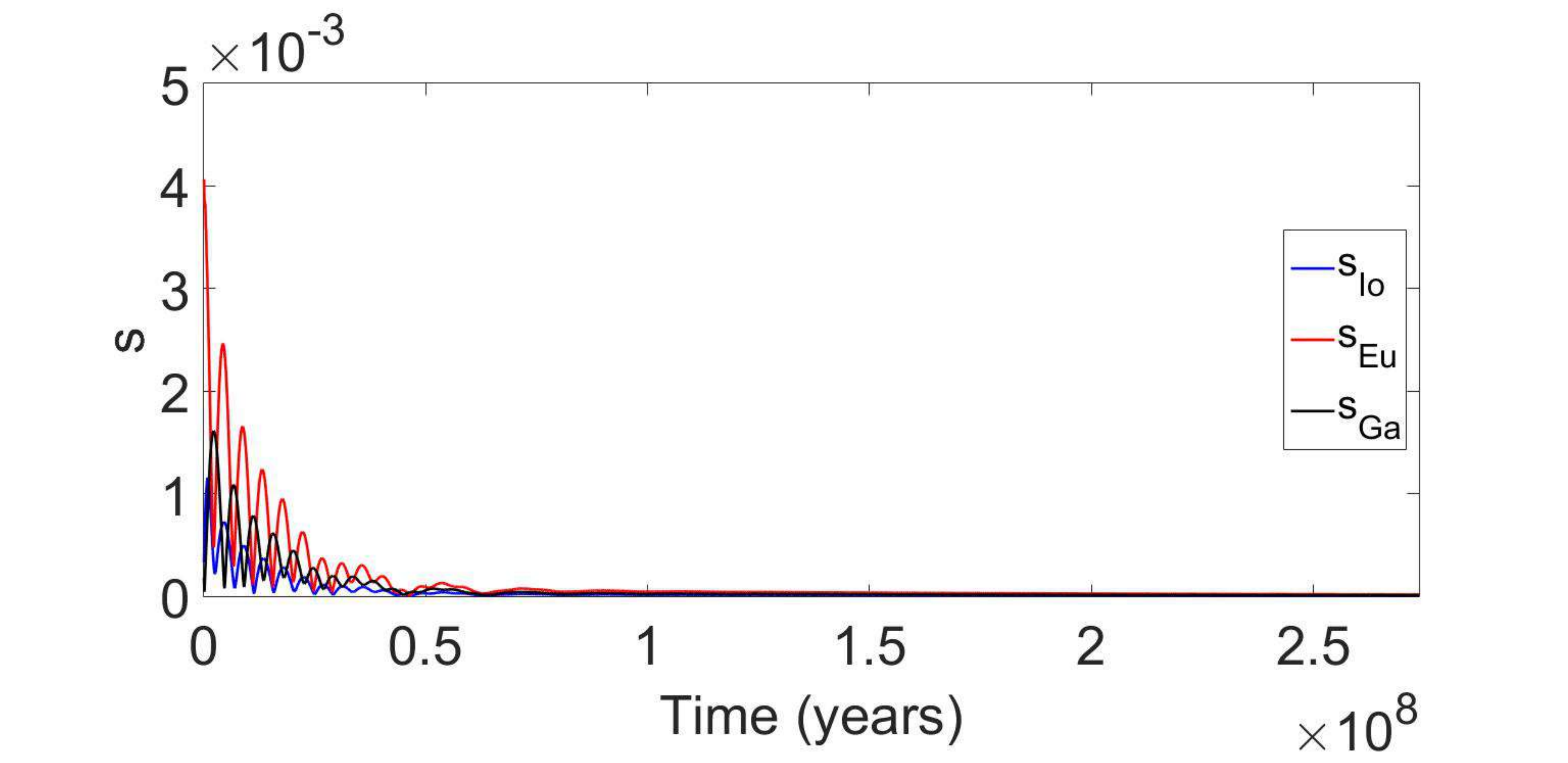}
\includegraphics[width=.99\linewidth]{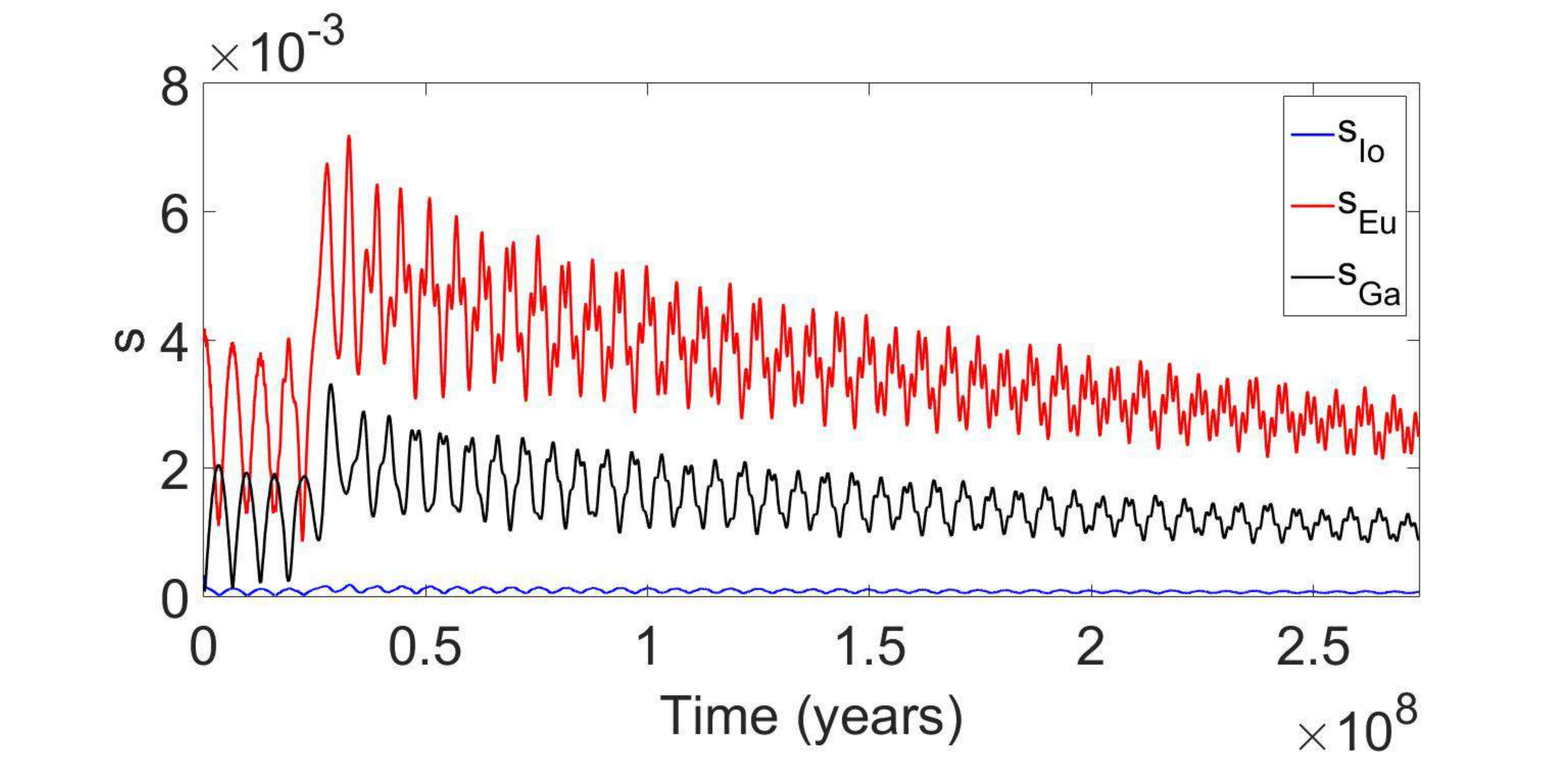}
\includegraphics[width=.99\linewidth]{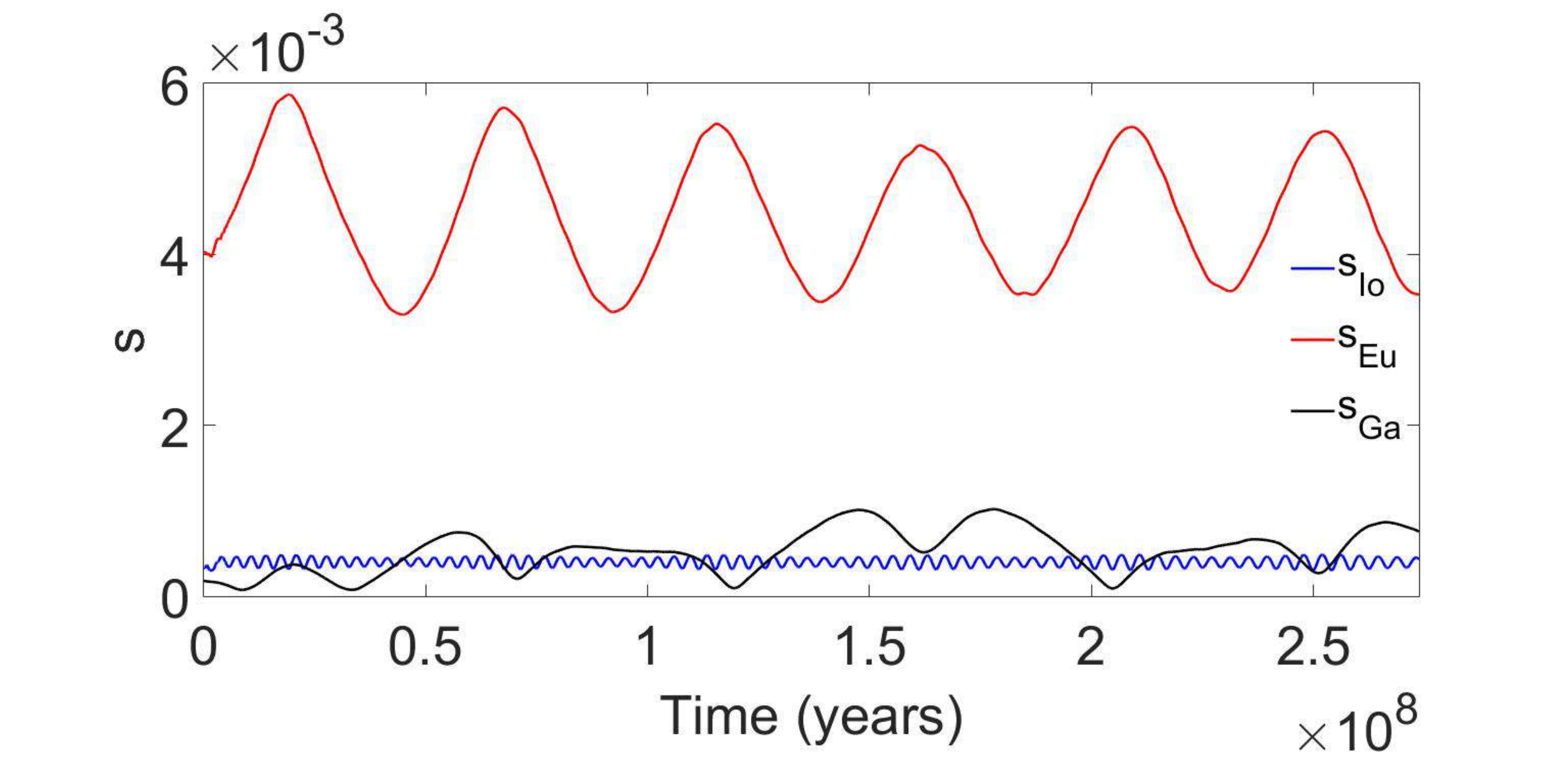}
\includegraphics[width=.99\linewidth]{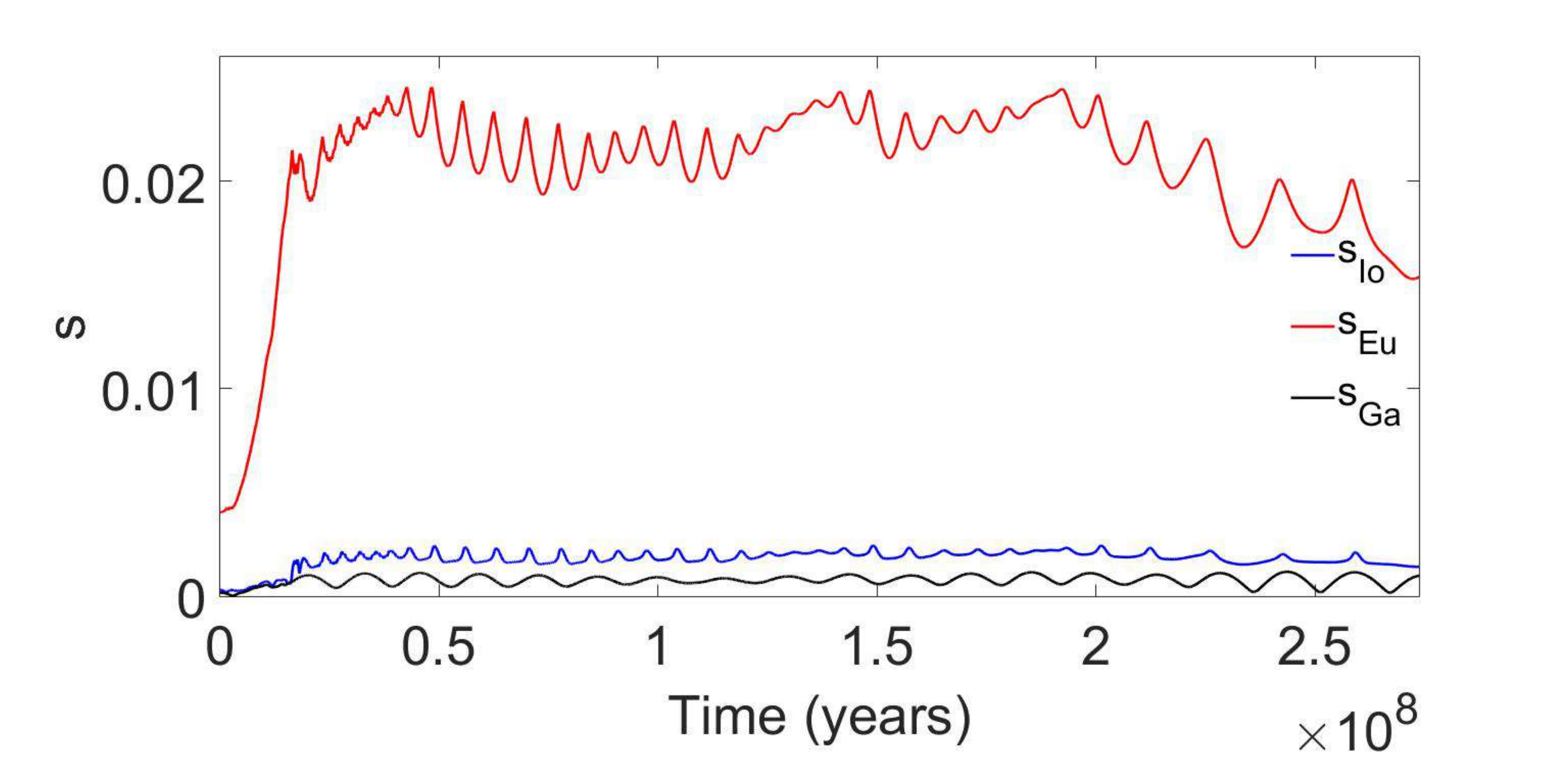}
    \caption{Variation in inclinations of the three satellites due to the dissipative
    effects between the planet and the inner satellite considering a fast-rotating central body. The four panels from top to bottom correspond to the 3:2\&3:2, 2:1\&3:2, 3:1\&3:1, and 2:1\&3:1 resonances. The tidal effects are multiplied by a factor of $\alpha = 10^5$. }\label{inclinations}
\end{figure}

The inclinations in the 2:1\&3:2 and 2:1\&3:1 resonances
oscillate randomly, but remain close to the initial ones. In the
3:2\&3:2 resonance, the inclinations converge to zero, while in
the 3:1\&3:1 resonance, they increase and are higher than the initial ones.
We can conclude that in the 3:2\&3:2 resonance, the planar model,
which is simpler, would give reliable results. Instead, in the mixed- and
second-order resonances and especially in the 2:1\&3:1 resonance, the
inclinations play an important role in the dynamics of the system
(Figure~\ref{inclinations}).

\subsubsection{Dependence on the $J_2$ value of the central body}

In the case of the 3:2\&3:2 resonance, the effect due to the
oblateness of the central body plays an important role in the
evolution of the resonant argument. It is worth noting that the effects due to the oblateness of the central body induce an additional precession of the pericenters. When the $J_2$ value of
the central planet is set equal to zero, the resonance is kept, but the
resonant argument librates around $180^{\circ}$ with a higher
amplitude than when the $J_2$ value is different from
zero. In the case of the 2:1\&3:2 resonance, the resonant
argument librates around $180^{\circ}$, but when the oblateness of
Jupiter is not taken into account, the resonant argument takes
values from 0 to $360^{\circ}$ , and after some time, the resonant angle
enters a librational regime.

\begin{figure}[h]
\centering
\includegraphics[width=.99\linewidth]{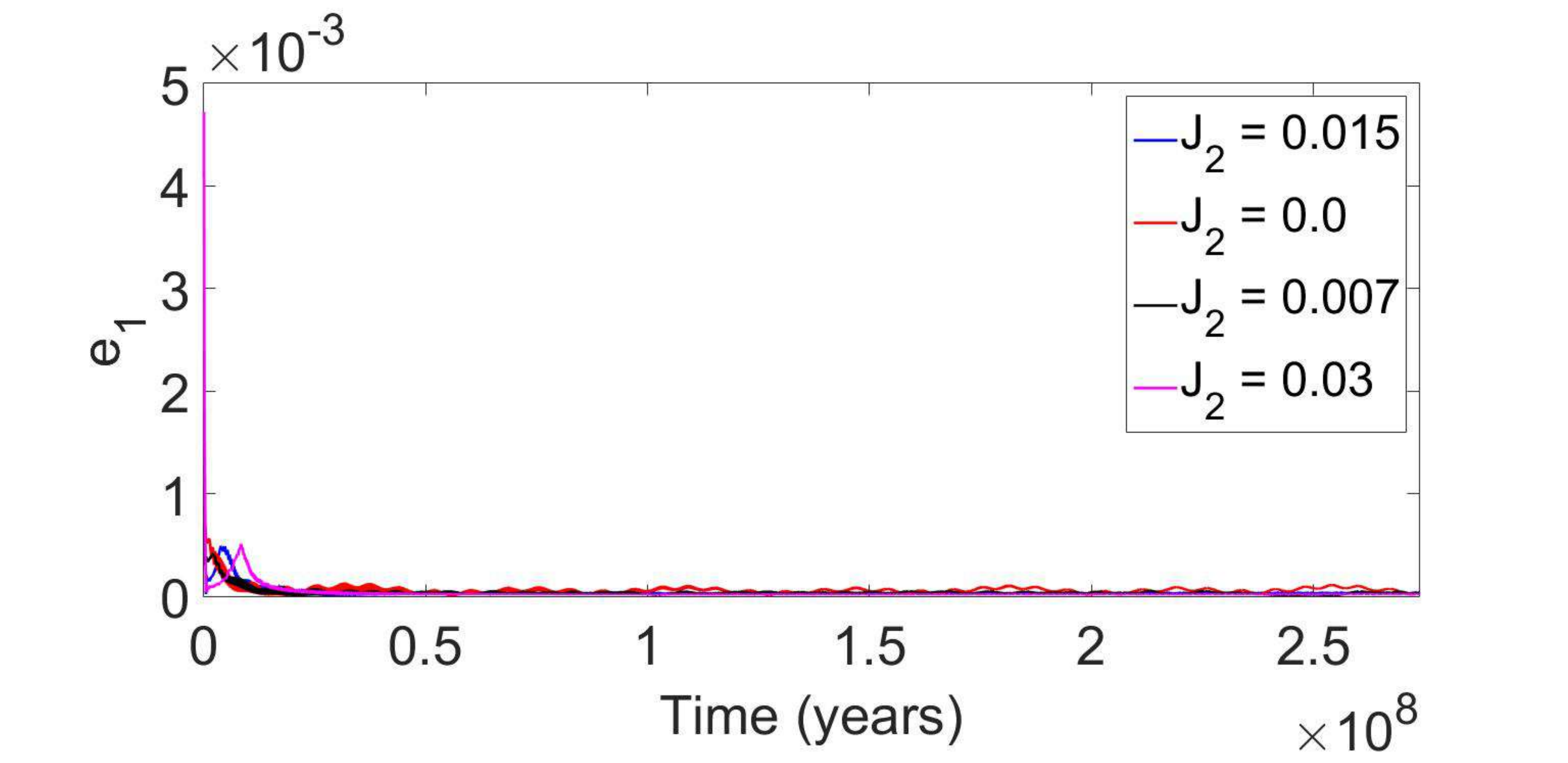}
\includegraphics[width=.99\linewidth]{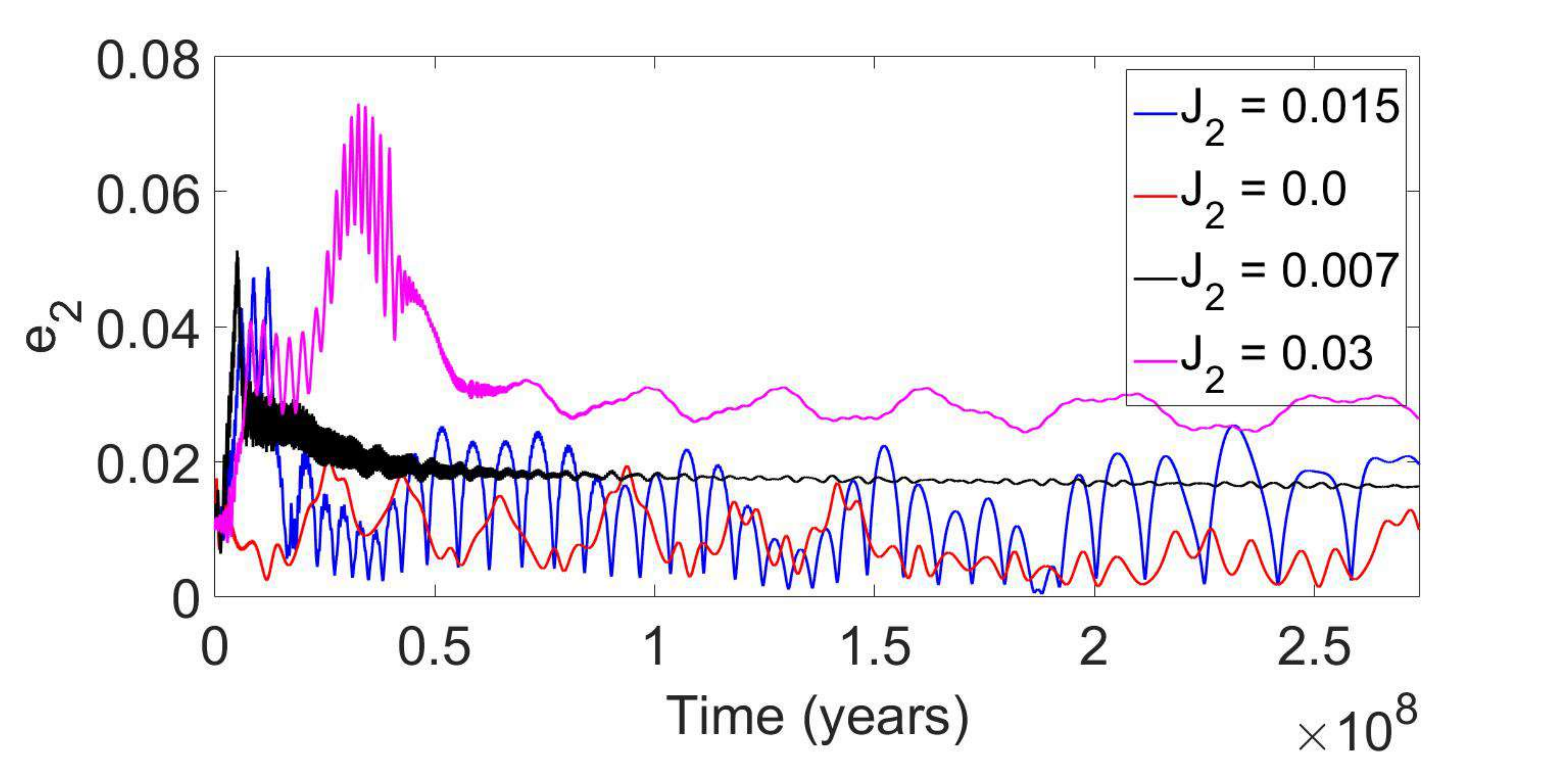}
    \caption{Evolution of the eccentricity of $S_1$ in the 3:1\&3:1 resonance (top panel) and $S_2$ in the 2:1\&3:1 (bottom panel) from the dissipative model using four different $J_2$ values of the central planet when the factor multiplying the tidal dissipation is $\alpha = 10^5$.}\label{J2test}
\end{figure}

In the case of the second-order resonances, the effect of the
oblateness of the central planet can be observed in the evolution of the
eccentricities of the three satellites. In the case of the
3:1\&3:1 resonance, the eccentricity of the inner satellite
converges to zero for all three values of $J_2$. When the $J_2$
value is equal to zero, an oscillation with a very small amplitude
is observed in the case of the eccentricity of $S_1$, while the
eccentricities of $S_2$ and $S_3$ oscillate with a larger
amplitude. When the $J_2$ value is different from zero, the
eccentricities of $S_2$ and $S_3$ oscillate around lower
values with shorter frequencies (see Figure~\ref{J2test}).

In the case of the 2:1\&3:1 resonance, the eccentricity of $S_1$ does not converge to zero when the $J_2$ value of the central body is different from the actual value. However, for all four values of $J_2$ we have taken as sample, the eccentricity oscillates around a value that is lower than the initial eccentricity. The eccentricities of $S_2$ and $S_3$ oscillate with a higher amplitude when the perturbation due to the oblateness of the central planet is not considered. 

\subsubsection{Dependence on the initial value of the semimajor axis}

In this section, we investigate the sensitivity in the evolution
of the orbital elements of the satellites on the initial value of
the semimajor axis of the inner satellite by varying it using the
relation $a_0 (1 \pm 10^{-3})$.

\begin{figure}[h]
\centering
\includegraphics[width=.99\linewidth]{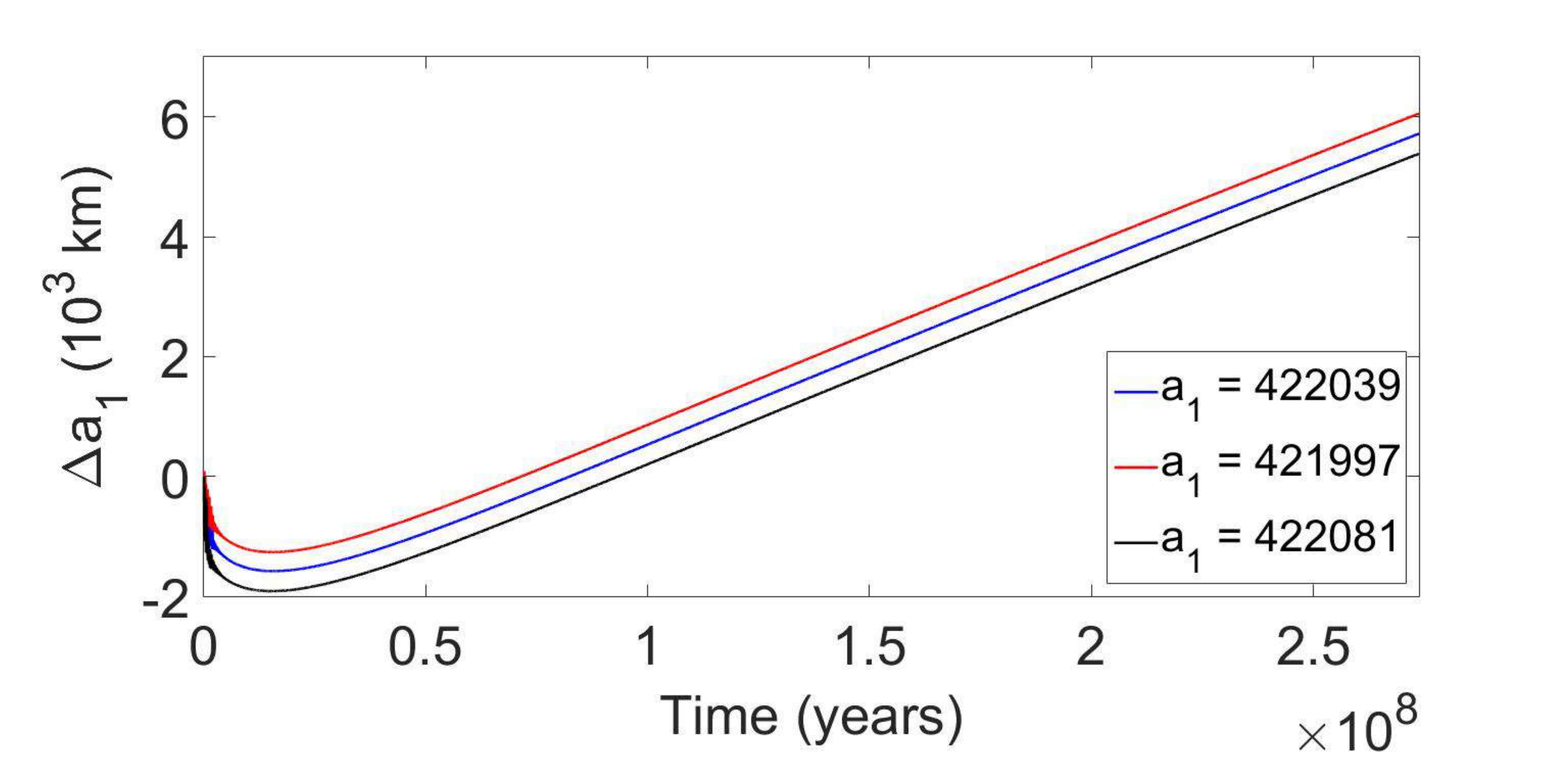}
\includegraphics[width=.99\linewidth]{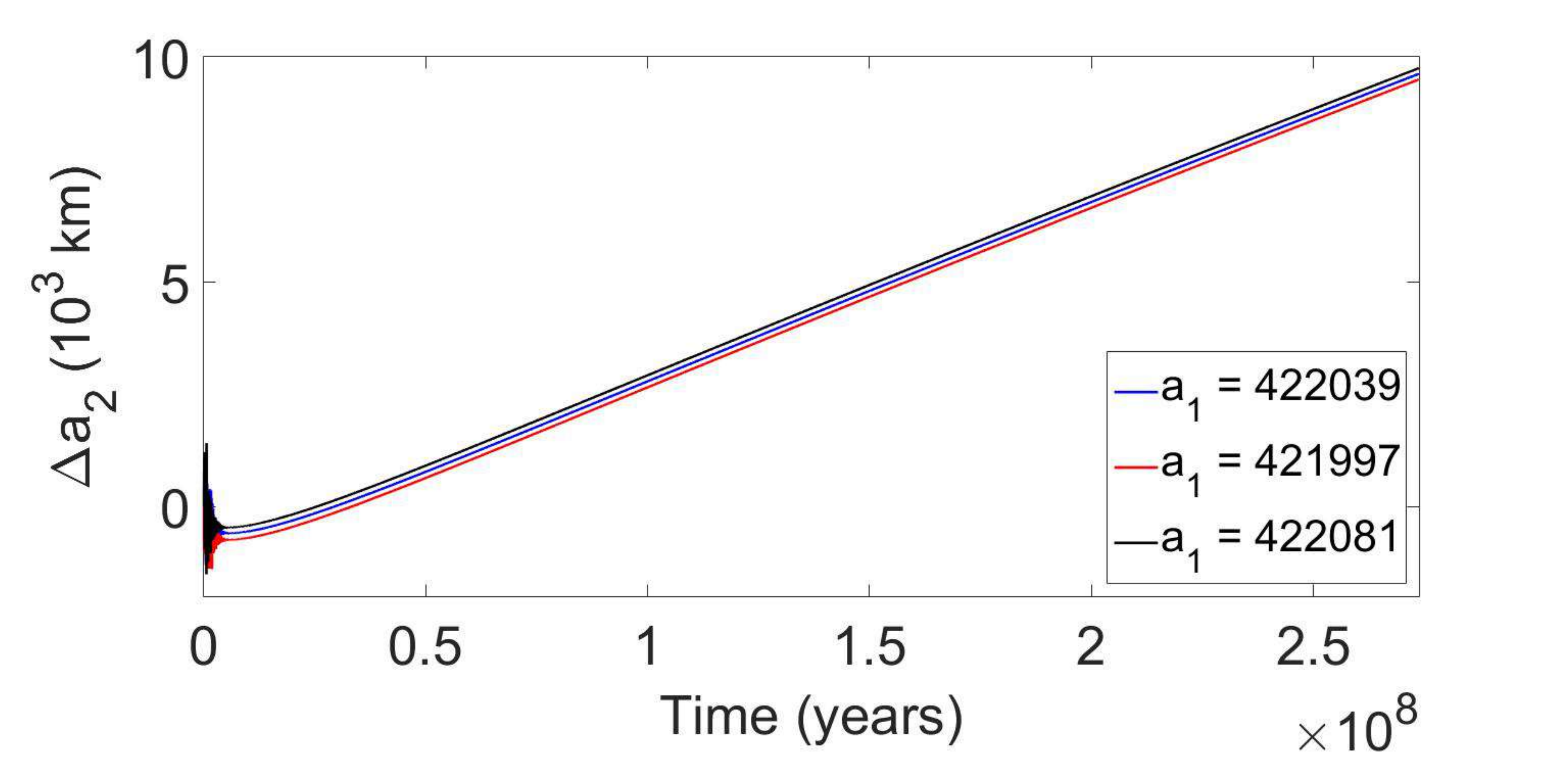}
\includegraphics[width=.99\linewidth]{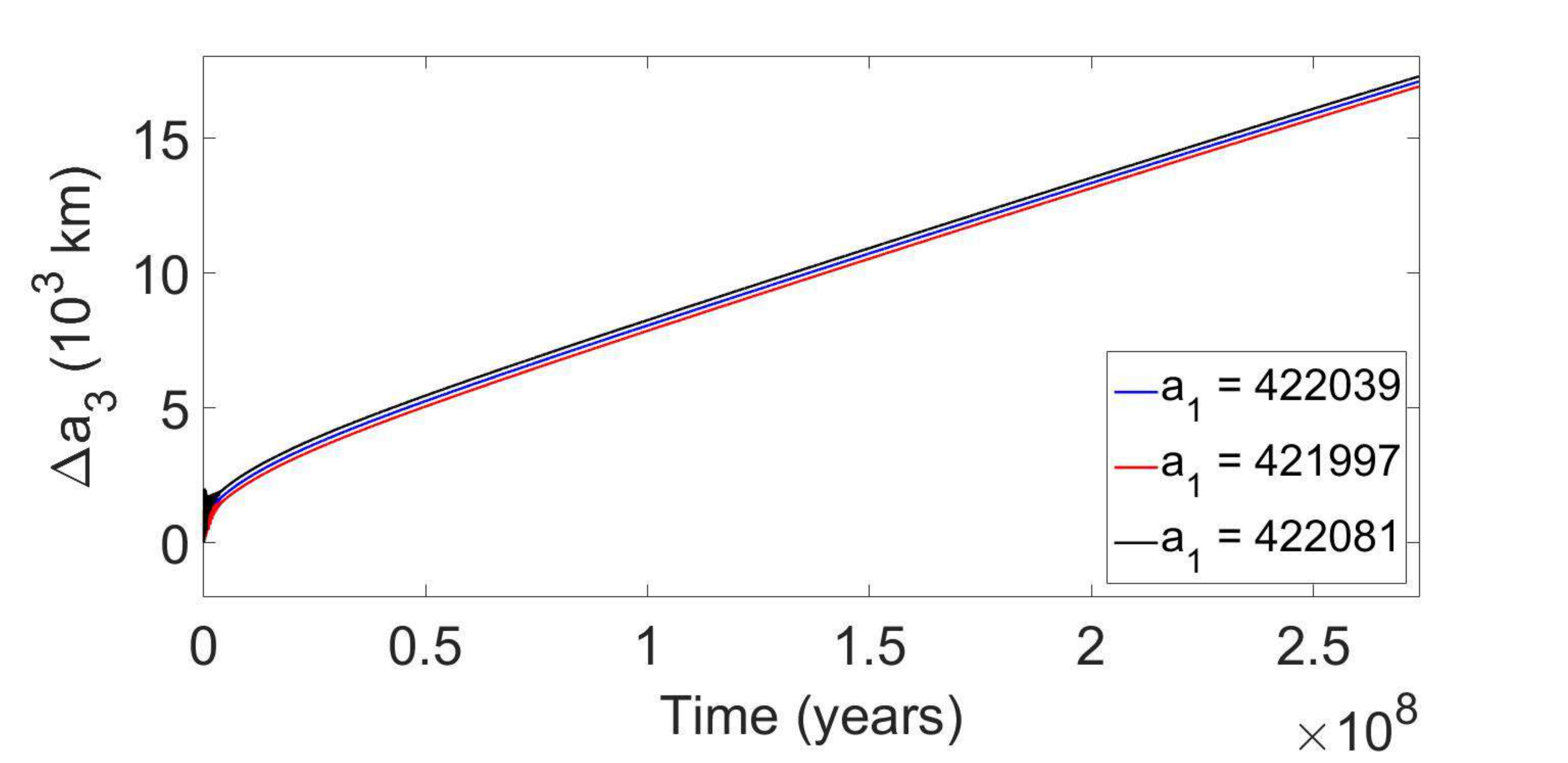}
    \caption{Variation in semimajor axes of $S_1$ (top panel),
    $S_2$ (middle panel), and $S_3$ (bottom panel) from the dissipative model assuming a
    3:2\&3:2 resonance and using three different initial values of semimajor axis of $S_1$
    when the factor multiplying the tidal dissipation is $\alpha = 10^5$. The tidal model assumes
    that the central body rotates fast. The different initial values of the semimajor axis of $S_1$ are $a_{10} = 422039$ (the nominal one), $a_{10} = 421997,$ and $a_{10} = 422081$.}\label{3232_a1test}
\end{figure}

In the case of the 3:2\&3:2 resonance, the semimajor axes of the three satellites increase, with the exception of the semimajor axis of the inner satellite, which decreases at the beginning and then increases. When the initial value of the semimajor axis of the inner satellite is lower, the initial decrease is larger and reaches lower values. On the other hand, the semimajor axes of $S_2$ and $S_3$ increase with a higher rate, using the lower initial value of the semimajor axis of the inner satellite (see Figure \ref{3232_a1test}).

The semimajor axes of the three satellites in the case of the 2:1\&3:2 resonance increase, and only the semimajor axis of the inner satellite shows a slight decrease at the beginning.

\begin{figure}[h]
\centering
\includegraphics[width=.99\linewidth]{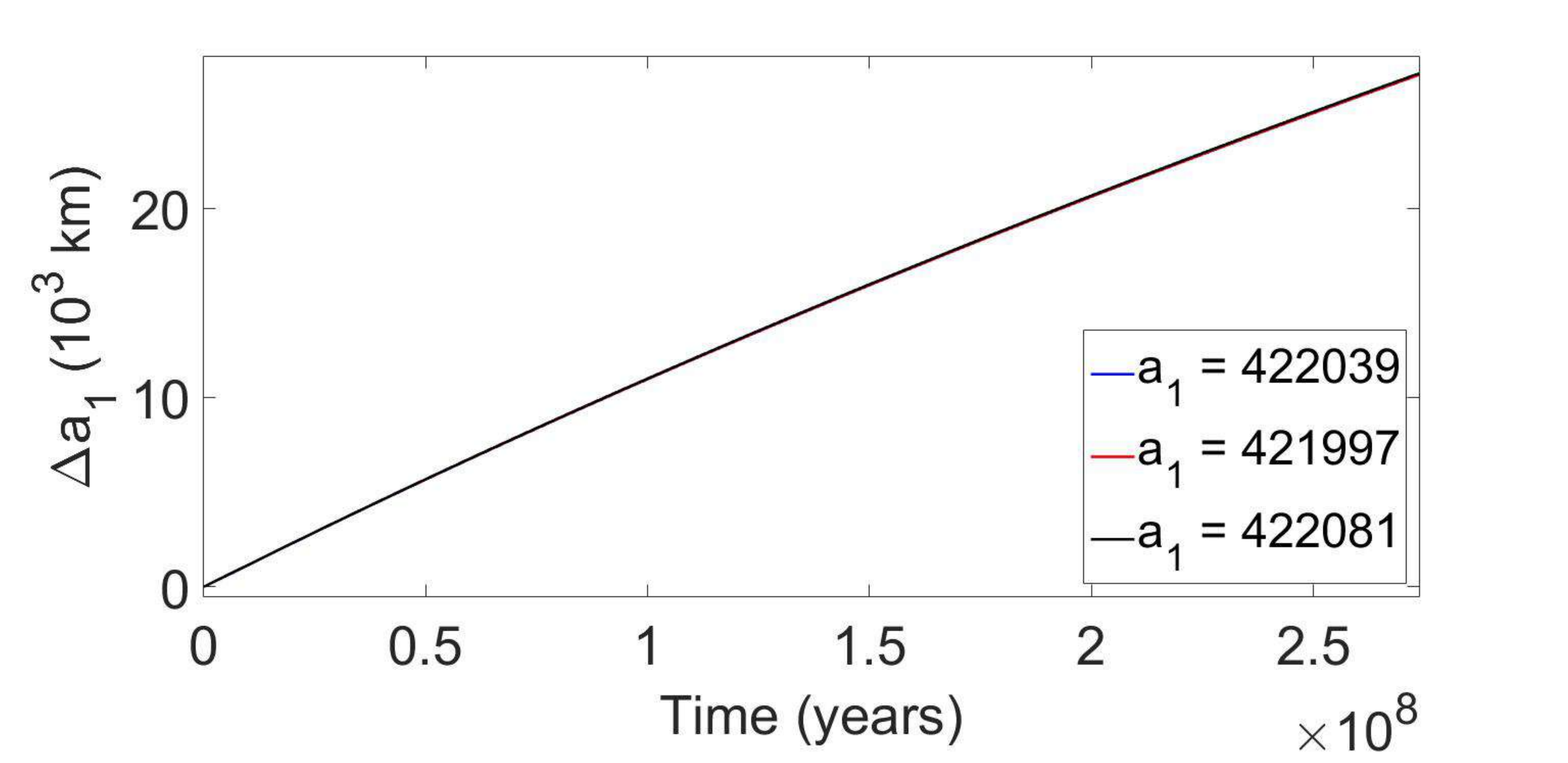}
\includegraphics[width=.99\linewidth]{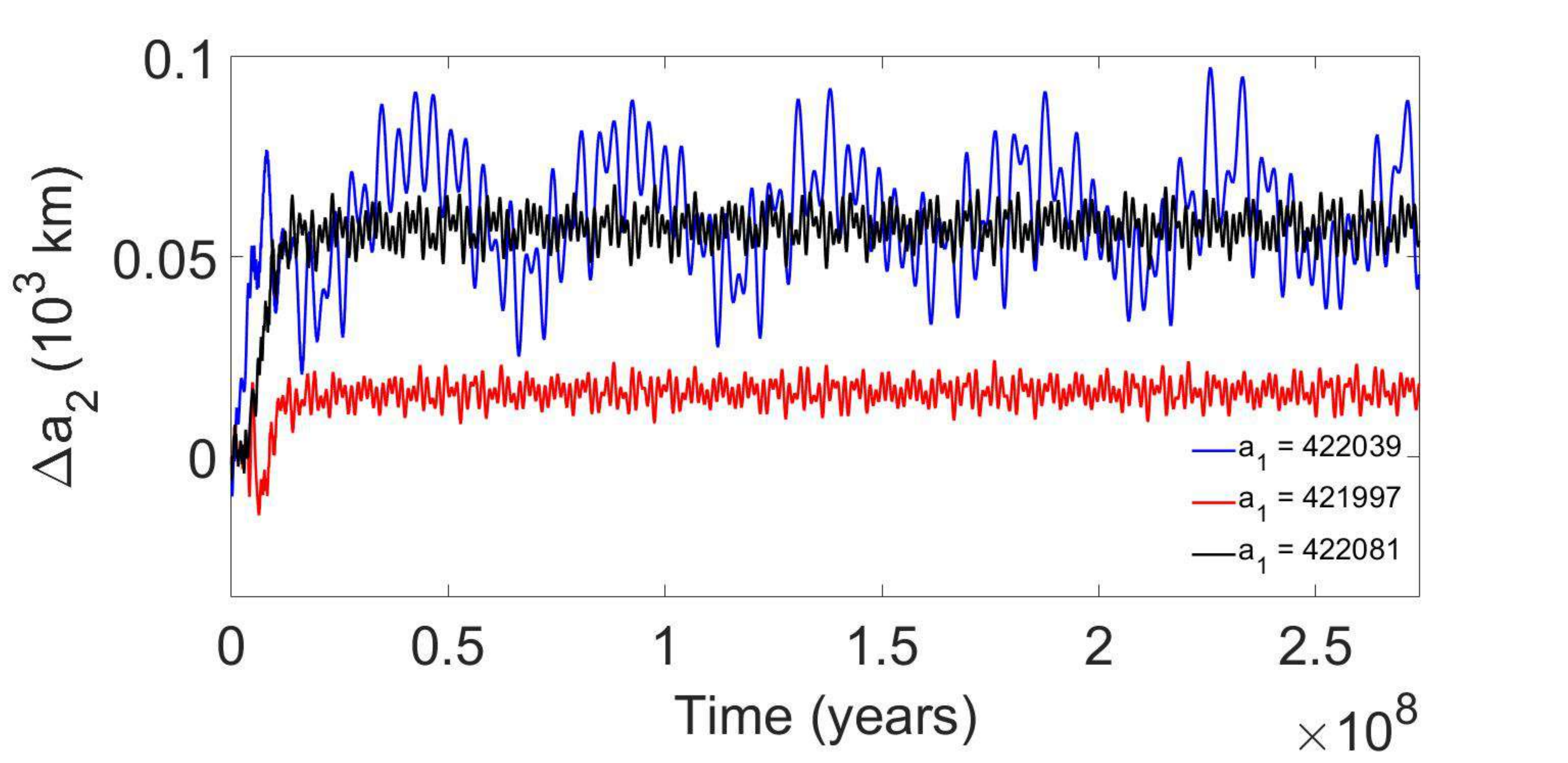}
\includegraphics[width=.99\linewidth]{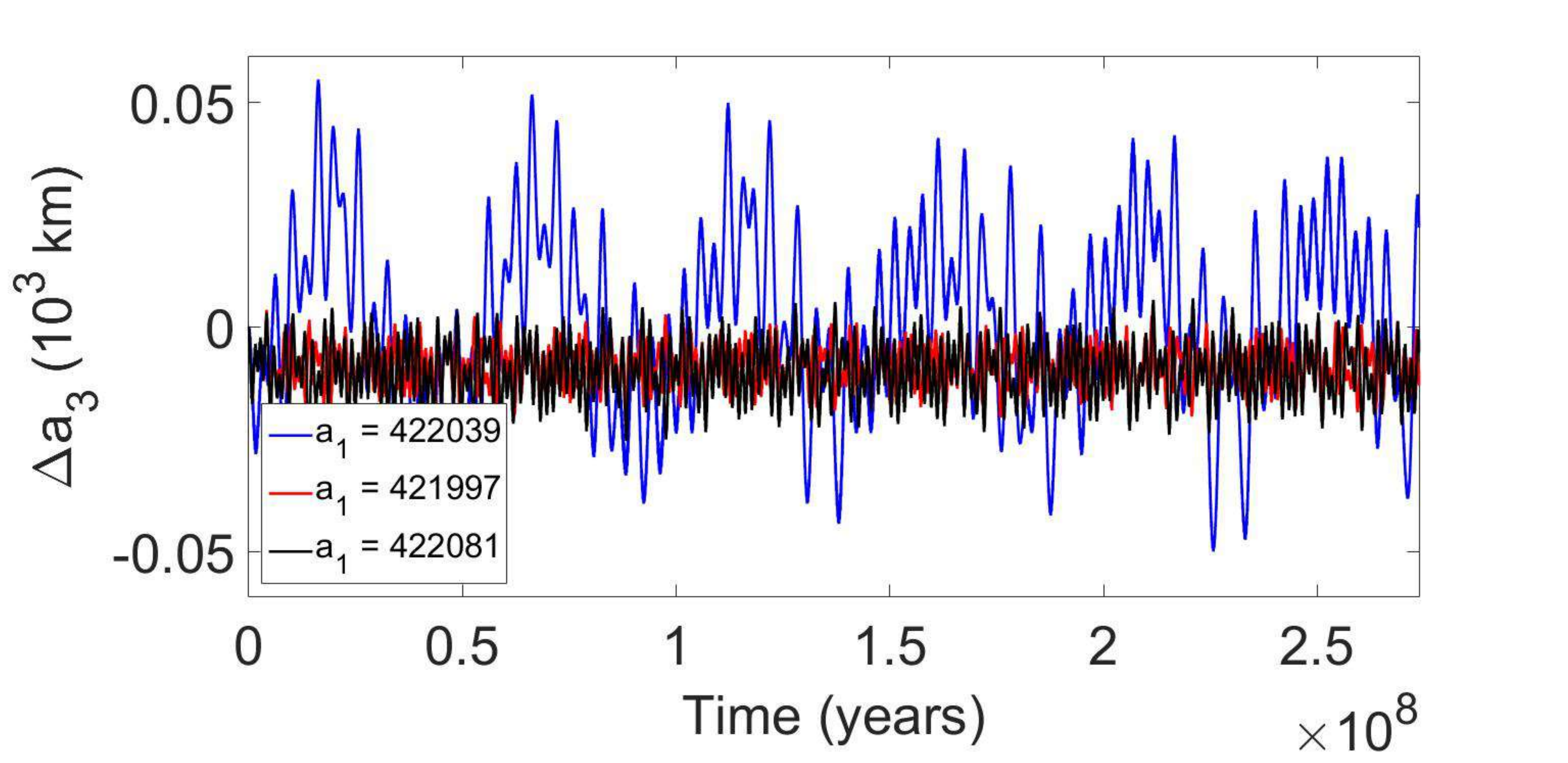}
    \caption{Variation in semi-major axes of $S_1$ (top panel), $S_2$ (middle panel), and $S_3$ (bottom panel) from the dissipative model assuming a 3:1\&3:1 resonance and using three different initial values of semimajor axis of $S_1$ when the factor multiplying the tidal dissipation is $\alpha = 10^5$. The tidal model assumes that the central body rotates fast. The different initial values of the semimajor axis of $S_1$ are $a_{10} = 422039$ (the nominal one), $a_{10} = 421997,$ and $a_{10} = 422081$.}\label{3131_a1test}
\end{figure}

A different behavior is observed for second-order resonances.
The semimajor axis of the inner satellite in the case of
the 3:1\&3:1 resonance increases, and the same effect is
observed even when the initial value is changed. The semimajor axes of
$S_2$ and $S_3$ remain constant and oscillate around a given
value. The amplitude of the oscillations is larger when the
initial value of the semimajor axis of the inner satellite is
larger (see Figure \ref{3131_a1test}).

The semimajor axis of the inner satellite in the case of the
2:1\&3:1 resonance increases even when the initial value of
$a_{1}$ is changed according to $a = a_0 (1 \pm 10^{-3})$. The semimajor
axes of $S_2$ and $S_3$ oscillate with a larger amplitude when
the initial value of $a_1$ is the nominal one.

\subsubsection{Dependence on the initial value of the eccentricity}

In this section, the sensitivity in the orbital evolution on the
initial value of the eccentricity of the inner satellite is
investigated. In the first-order resonances, all three
eccentricities converge to a value or oscillate around a value
with a very small amplitude. Instead, in the second-order
resonances, the eccentricity of $S_1$ tends to zero, while the
eccentricities of the other two satellites oscillate around
certain values. The initial value of the eccentricity of the inner
satellite is varied according to the following relation:
$e = e_0 (1 \pm 10^{-2})$.

In the case of the 3:2\&3:2 resonance, the eccentricities of the
three satellites converge to certain values. When the initial
value of the eccentricity of the first satellite is lowest,
the eccentricities oscillate for a longer period of time.
However, the evolution is the same in the three cases on a long
timescale.

In the case of the 2:1\&3:2 resonance, the eccentricities of the
three satellites oscillate with a high amplitude when the initial
eccentricity of the inner satellite is different from the nominal
one of $S_1$.

\begin{figure}[h]
\centering
\includegraphics[width=.99\linewidth]{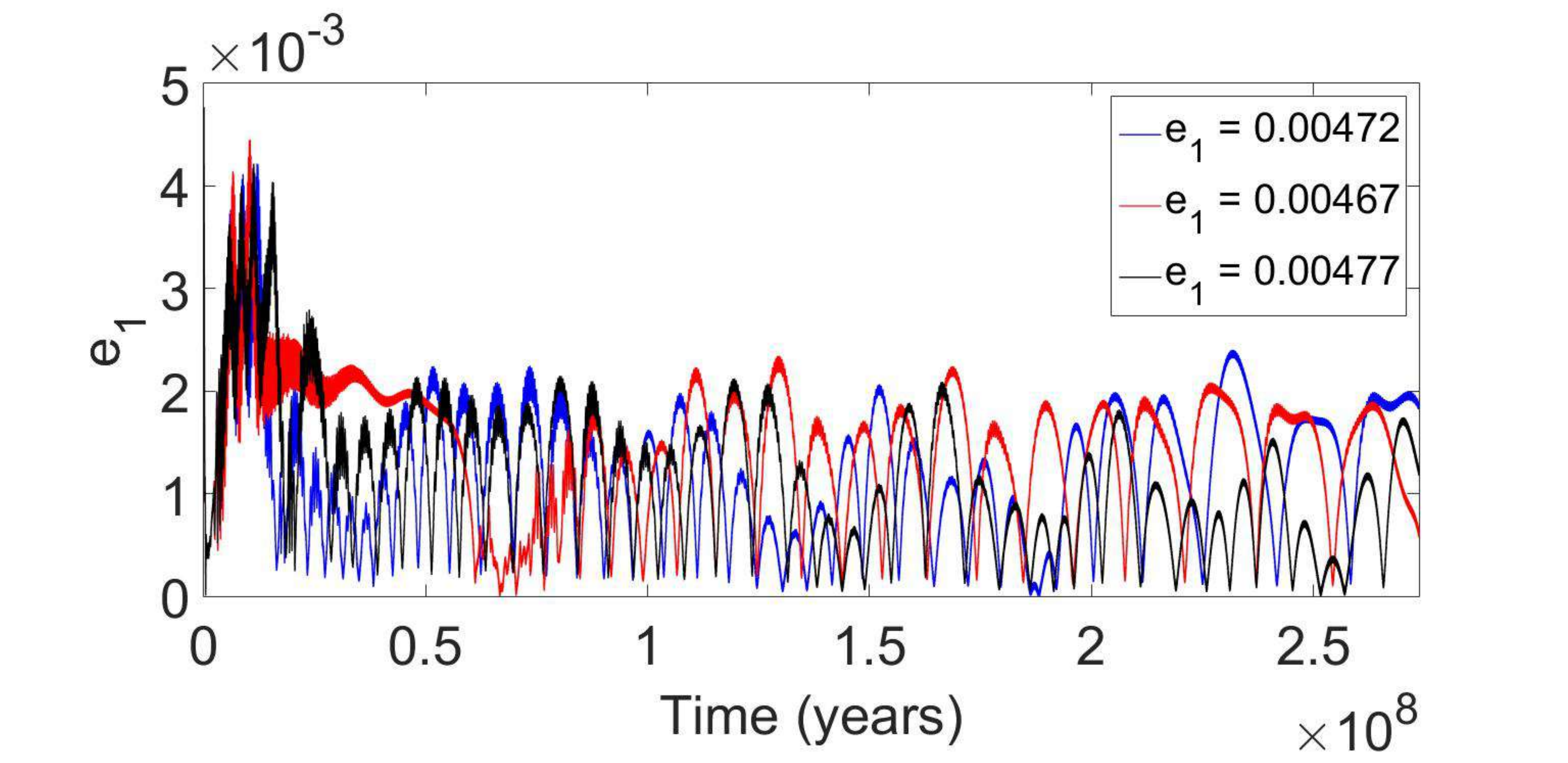}
\includegraphics[width=.99\linewidth]{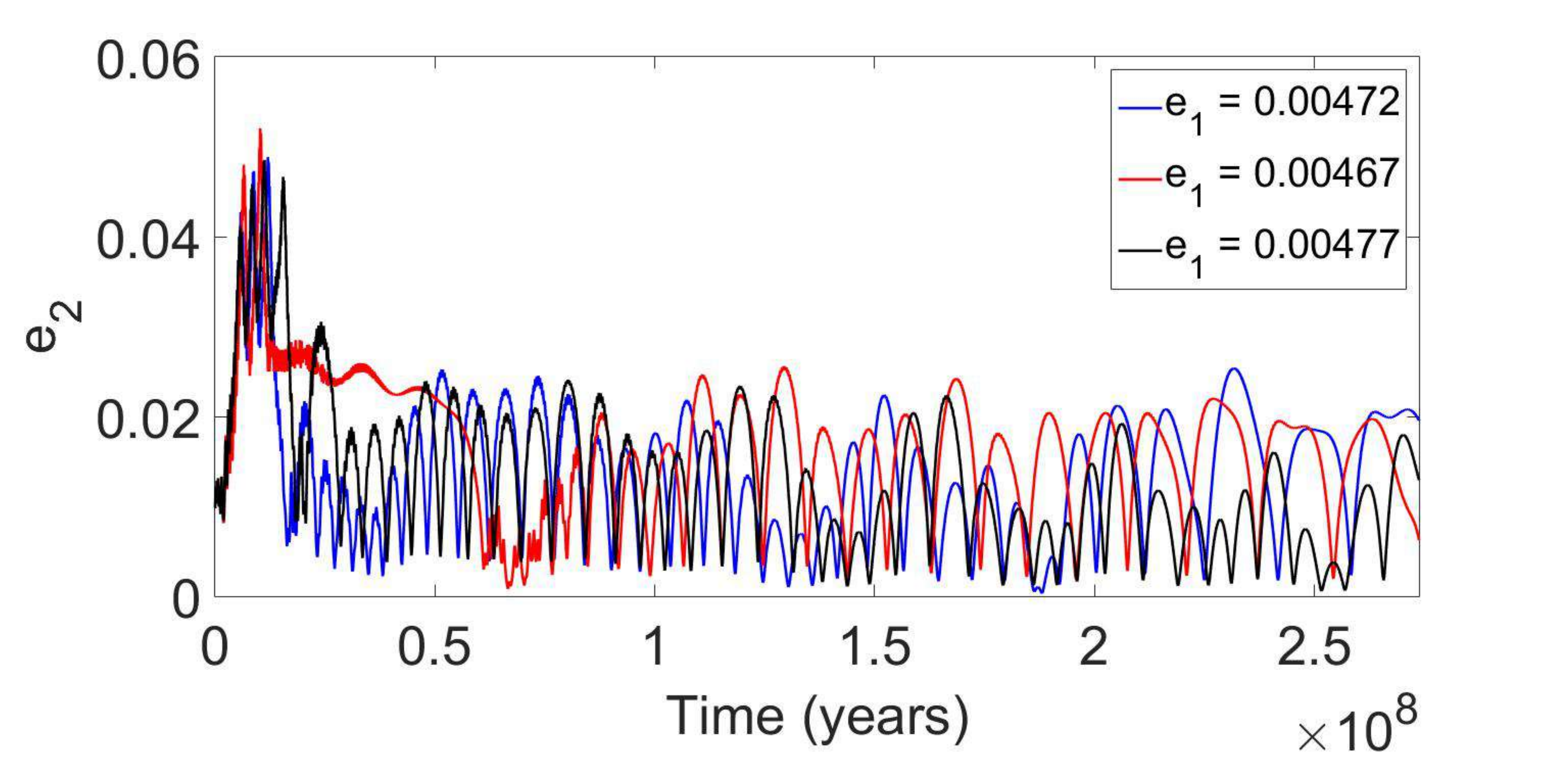}
\includegraphics[width=.99\linewidth]{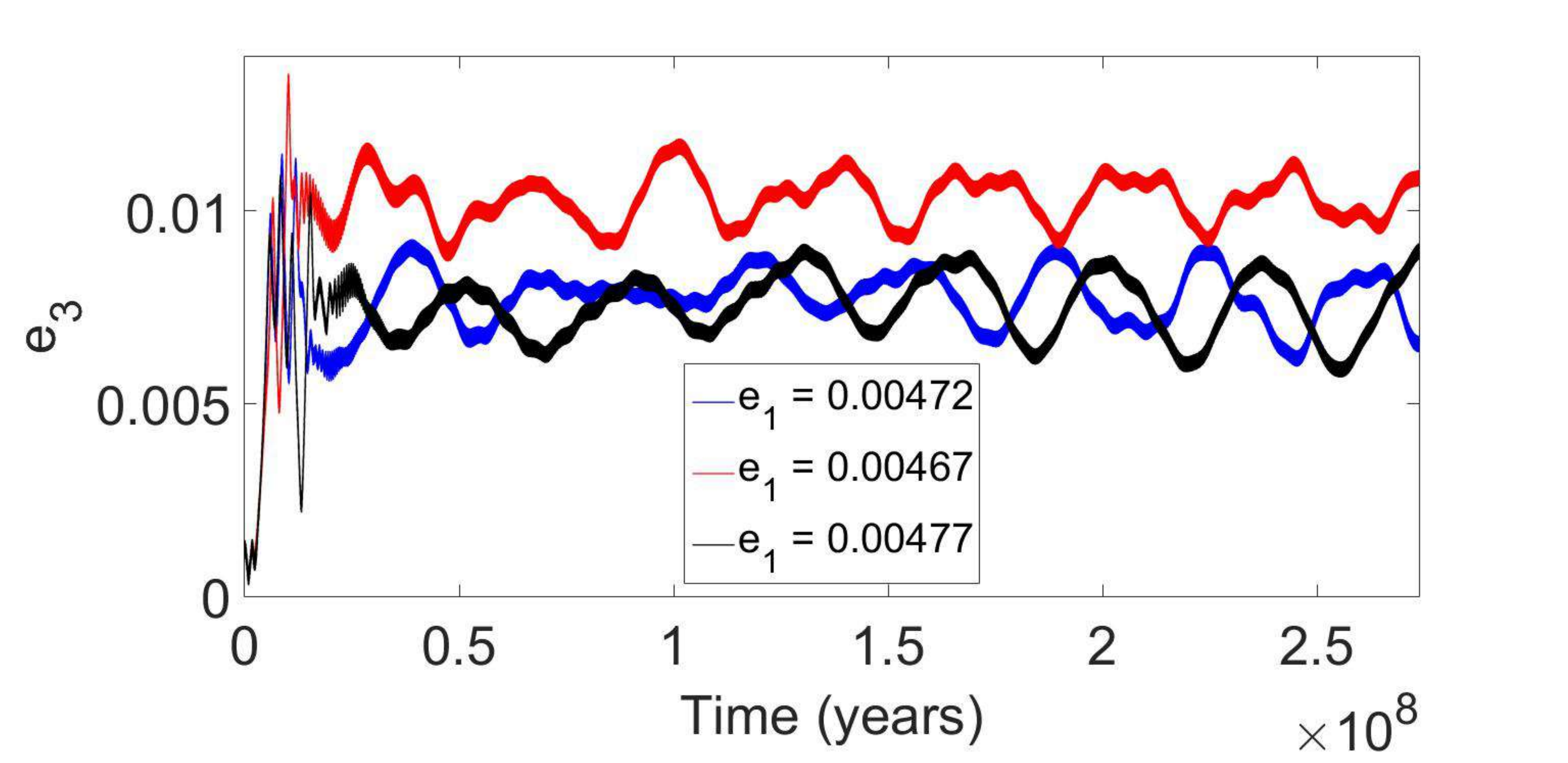}
    \caption{Evolution of the eccentricities of $S_1$ (top panel), $S_2$ (middle panel), and $S_3$ (bottom panel) from the dissipative model assuming 2:1\&3:1 resonance and using three different initial values of the eccentricity of $S_1$ when the factor multiplying the tidal dissipation is $\alpha = 10^5$. The different values of the eccentricity of $S_1$ are $e_1 = 0.00472$ (the nominal one), $e_1 = 0.00467,$ and $e_1 = 0.0477$ } \label{2131_e1test}
\end{figure}

The eccentricity of the inner satellite in the case of the 2:1\&3:1
resonance converges to zero using all three different initial
values of $e_1$. The eccentricities of $S_2$ and $S_3$
oscillate around the same values with a higher amplitude when the
lowest initial value of $e_{1}$ is used (see Figure
\ref{2131_e1test}).

\section{Phase-space geometry and linear stability}\label{sec:linear}

In this section, we focus on the geometry of the phase space and the linear
stability in the vicinity of exact resonant solutions including the tidal
effect.  We limit our study to the planar case with focus on first- and second-order resonances. We work with the resonant variables provided
in Table~\ref{tab:q}; we note that $(q_7,q_8,q_9)$ are not considered because
we consider the planar problem alone.

Let $H_r=H_r(\vec{q},\vec{Q})$ be the Hamiltonian
\eqref{final-Ham} in resonant action-angle variables
$\vec{q}=(q_1,\dots,q_6)$, $\vec{Q}=(Q_1,\dots,Q_6)$, with $\vec{Q}$ being the
conjugated actions obtained from a suitable generating
function in the conservative problem. The dynamics in these
variables is determined by the system of equations:

\begin{eqnarray}
\label{rsB}
\dot q_k &=&\frac{\partial H_{r}}{\partial Q_k} \ , \quad k=1\dots6 \ , \nonumber \\
\dot Q_1 &=& \dot P_1 = -\frac{\partial H_r}{\partial p_1} + \frac{dP_1}{dt}\bigg\vert_T \nonumber \\
\dot Q_k &=& \dot P_k = -\frac{\partial H_r}{\partial p_k} \ , \quad k=2,3 \ , \nonumber \\
\dot Q_k &=& F_k(\vec{q},\vec{Q})  \ .
\quad k=4,5,6 \ ,
\end{eqnarray}
where $\frac{dP_1}{dt} \big|_T$ denotes the counterpart of the dissipative
contributions \eqref{fast-rotating} or \eqref{slow-rotating} expressed in modified
Delaunay variables.  The functions $F_k=F_k(\vec{q},\vec{Q})$ are defined
for first-order resonances as listed below:

\vskip.1in

{\bf 2:1 \& 2:1}:
\begin{eqnarray}
\label{rsC1}
F_4 &=& \frac{1}{3}\left(2\dot L_1+\dot L_2-\dot L_3\right) \ , \nonumber \\
F_5 &=& \frac{1}{3}\left(\dot P_1+\dot P_2+\dot P_3+3\dot L_1+\dot L_2\right) \ , \nonumber \\
F_6 &=& \dot L_1+\dot L_2+\dot L_3-\dot P_1-\dot P_2-\dot P_3 \ ,
\end{eqnarray}

{\bf 2:1 \& 3:2}:
\begin{eqnarray}
\label{rsC2}
F_4 &=& \frac{1}{4}\left(2\dot P_1+2\dot P_2-2\dot P_3 - \dot L_2\right) \ , \nonumber \\
F_5 &=& \frac{1}{4}\left(2\dot P_1+2\dot P_2+2\dot P_3+4\dot L_1+\dot L_2\right) \ ,
\nonumber \\
F_6 &=& \dot L_1+\dot L_2+\dot L_3-\dot P_1-\dot P_2-\dot P_3 \ .
\end{eqnarray}

{\bf 3:2 \& 3:2}:
\begin{eqnarray}
\label{rsC3}
F_4 &=& \frac{1}{5}\left(-3\dot P_1-3\dot P_2+2\dot P_3 + \dot L_2 \right) \ , \nonumber \\
F_5 &=& \frac{1}{5}\left(4\dot P_1+4\dot P_2 + 4\dot P_3+5\dot L_1+2\dot L_2 \right) \ ,
\nonumber \\
F_6 &=& \dot L_1+\dot L_2+\dot L_3-\dot P_1-\dot P_2-\dot P_3 \ .
\end{eqnarray}

In the above equations, \eqref{rsC1} - \eqref{rsC3},
the derivatives $\dot L_1$, $\dot L_2$, $\dot L_3$, $\dot P_1$, $\dot P_2$, and $\dot P_3$
are obtained from the Hamiltonian flow, including the tidal effect on the first satellite:
\begin{eqnarray}
\label{rsD}
\dot L_1 &=& -\frac{\partial H_r}{\partial \lambda_1} + \frac{dL_1}{dt}\bigg\vert_T \nonumber \\
\dot L_k &=& -\frac{\partial H_r}{\partial \lambda_k} , \quad k=2,3 \ , \nonumber \\
\dot P_1 &=& -\frac{\partial H_r}{\partial p_1} + \frac{dP_1}{dt}\bigg\vert_T , \nonumber \\
\dot P_k &=& -\frac{\partial H_r}{\partial p_k} \ , \quad k=2,3\ ,
\end{eqnarray}
where, again, $\frac{dL_1}{dt}\bigg\vert_T$ denotes the counterpart of the dissipative
contributions \eqref{fast-rotating} or \eqref{slow-rotating} in the variables
$(\lambda_k,p_k,L_k,P_k)$.

We remark that the derivatives of terms with respect to the modified Delaunay variables in Eqs.
\eqref{rsB}-\eqref{rsD} have to be expressed in the suitable set of
resonant variables (provided in Table~\ref{tab:q}). We also note that the resonant
system, including the nonconservative terms, is again independent of the resonant
angles $q_5$, $q_6$ because the nonconservative contributions only consist of secular
terms that are independent of these variables, as in the conservative case. Thus,
it suffices to investigate the reduced phase space $(q_1,\dots,q_4)$,
$(Q_1,\dots,Q_4)$. However, we note that in presence of tides, the
variables $Q_5$, $Q_6$ are not conserved quantities anymore, unless we
solve for initial conditions with $\dot Q_5 = \dot Q_6 = 0$, as is the 
case for fully resonant initial conditions.

\subsection{Equilibria of the system and linear stability}

An equilibrium of the vector field \eqref{rsB} in resonant
variables is determined by the system of equations
\begin{eqnarray}
\label{rsE}
\dot q_1 = \dot q_2 = \dot q_3 = \dot q_4 =
\dot Q_1 = \dot Q_2 = \dot Q_3 = \dot Q_4 = \dot Q_5 = \dot Q_6 = 0 \ .
\end{eqnarray}
We solve it by using a root finding algorithm with the starting values close to
the commensurability of the mean motions of the satellites.
We remark that 
the additional conditions $\dot Q_5 = \dot Q_6 = 0$ allow us to freeze 
the dynamics in phase space, which is done on purpose to reveal the 
structure of the phase space during capture, close to exact
resonant conditions, that is, valid only for a frozen moment in time.
We note that from a physical point of view, conditions (\ref{rsE}) are never exactly fulfilled and $\dot Q_5, \dot Q_6\neq0$
in general.

Let
$\vec{q}^{\,*}=(q_1^*,\dots,q_4^*)$, ${\vec{Q}}^*=(Q_1^*,\dots,Q_6^*)$ be quantities that solve Eq. \eqref{rsE}
for fixed values of the system parameters. The equilibria \eqref{rsE} depend on i)
the choice of the parameters (tidal effect, masses, etc.) and on ii)
the choice of the resonant variables. In the following, we also investigate the effect of the system parameters on the linear stability indices at the equilibrium value for each
resonance. If we denote the vector field \eqref{rsB} by
\begin{eqnarray}
\label{rsF}
\dot {\vec X} = \vec F(\vec X)
,\end{eqnarray}
with the vector $\vec X = (\vec{q},\vec{Q})$, the linear stability around the equilibrium
$\vec X^* = (\vec{q}^{\,*},{\vec{Q}}^*)$ is given by the eigenvalues of the Jacobian matrix
$\mathbf J$ evaluated at $\vec X^*$,

\begin{eqnarray}
\label{rsG}
\mathbf J = \left[
\frac{\partial \vec F}{\partial X_1} \dots
\frac{\partial \vec F}{\partial X_{12}}
\right]\bigg\vert_{\vec X = \vec X^*} .
\end{eqnarray}

We evaluate $\mathbf J$ at the equilibrium $ (\vec{q}^{\,*},{\vec{Q}}^*)$ and
determine numerically the eigenvalues for different system parameters.  In the
purely conservative case (without tides), we find complex conjugated
eigenvalues, with zero real parts in all first-order
resonant cases. Including the tidal effects, we find nonzero real parts due to
the nonconservative effects.

We report projections of the phase space in the planes $(q_1,Q_1)$,
$(q_2,Q_2)$, $(q_3,Q_3)$, and $(q_4,Q_4)$ in Figure~\ref{f:pha} for the 2:1\&2:1 resonant case. The equilibria coincide with $ (\vec{q}^{\,*},{\vec{Q}}^*)$
obtained from Eq. \eqref{rsE}. Close to the equilibrium, the system oscillates around
the centers with increasing amplitudes and increasing distance from $
(\vec{q}^{\,*},{\vec{Q}}^*)$.

To obtain the limiting librational curves (in red) in the projections 
$(q_\ell, Q_\ell)$, with $\ell=1,..,4$ we make use 
of the following iterative approach: we start by integrating sets of initial 
conditions with $(q_k,Q_k)=(q_k^{\,*},Q_k^{\,*})$ with $k=1,...,4$, $k\neq\ell$ and 
$q_\ell=q_\ell^{\, *}$, $Q_\ell=Q_\ell^{\,*}+\delta Q_\ell$ with one $\delta_a=\delta Q_\ell$
to obtain a librational curve and a second $\delta_b=\delta Q_\ell$ that yields a solution
with $q_\ell\in[0,2\pi]$. Next we choose a fixed number of equally spaced initial conditions 
within the interval $\delta Q_\ell\in[\delta_a,\delta_b]$ and keep at each iteration 
i) the largest $\delta Q_\ell$ that still yields librational motion and identify it with 
the new $\delta_a$, and ii) the smallest $\delta Q_\ell$ that still yields rotational motion and 
identify it with the new $\delta_b$. We iterate the process until $\delta_b-\delta_a$
becomes sufficiently small and define the librational half-width to be 
$\delta Q_\ell=\delta_a$. We note that this method yields a numerical estimate of
the separatrix half-width in the plane $(q_4,Q_4)$, while for $\ell=1,2,3$ we obtainan
estimate of the distance of the last paradoxal curve (see, e.g., \cite{BR2001}) that 
does not take all values from zero to $2\pi$.

\subsection{Effect of tides and mass of $S_1$}

\begin{figure*}[h]
\centering
\includegraphics[width=.49\linewidth]{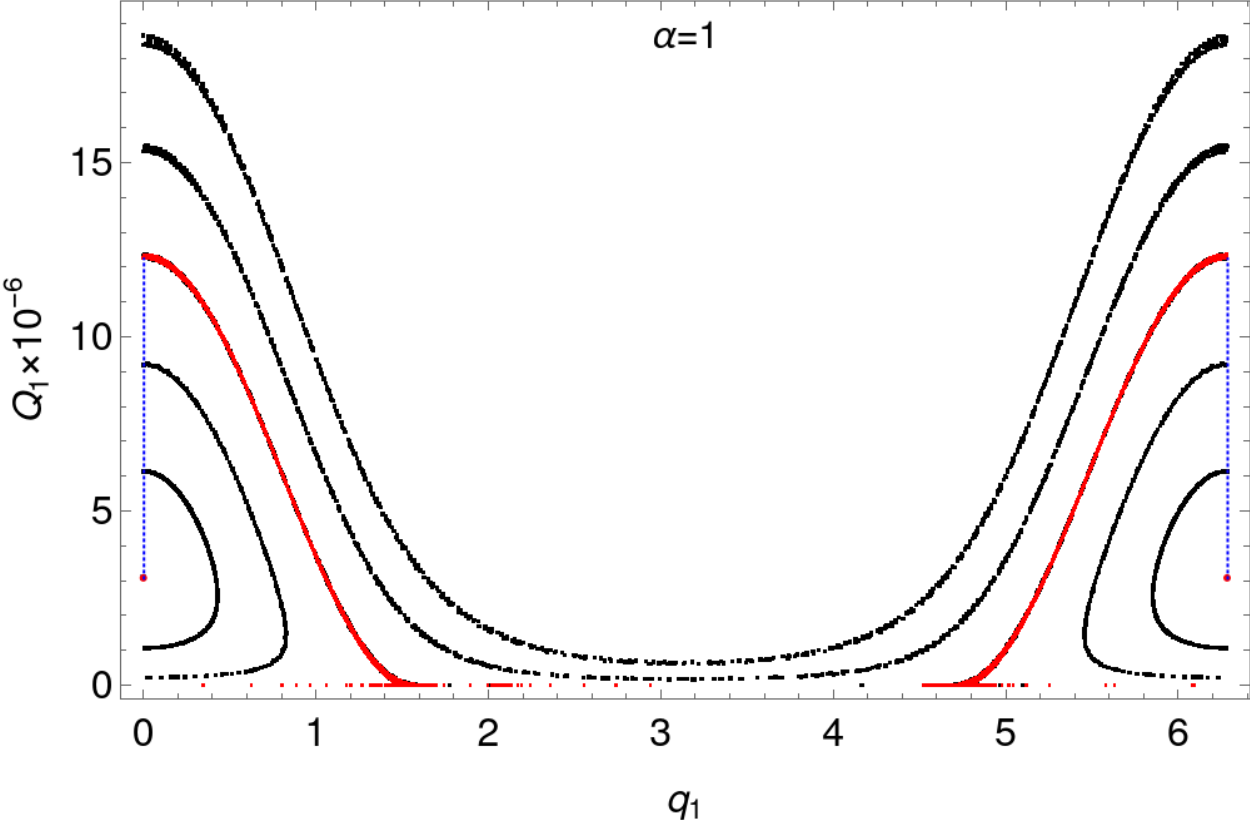}
\includegraphics[width=.49\linewidth]{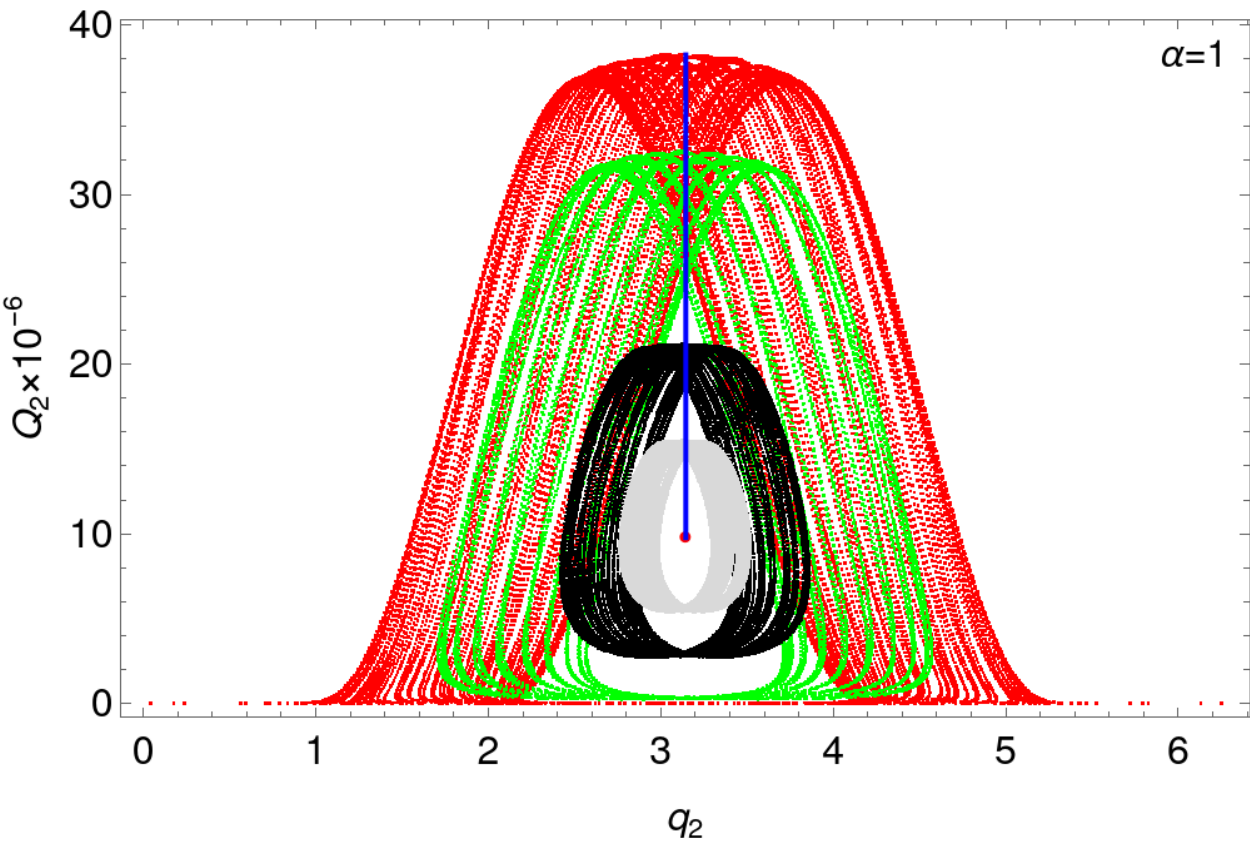}
\includegraphics[width=.49\linewidth]{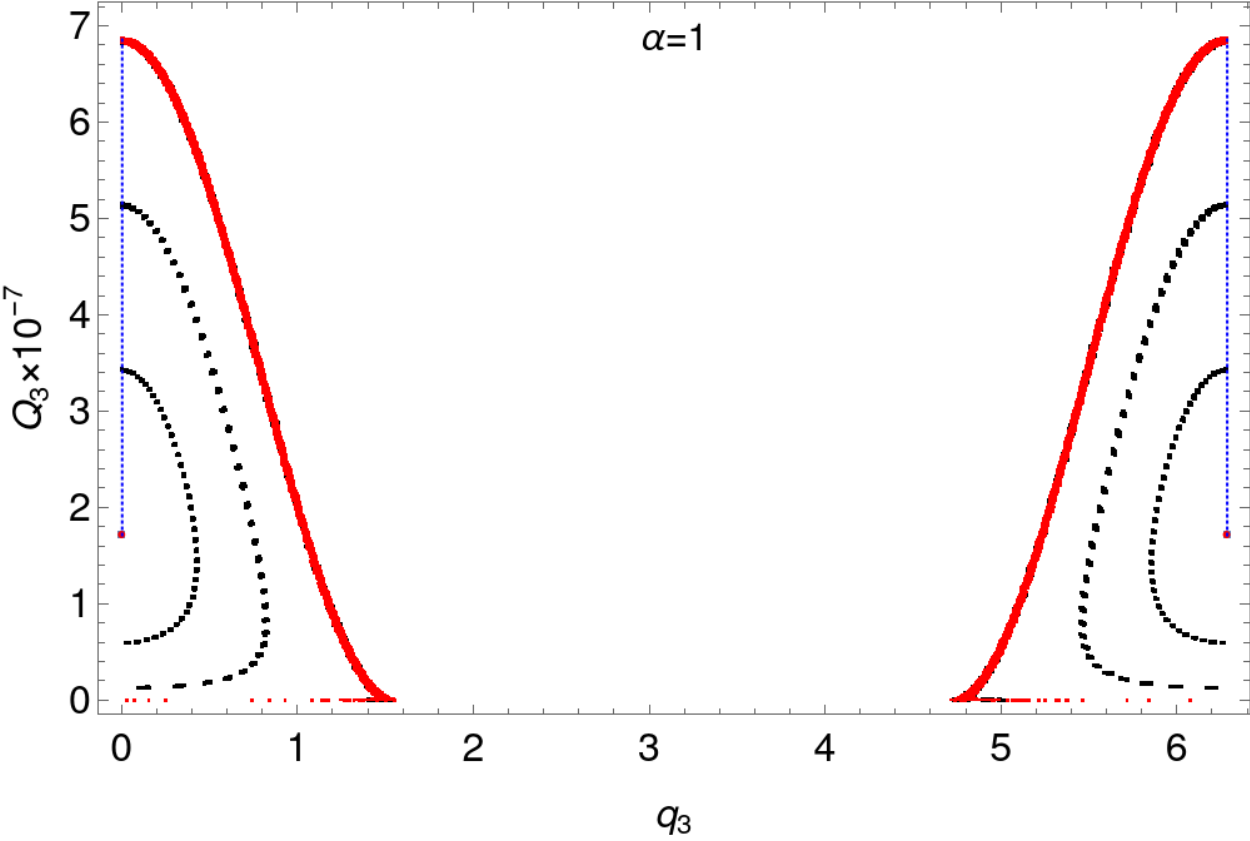}
\includegraphics[width=.49\linewidth]{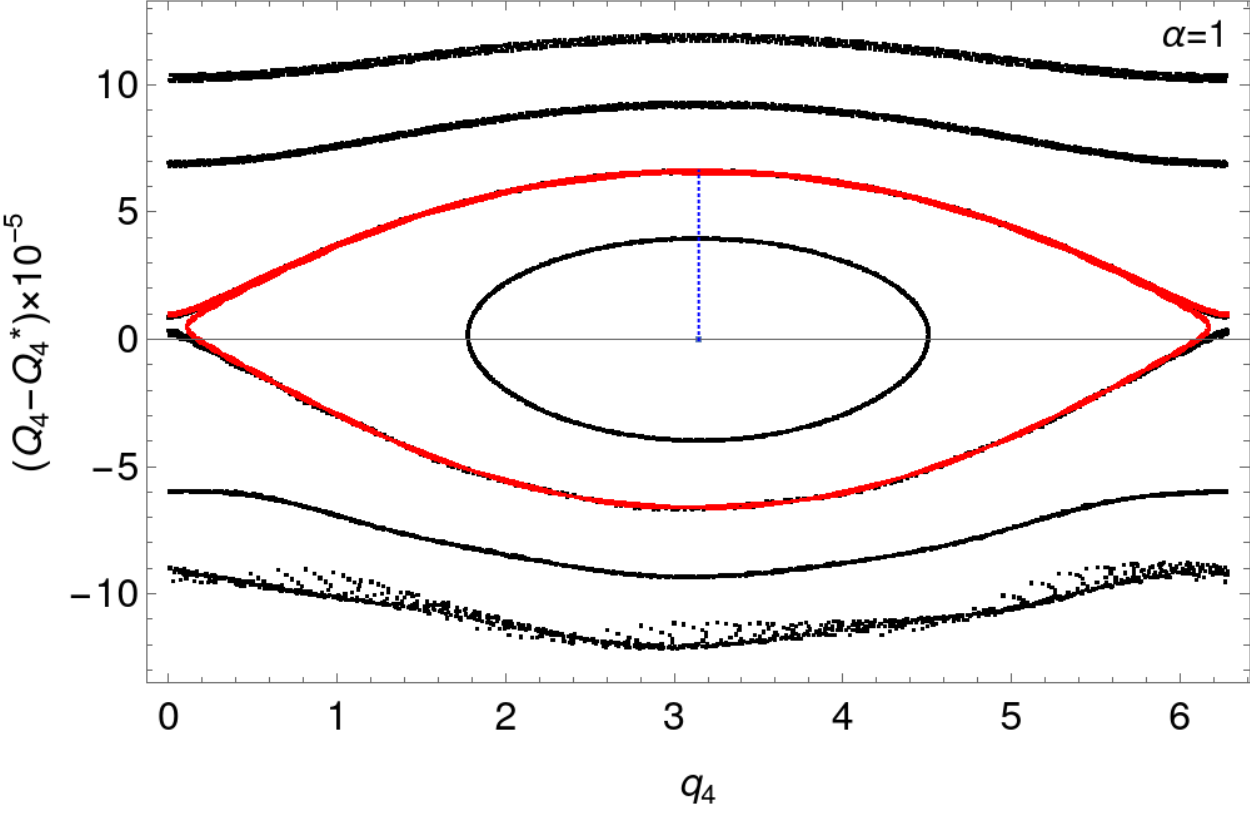}
\caption{Projection of the phase space onto the plane $(q_k,Q_k)$ with $k=1,\dots,4$
        for the 2:1\&2:1 resonant case for $\alpha=1$. Equilibria, separatrix, 
        and last
        paradoxal librational curve (see \cite{BR2001}) are shown in red, and the libration 
        half-width is shown as a dotted blue line.}\label{f:pha}
\end{figure*}

In order to reduce the integration time but at the same time
observe the tidal effects for a long physical time, we multiply
the variable $c$ in \eqref{eq:c} by a factor $\alpha$. This technique is used in
\cite{sho97} and \cite{lar20}, see also \cite{cklpv}. 
We confirmed that the real parts of the eigenvalues, associated
with Eq. (\ref{rsG}), do not change sign with increasing value of $\alpha\in(1,10^5)$.
Moreover, we verified that the equilibria are only slightly shifted 
by magnification of the tidal effect. We conclude that a change in $\alpha$
does not change the topology of the phase space from a qualitative point of
view. 

The study to obtain the equilibria and stability indices was done for
the 2:1\&2:1, 2:1\&3:2 and 3:2\&3:2 resonant cases,
taking the suitable resonant variables, see Table~\ref{tab:q} and Eq. \eqref{rsC2}
for the definition of the functions $F_4$--$F_6$ in \eqref{rsB}. We solve Eq.
\eqref{rsE} for the equilibria ($\vec{q}^{\,*},{\vec{Q}}^*$) and make use of Eq.
\eqref{rsG} to obtain the eigenvalues at equilibrium. We validate the results by a numerical integration of Eq.
\eqref{rsB} for different initial conditions and also calculate the $\delta
Q_k$ with $k=1,...,4$.  The projections on the planes $(q_1, Q_1), \dots, (q_4,
Q_4)$ are similar to those of Figure~\ref{f:pha} (not shown here).

\begin{figure}[h]
\centering
\includegraphics[width=.95\linewidth]{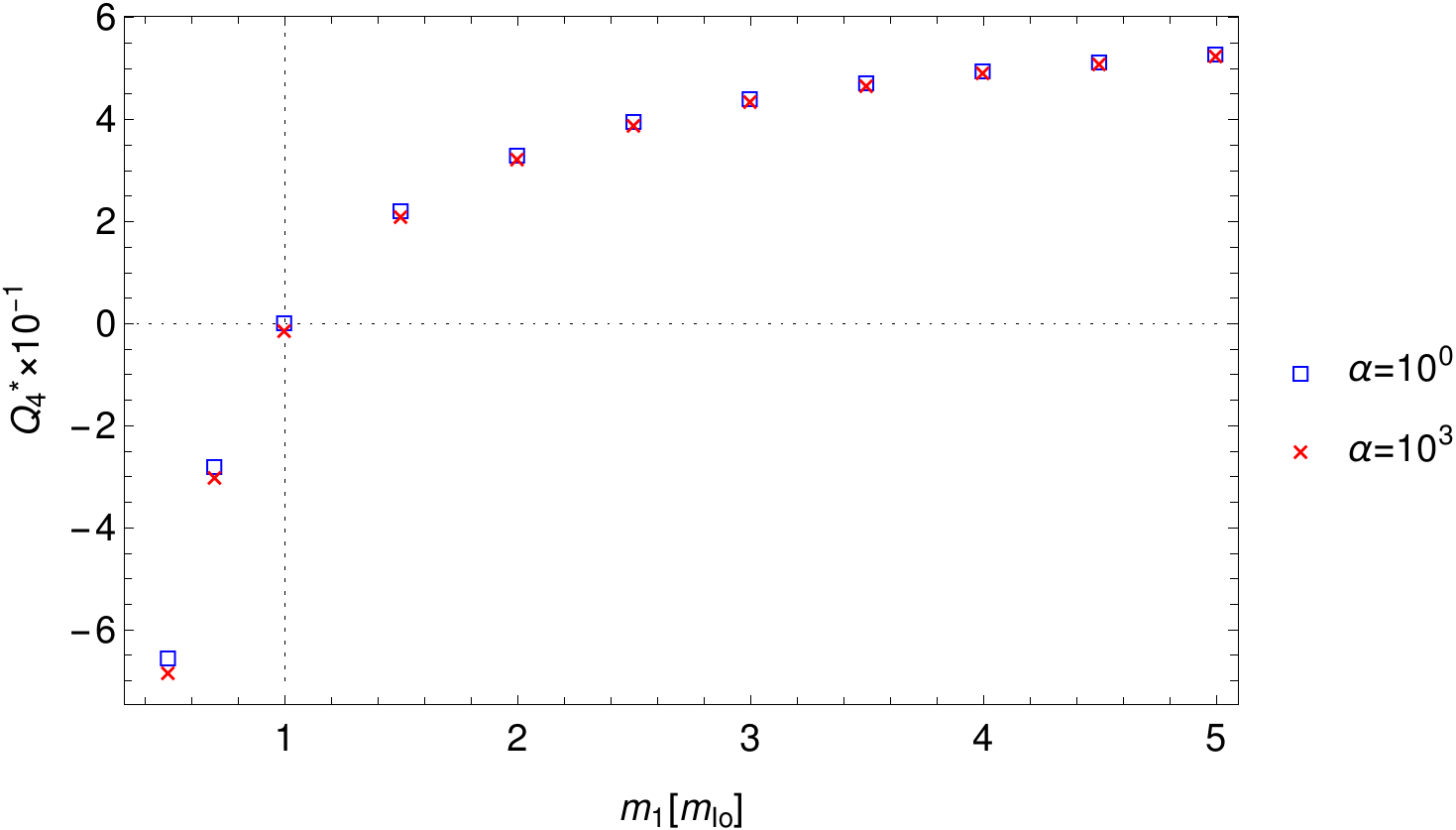}
    \caption{Equilibrium value $Q_4^*$ vs. $m_{1}$ (in units of Io's mass)
    for different $\alpha$ (2:1\&2:1 resonance). The reference values for the
        actual Galilean system are indicated by dashed lines.}\label{f:mIo}
\end{figure}

We provide the results in phase space in dependence on the
mass parameter of $S_1$. We repeat the study to
solve Eq. \eqref{rsE} with varying values of $m_{1}$. The results
are shown in Figure~\ref{f:mIo}, where we report the shift
in equilibrium value $Q_4^*$ versus the mass
$m_{1}$ ranging from $0.5\ m_1$ to $5\ m_1$. We clearly see
a dependence on the location of $Q_4^*$ in the projections
of the phase space $(q_4,Q_4)$ with increasing mass of
the innermost moon. We provide the study for case
$\alpha=10^0$ (in blue) and $\alpha=10^3$ (in red), the shift following
the same dependence on $m_{1}$ with slightly larger
deviations from each other for lower masses of $S_1$.

\begin{figure*}[h]
\centering
\includegraphics[width=.51\linewidth]{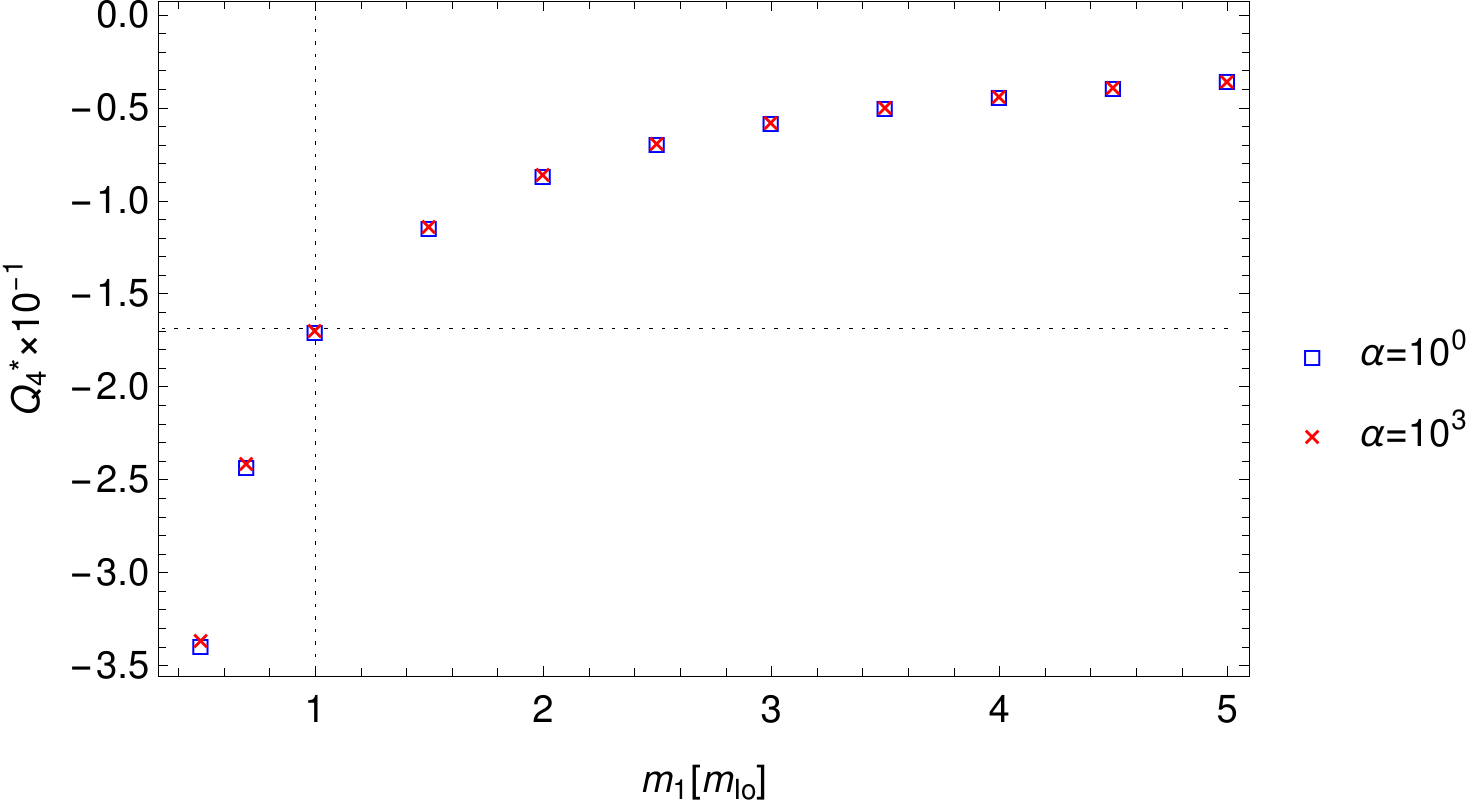}
\includegraphics[width=.48\linewidth]{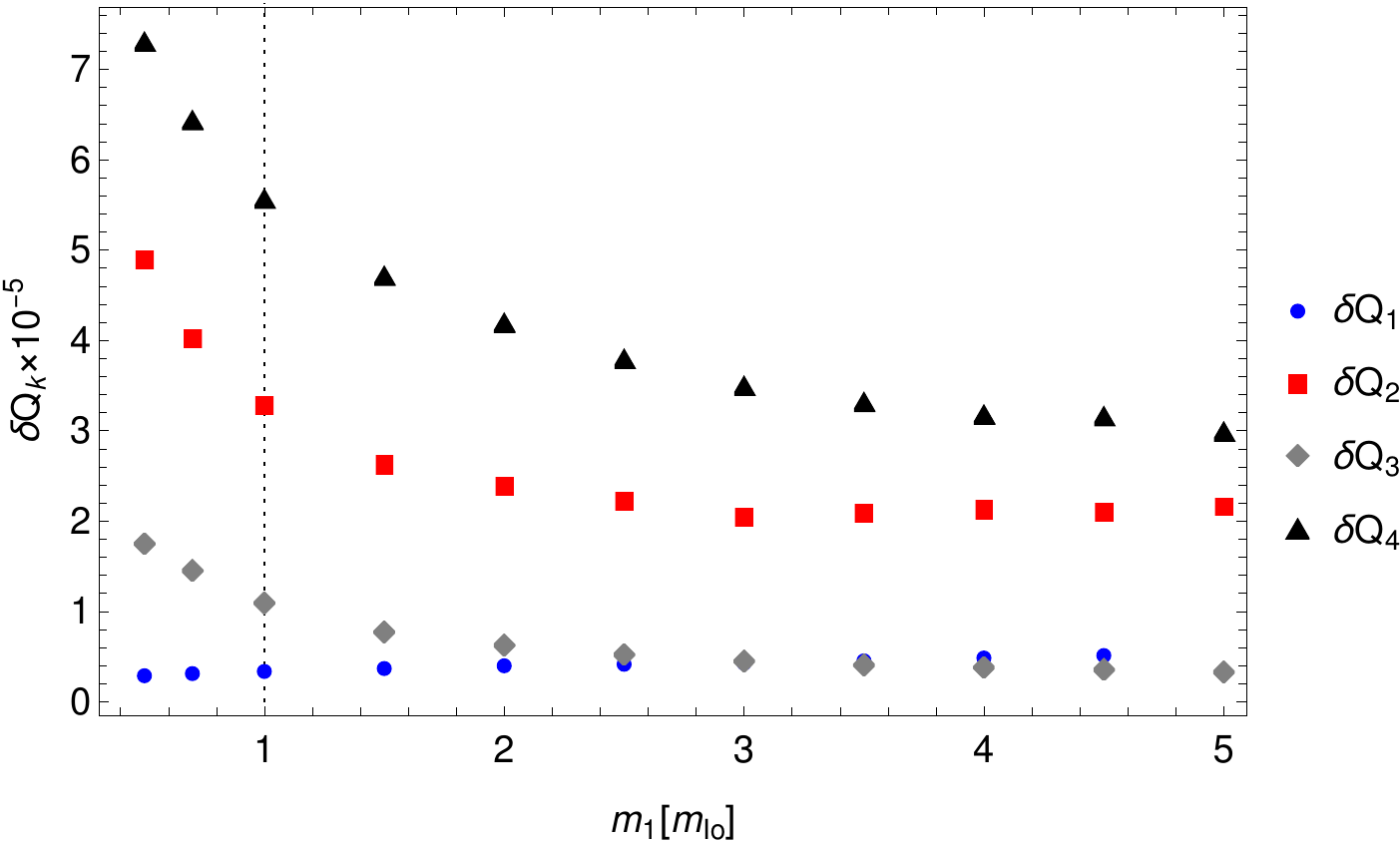}
\includegraphics[width=.51\linewidth]{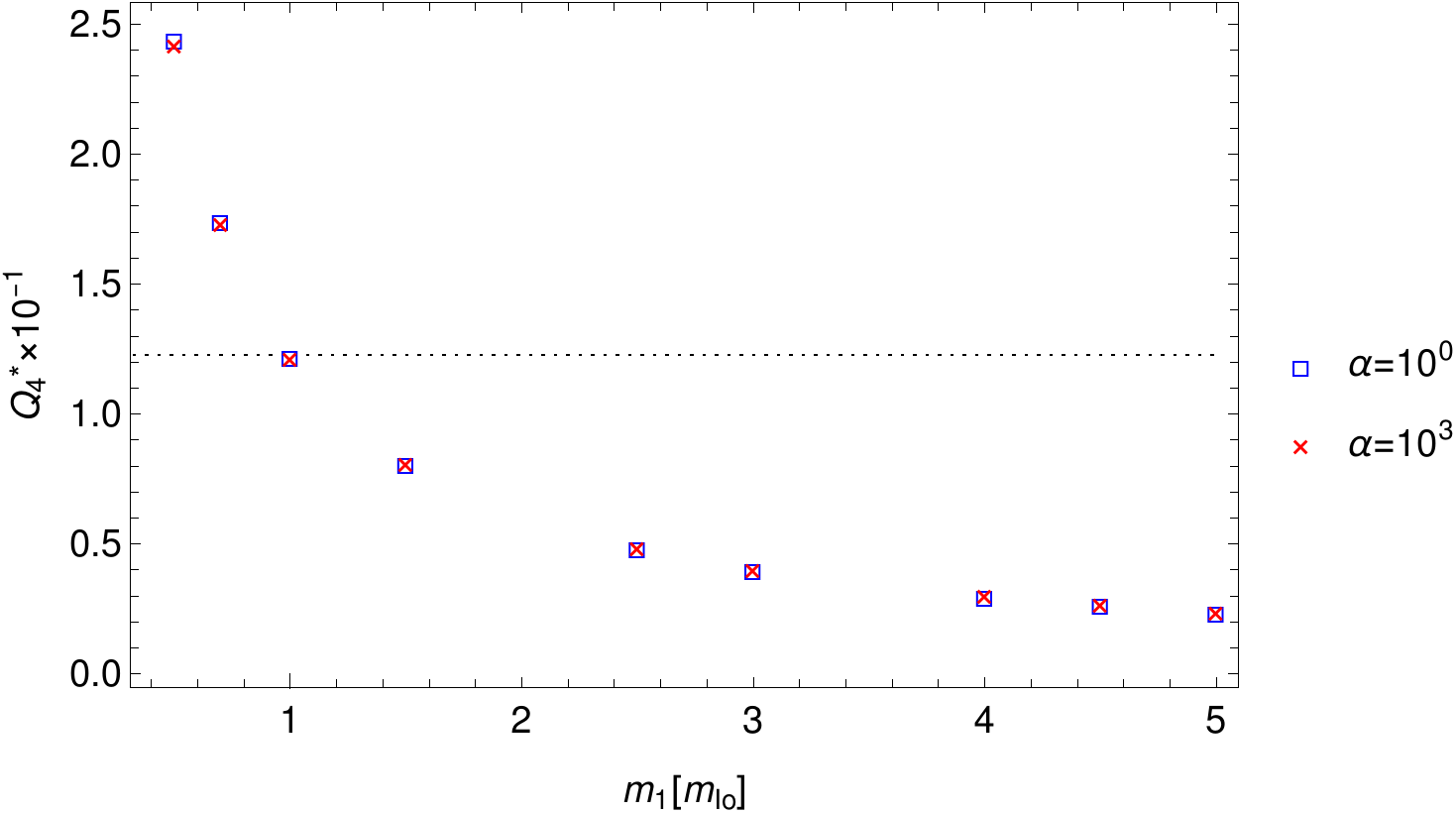}
\includegraphics[width=.48\linewidth]{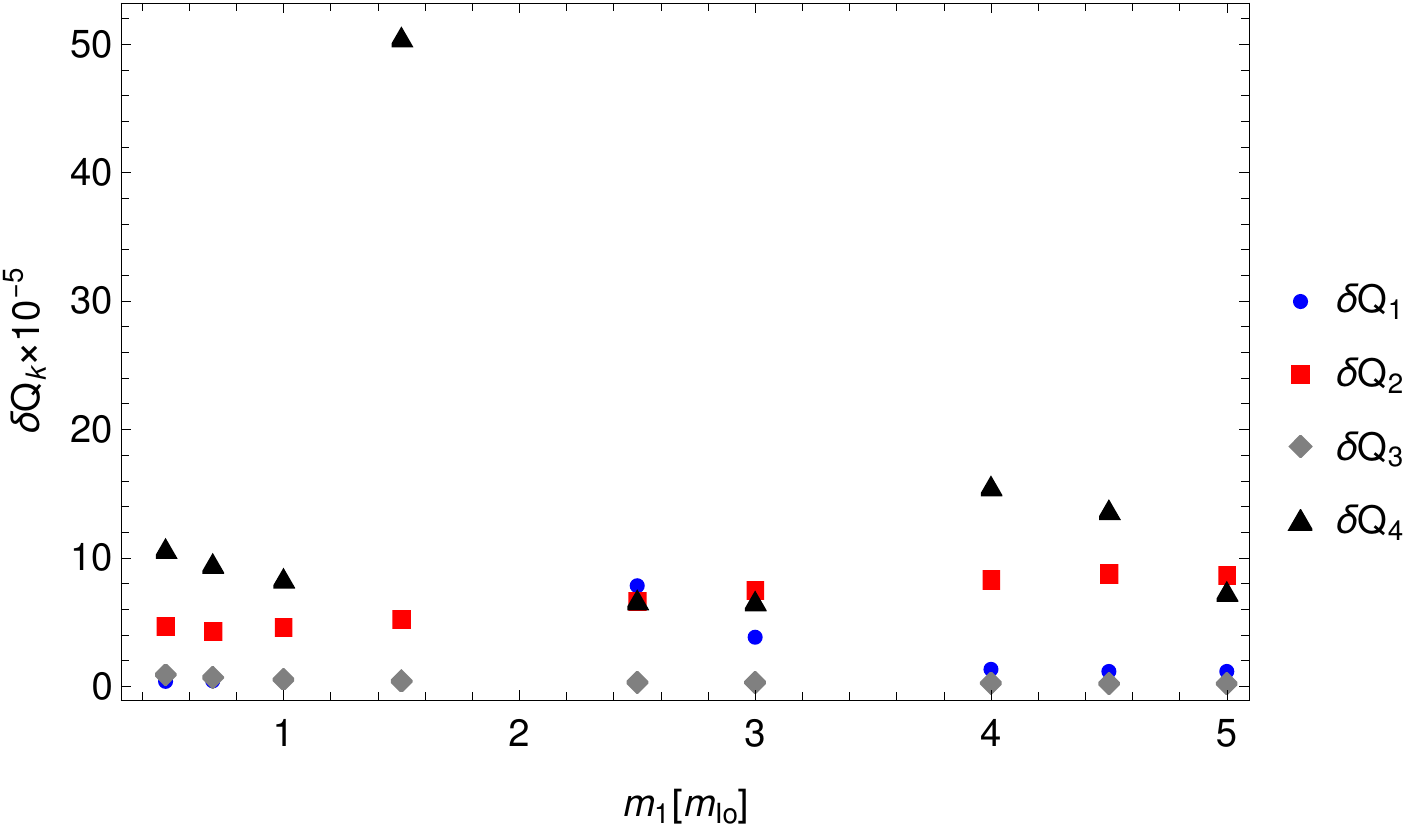}
    \caption{Equilibrium value $Q_4^*$ (left) and libration half-width
    $\delta Q_1$ ... $\delta Q_4$ (right) vs. mass of $S_1$ - $m_{1}$ (in units of the mass of Io) for different $\alpha$ (left column). The top row refers to the 2:1\&3:2 resonance, and the bottom row shows the 3:2\&3:2 resonance. The reference values for the actual Galilean system are indicated by dashed lines.}\label{f:mIo2}
\end{figure*}

We repeat the study for the 2:1\&3:2 case.
The results are shown in the top row of Figure~\ref{f:mIo2}. In the left panel of Figure~\ref{f:mIo2} we see a comparable
behavior of $Q_4^*$ with changing $m_{1}$ as for the 2:1\&2:1 resonant case, but
on different orders of magnitudes. The equilibrium $Q_4^*$
is shifted toward lower values for lower values of $m_{1}$,
and higher if the mass of $S_1$ is increased.

Next, we investigate the width of the librational motions in
dependence on the mass of $S_1$. The widths are defined as the
distance $\delta Q_k$ from the equilibrium toward the last
librational curve projected onto the planes, $(q_k,Q_k)$, found
numerically by fixing the $(q_\ell,Q_\ell)$, and varying ${Q_k}^*$
with $k\neq \ell$ and $k = 1, \dots, 4$ (see end of Section 4.1 for more details).
We clearly see that the librational widths decrease with increasing $m_1$ and keeping the
order $\delta Q_4 <  \delta Q_2 < \delta Q_3 < \delta Q_1$.
We remark that this result is of particular importance for understanding
the capture probabilities that are also related to the width of the resonance
in phase space. We conclude that for higher masses, the width (and capture
probabilities) decreases.

We make a similar computation for the 3:2\&3:2 resonance. The results
are provided in the bottom row of Figure~\ref{f:mIo2}. We note the following
difference in comparison with the 2:1\&2:1 and 2:1\&3:2
resonances: i) the equilibrium value of $Q_\ell$ decreases with increasing $m_1$ (numerical
problems occur at $m_1=2$ and $m_1=3.5$ that are left out); ii) the separatrix
half-widths change in a much more irregular way than in the previous
cases. We stress that in this case, the dynamics is heavily affected by the mass
$m_1$ in the sense that the evolution displays a chaotic behavior when $m_1$
increases. 

The results shown
in the bottom right panel of Figure~\ref{f:mIo2} should therefore be seen with caution because the numerical method  with which $\delta Q_4$ was obatained is strongly affected by the
irregular behavior of the dynamics.

\subsection{Higher-order resonances}

\begin{figure}[h]
        \centering
        \includegraphics[width=0.95\linewidth]{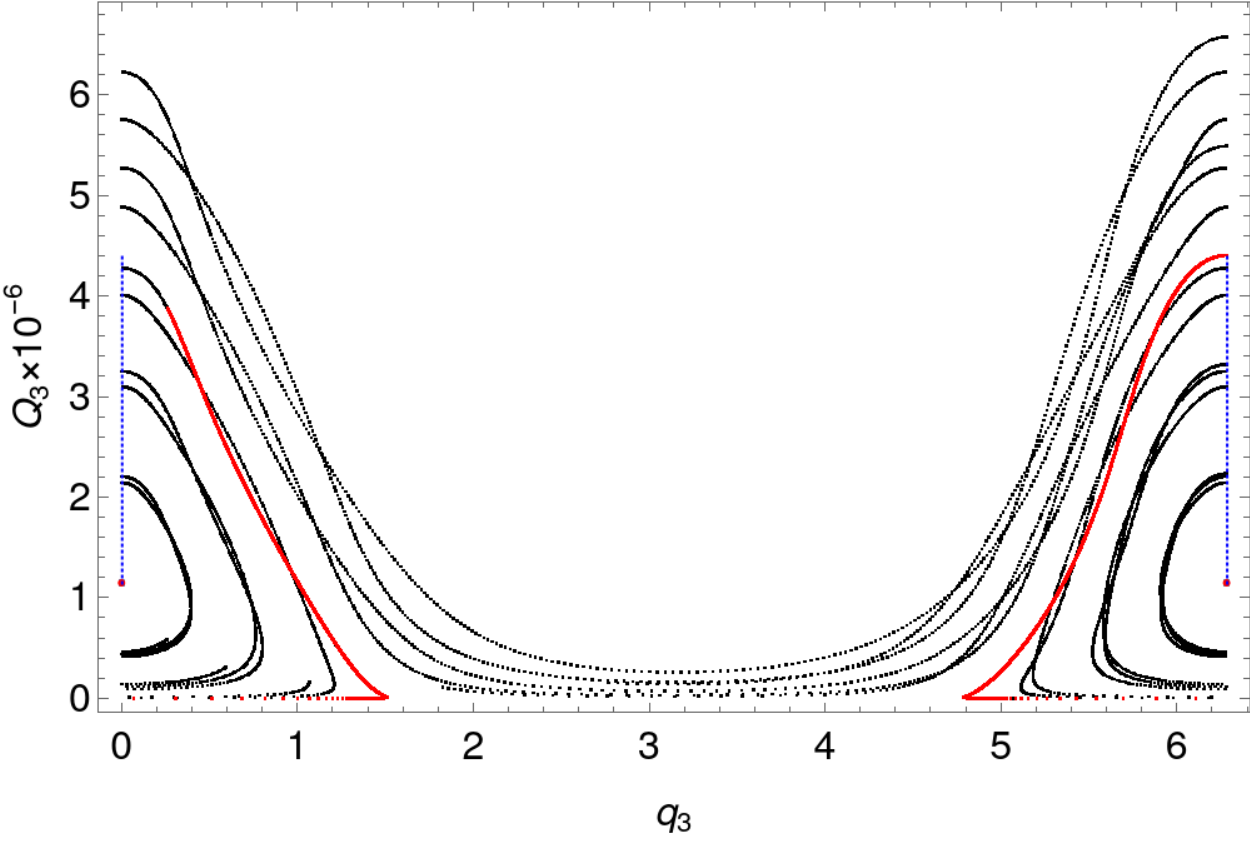}
        \includegraphics[width=0.95\linewidth]{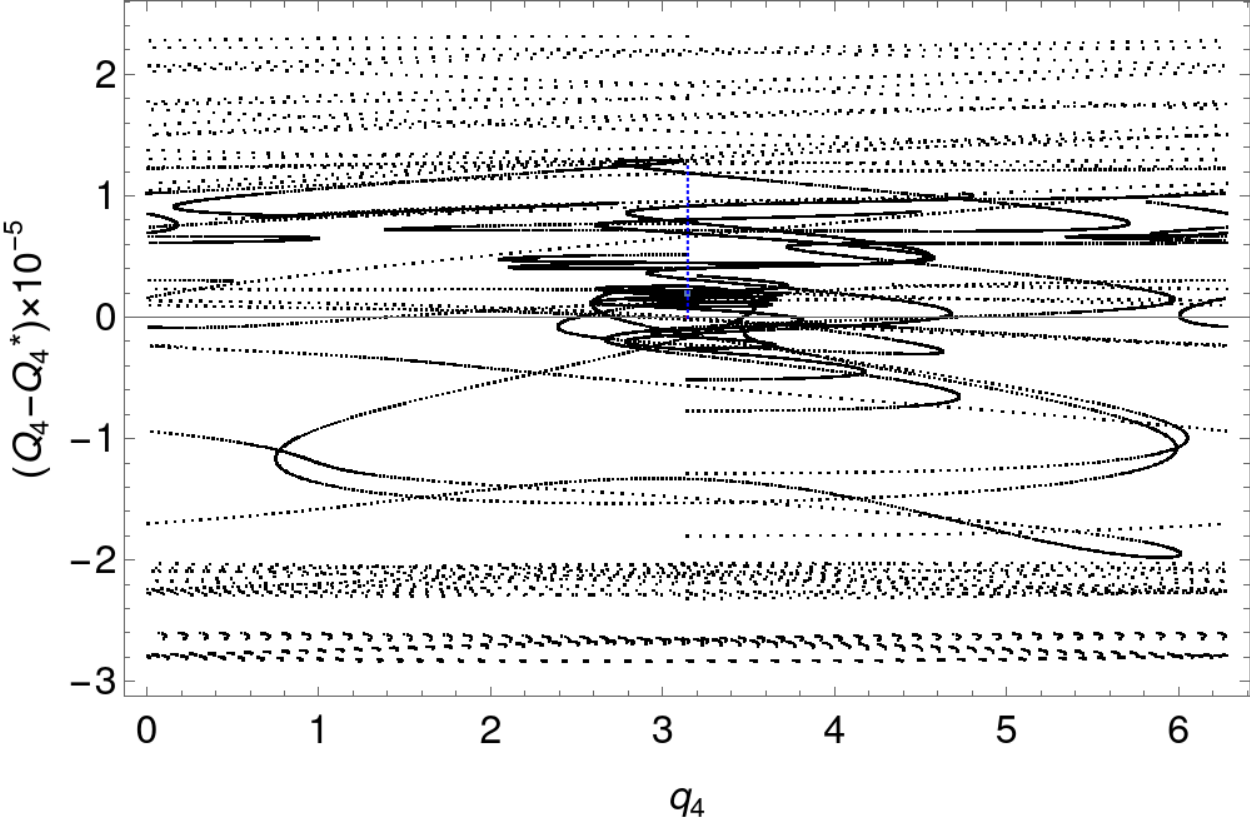}
        \caption{Strongly perturbed dynamics close to an equilibrium of the 
        3:1\&3:1 resonance, projected onto the plane $(q_3,Q_3)$ (top)
        and $(q_4,Q_4-{Q_4}^*)$ (bottom).}
        \label{f:second}
\end{figure}

We also performed similar calculations for the second- and mixed-order resonant
cases.  However, as already indicated by the numerical simulations in Section 3,
the motion close to the exact resonant configuration becomes much more complex.
In case of the second-order 3:1\&3:1 resonance, when tides are incldued, we find
complex conjugated pairs of eigenvalues with nonzero real parts. The librational
half-widths for Galilean mass parameters are smaller by a 
factor 1-10 ($\delta Q_1,\delta Q_2\propto10^{-5}$, $\delta Q_3\propto10^{-6}$),
and $\delta Q_4\simeq1.3-1.4\times10^{-5}$ than for the first-order resonances,
for which $\delta Q_4\simeq7-10\times10^{-5}$, for instance. Moreover, in addition to the reduced size of the resonant domain of 
motions, we also find a much stronger distortion of near resonant orbits when projected onto 
the $(q_k,Q_k), \; k=1,\dots,4$ planes. As an example, we provide the projection of orbits onto 
the sections $(q_3,Q_3)$ (top) and $(q_4,Q_4)$ (bottom) plotted in Figure~\ref{f:second}.  
We clearly observe that projections onto the plane $(q_3,Q_3)$ (for initial
conditions starting at exact resonant values in the other planes) look quite regular,
while deviations from exact resonant conditions in the plane $(q_4,Q_4)$ result in
strongly perturbed orbits, where the resonant structure of the phase space is 
essentially destroyed. We notice that a series of numerical simulations revealed that 
the perturbations strongly depend on the mass ratio of the innermost
two moons. For example, decreasing the mass of the second satellite by several orders of magnitudes
results in less perturbed orbits. A similar phenomenon can be 
observed in the case of mixed-order resonances. If we repeat the study in the case of
the 2:1\&3:1 resonance, we find considerable distortions of the phase space
close to the exact resonant configuration, which is strongly related to the mass
of the second moon.

We note that more regular configurations of 
Laplace-like resonances may exist. However, a complete parameter study
extended to cases with arbitrary mass ratios is beyond the scope of the current
investigations. To conclude, trapping in mixed- and higher-order
resonances could not be found for parameters close to the Galilean
satellite system. The perturbations, mainly due to the second satellite, together
with the reduced size of the librational regimes in phase space, are a possible explanation 
to justify the phenomenon found by pure numerical integrations. An analytical 
test of these results can be devised by computing resonant normal forms 
along the lines followed in \cite{He84}; \cite{puc21} suitably extended to higher orders: a task for 
investigations in the near future.

\section{Fate of the fourth satellite}\label{sec:callisto}

In this section, we integrate numerically the equations of motion described in
Section 2 including the tidal interaction between the central body and the
first satellite for a long timescale. The numerical integrations of
this section are perfomed using a program in C language that implements a
Runge-Kutta fourth-order scheme with a fixed step size. The model was modified in
order to include the mutual gravitational interactions due to the fourth
satellite and not only the secular part. In this way, the possibility of a capture
of $S_4$ is investigated in the case of the Laplace-like resonances. According
to \cite{lar20}, in the Galilean system, Callisto is captured into resonance
with Ganymede after about 1.5 Gyr. Once again, the behavior of first- and
second-order resonances is markedly different: in the cases of first-order
resonances, $S_4$ is captured into resonance, while in mixed- and second-order resonances, this effect is not observed.

\begin{figure}[h]
\centering
\includegraphics[width=.99\linewidth]{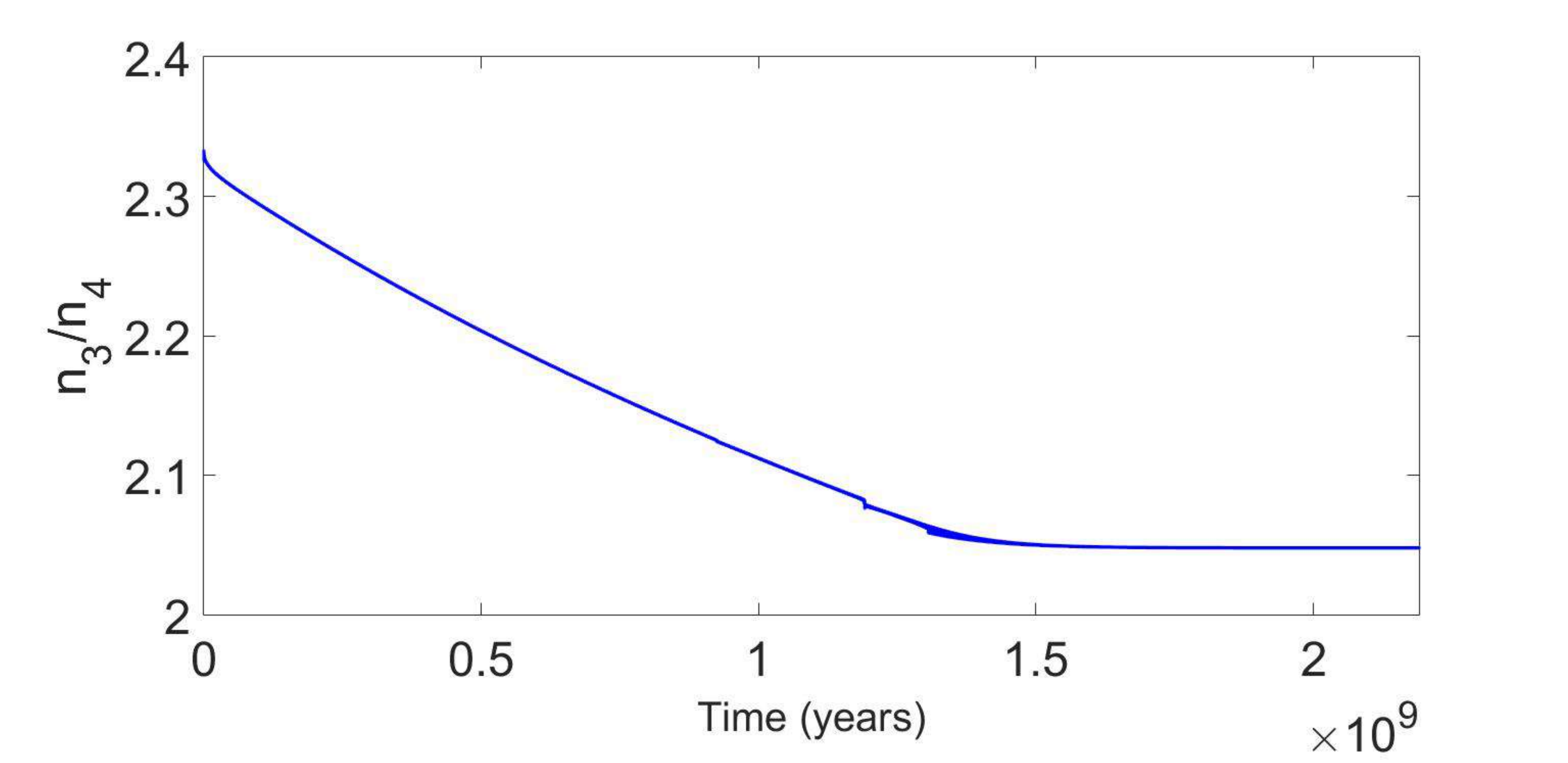}
\includegraphics[width=.99\linewidth]{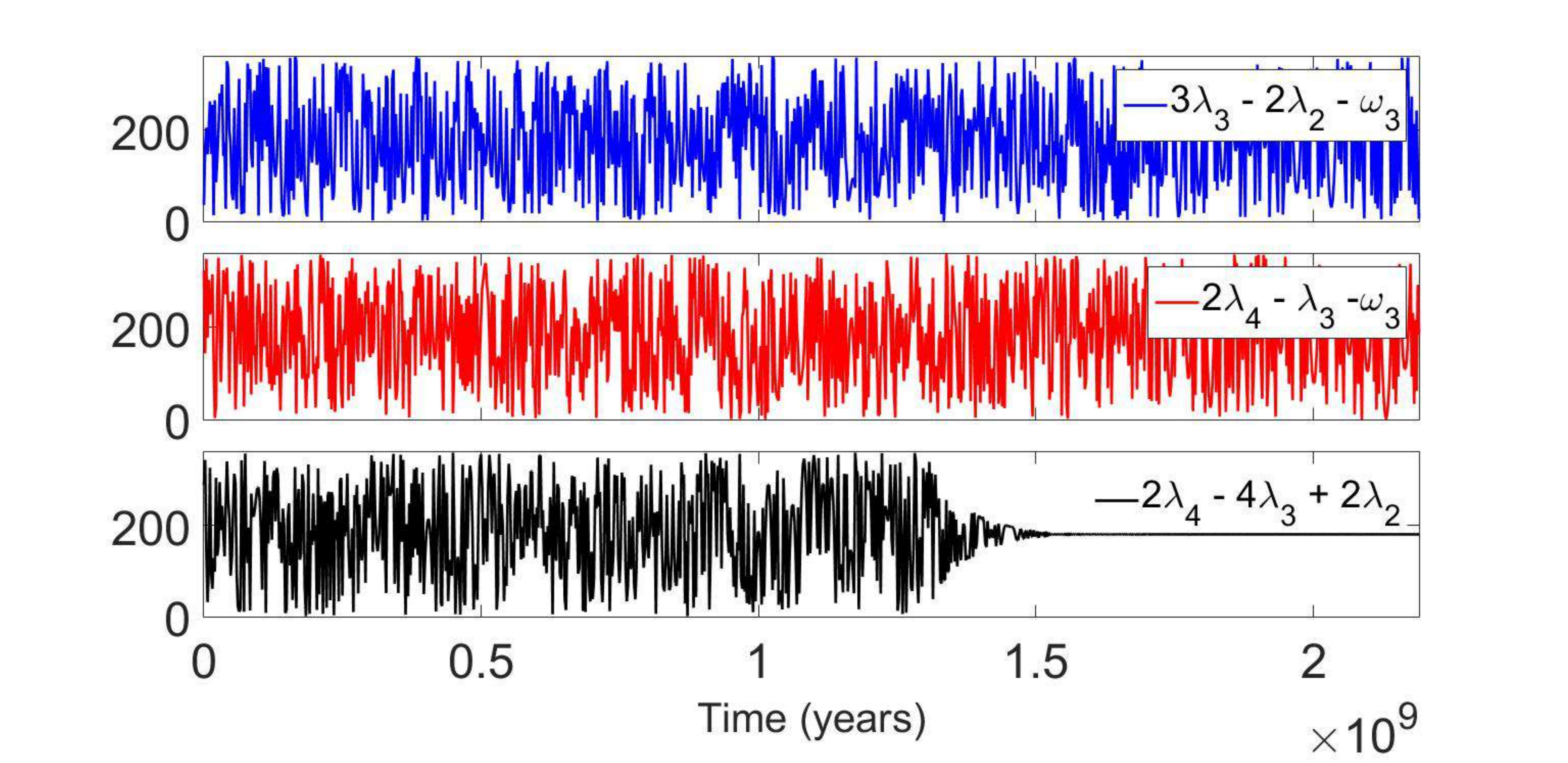}
    \caption{Evolution of the ratio of the mean motions of $S_3$ and $S_4$ (top panel) and the evolution of the resonant angles involving the mean longitudes of $S_2$, $S_3$ , and $S_4$ (bottom panel) in the case of the 3:2\&3:2 resonance between the three innermost satellites, considering the tidal dissipation assuming a fast-rotating central body. The tidal effects are multiplied by a factor $\alpha = 10^4$.} \label{3232_capture}
\end{figure}

In the case of the 3:2\&3:2 resonance, we note the capture of $S_4$ in a 2:1 resonance with $S_3$. We studied 100 different values for the initial mean longitude of $S_4$ , and in all the cases, the result was the same. The capture is obvious from the top panel of Figure \ref{3232_capture}, showing the ratio of the mean motions of $S_3$ and $S_4$. The resonant angle that involves the mean longitudes of the three outer satellites $S_2$, $S_3$ , and $S_4$, namely $2\lambda_4 - 4\lambda_3 + 2 \lambda_2$, rotates at the beginning; after almost 1.2 Gyr, we note that this angle librates around $180^\circ$. The individual resonant angles $3\lambda_3 - 2 \lambda_2 - \omega_3$ and $2\lambda_4 - \lambda_3 - \omega_3$ rotate before and after the trapping of $S_4$ (bottom of Figure \ref{3232_capture}).

A similar behavior is observed for the 2:1\&3:2 resonance testing 100 different initial values for the mean longitude of $S_4$. The difference between the two resonances is that in the 2:1\&3:2 case, both the individual resonant angles and the angle that involves all three mean longitudes librate after the trapping of $S_4$.

\begin{figure}[h]
\centering
\includegraphics[width=.99\linewidth]{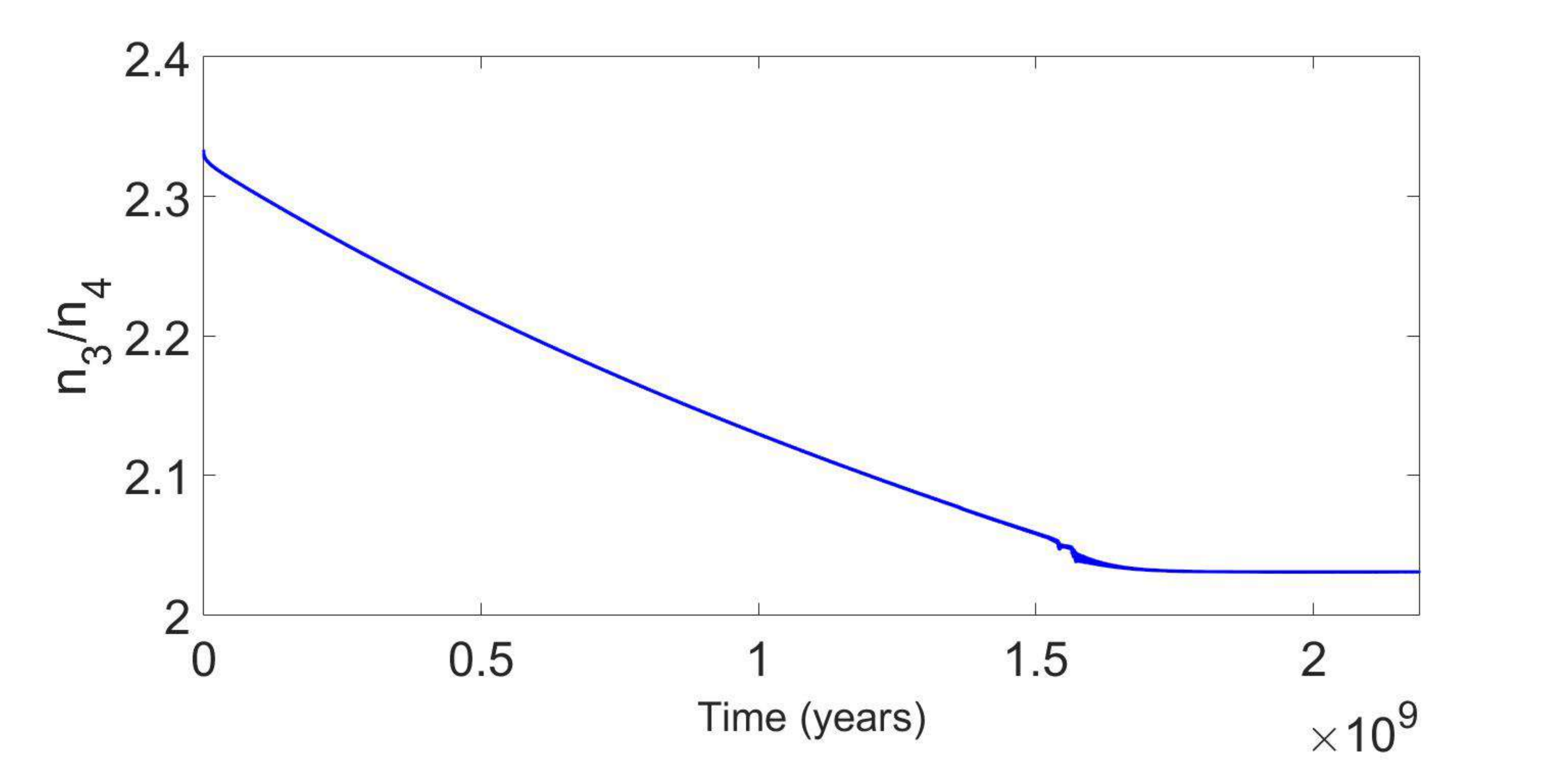}
\includegraphics[width=.99\linewidth]{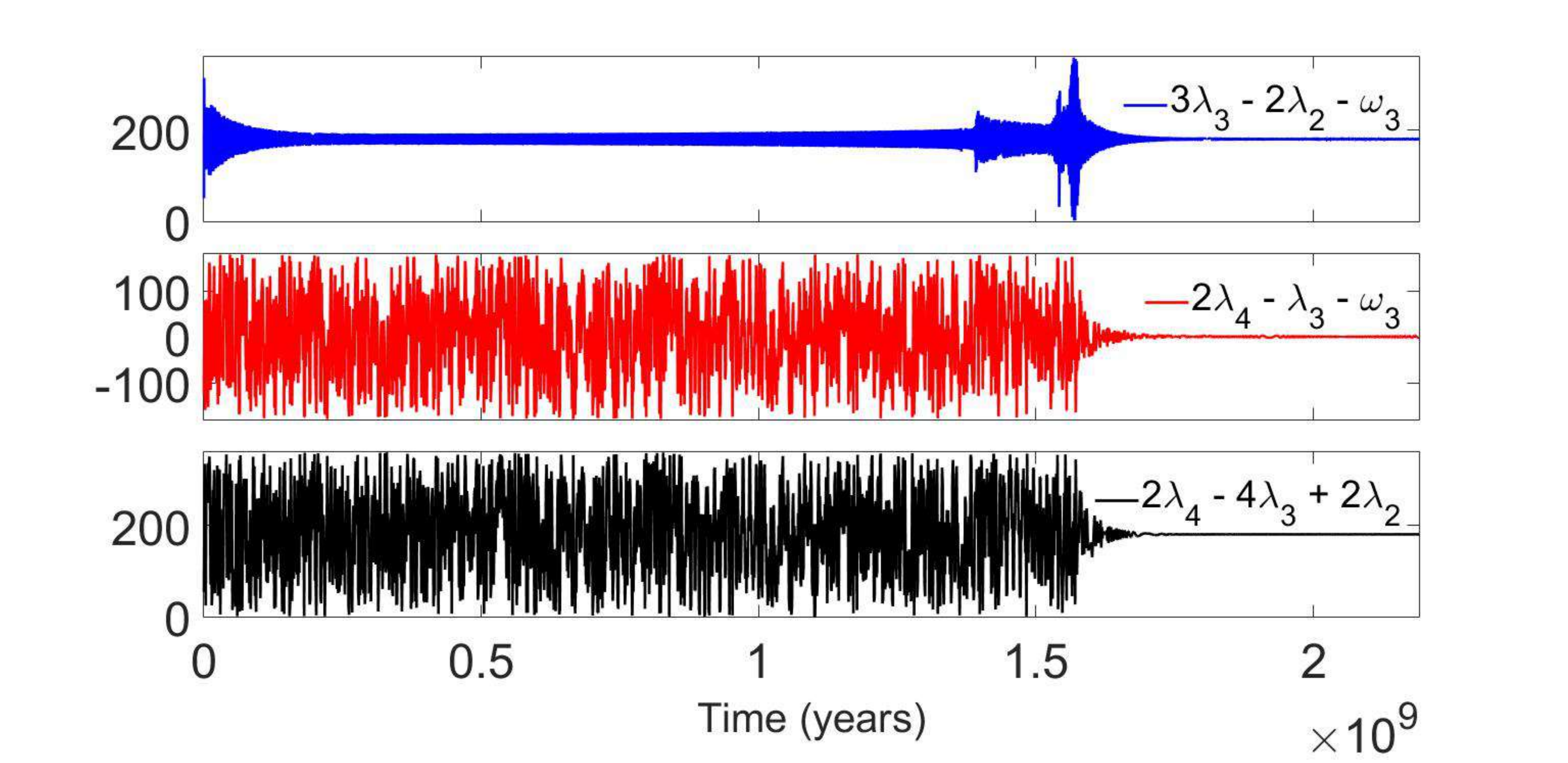}
    \caption{Evolution of the ratio of the mean motions of $S_3$ and $S_4$ (top panel) and the evolution of the resonant angles involving the mean longitudes of $S_2$, $S_3$ , and $S_4$ (bottom panel) in the case of the 2:1\&3:2 resonance of the three innermost moons considering the tidal dissipation assuming a fast-rotating central body. The tidal effects are multiplied by a factor $10^4$.} \label{2132_capture}
\end{figure}

\section{Summary and conclusions}\label{sec:conclusions}

We studied Laplace-like resonances generalized
to different commensurabilities in the frequencies of
the orbital longitudes of three celestial bodies that revolve
around a central body, and investigated the possibility of
capture of a fourth body in resonant configuration with
the other three bodies. Our model includes the mutual 
gravitational interaction of the celestial bodies, 
the secular effects of oblateness, and a distant star such as the Sun, 
as well as the tidal effect on the innermost celestial 
body close to the central body. A typical application
of this model is the Galilean satellite system of
planet Jupiter, where Io, Europa, and Ganymede are
already found in a 2:1\&2:1 resonant configuration, 
and the moon Callisto may be trapped in resonance with
these moons in future times. The outcome of this
system during the formation of the Solar System 
may have been different, that is, the three moons may have been in 
different resonant configurations. In addition, 
the results of our study may be applied to
other planetary moon systems.

From the numerical integrations performed in this work, a significant difference between the first-, second- and mixed-order resonances can be highlighted. Under the tidal effects between the central body and the innermost body, the three inner satellites move outward in the case of first-order resonances. In the second- and mixed-order resonances, the evolution of the system is different, and only the closest or the two closest satellites move outward, respectively. Moreover, the resonant angle that involves the mean longitudes of the three inner bodies librates only in the first-order resonances, but circulates in the remaining resonances. 

The evolution of a satellite system is dominated by the tidal interaction of the central body and the innermost satellite. However, the rotation of the central body plays a crucial role in the evolution of the orbital elements of the satellites. While in the case of the fast-rotating central body, the resonant arguments librate in first-order resonances and circulate in mixed- and second-order resonances, when a slowly rotating central body is assumed, the resonant arguments circulate. 

Moreover, the geometry of the phase space of the first-order resonances was studied. The system oscillates around the equilibria, and the separatrix and last paradoxal librational curve were computed. The computation of the equilibrium values and especially ${Q_4}^*$ was performed using different values of the mass of the innermost satellite pointing out the dependence on this parameter. Specifically, the maximum value of the eigenvalues decreases in magnitude as the value of $m_1$ increases. 

Finally, the possibility of capture of the fourth satellite was studied in the case of first-order resonances. In the two first-order resonances we included, the fourth satellite is captured into resonance using 100 different values for the initial mean longitude of the fourth satellite. In the 3:2\&3:2 resonance, the system develops two three-body resonances, a  3:2\&3:2 among the inner three satellites and a 3:2\&2:1 resonance among $S_2$, $S_3$ and $S_4$. In the 2:1\&3:2 resonance,  a chain of two-body resonances 2:1, 3:2, and 2:1 develops. This effect is not observed in the mixed- and second-order resonances. 


\begin{acknowledgements}
We acknowledge constant support and advice from S. Ferraz-Mello.
A.C., C.L., G.P. acknowledge EU H2020 MSCA ETN Stardust-R Grant Agreement 813644.
A.C. (partially), C.L. acknowledge the MIUR Excellence Department Project
awarded to the Department of Mathematics, University of Rome Tor
Vergata, CUP E83C18000100006.
A.C. (partially), C.L., G.P. acknowledge MIUR-PRIN 20178CJA2B
``New Frontiers of Celestial Mechanics: theory and Applications'',
ASI Contract n. 2018-25-HH.0 (Scientific Activities for JUICE, C/D phase).
C.L., G.P., M.V. acknowledge the GNFM/INdAM.
E.K., M.V. acknowledge the ASI Contract n. 2018-25-HH.0 (Scientific
Activities for JUICE, C/D phase).
G.P. is partially supported by INFN (Sezione di Roma II).
\end{acknowledgements}


\bibliographystyle{aa} 
\bibliography{Laplace-like}

\begin{thebibliography}{27}
\expandafter\ifx\csname natexlab\endcsname\relax\def\natexlab#1{#1}\fi

\bibitem[{{Beaug{\'e}} \& {Roig}(2001)}]{BR2001}
{Beaug{\'e}}, C. \& {Roig}, F. 2001, \icarus, 153, 391

\bibitem[{{Celletti} {et~al.}(2021){Celletti}, {Karampotsiou}, {Lhotka},
  {Pucacco}, \& {Volpi}}]{cklpv}
{Celletti}, A., {Karampotsiou}, E., {Lhotka}, C., {Pucacco}, G., \& {Volpi}, M.
  2021, Preprint

\bibitem[{{Celletti} {et~al.}(2019){Celletti}, {Paita}, \& {Pucacco}}]{cel19}
{Celletti}, A., {Paita}, F., \& {Pucacco}, G. 2019, Chaos, 29, 033111

\bibitem[{{Christiansen} {et~al.}(2018){Christiansen}, {Crossfield},
  {Barentsen}, {Lintott}, {Barclay}, {Simmons}, {Petigura}, {Schlieder},
  {Dressing}, {Vanderburg}, {et~al.}}]{chr18}
{Christiansen}, J.~L., {Crossfield}, I.~J., {Barentsen}, G., {et~al.} 2018,
  \aj, 155, 57

\bibitem[{{David} {et~al.}(2019){David}, {Petigura}, \& {Luger}}]{dav16}
{David}, T.~J., {Petigura}, E.~A., \& {Luger}, R.~e. 2019, \apjl, 885, L12

\bibitem[{de~Sitter(1928)}]{desitter1928}
de~Sitter, W. 1928, Annalen van de Sterrewacht te Leiden, 16, B1

\bibitem[{{Delisle}(2017)}]{del17}
{Delisle}, J.~B. 2017, \aap, 605, A96

\bibitem[{{Ellis} \& {Murray}(2000)}]{ell00}
{Ellis}, K.~M. \& {Murray}, C.~D. 2000, \icarus, 147, 129

\bibitem[{Ferraz-Mello(1979)}]{fer81}
Ferraz-Mello, S. 1979, CEP, 5508, 090

\bibitem[{Ferraz-Mello {et~al.}(2008)Ferraz-Mello, Rodr{\'\i}guez, \&
  Hussmann}]{fer07}
Ferraz-Mello, S., Rodr{\'\i}guez, A., \& Hussmann, H. 2008, Celestial Mechanics
  and Dynamical Astronomy, 101, 171

\bibitem[{Hara {et~al.}(2020)Hara, Bouchy, Stalport, Boisse, Rodrigues,
  Delisle, Santerne, Henry, Arnold, Astudillo-Defru, {et~al.}}]{har20}
Hara, N., Bouchy, F., Stalport, M., {et~al.} 2020, Astronomy \& Astrophysics,
  636, L6

\bibitem[{Henrard(1984)}]{He84}
Henrard, J. 1984, Celestial Mechanics, 34, 255

\bibitem[{Lainey {et~al.}(2009)Lainey, Arlot, Karatekin, \& Van~Hoolst}]{lai09}
Lainey, V., Arlot, J.-E., Karatekin, {\"O}., \& Van~Hoolst, T. 2009, Nature,
  459, 957

\bibitem[{Lari {et~al.}(2020)Lari, Saillenfest, \& Fenucci}]{lar20}
Lari, G., Saillenfest, M., \& Fenucci, M. 2020, Astronomy \& Astrophysics, 639,
  A40

\bibitem[{Lieske(1998)}]{lie97}
Lieske, J. 1998, Astronomy and Astrophysics Supplement Series, 129, 205

\bibitem[{Luger {et~al.}(2017)Luger, Sestovic, Kruse, Grimm, Demory, Agol,
  Bolmont, Fabrycky, Fernandes, Van~Grootel, {et~al.}}]{lug17}
Luger, R., Sestovic, M., Kruse, E., {et~al.} 2017, Nature Astronomy, 1, 1

\bibitem[{Malhotra(1991)}]{mal91}
Malhotra, R. 1991, Icarus, 94, 399

\bibitem[{Murray \& Dermott(1999)}]{mur99}
Murray, C.~D. \& Dermott, S.~F. 1999, Solar system dynamics (Cambridge
  university press)

\bibitem[{Paita {et~al.}(2018)Paita, Celletti, \& Pucacco}]{pai18}
Paita, F., Celletti, A., \& Pucacco, G. 2018, Astronomy \& Astrophysics, 617,
  A35

\bibitem[{Pichierri {et~al.}(2019)Pichierri, Batygin, \& Morbidelli}]{Pic19}
Pichierri, G., Batygin, K., \& Morbidelli, A. 2019, Astronomy \& Astrophysics,
  625, A7

\bibitem[{{Pucacco}(2021)}]{puc21}
{Pucacco}, G. 2021, Celestial Mechanics and Dynamical Astronomy, 133, 11

\bibitem[{Showalter \& Hamilton(2015)}]{sho15}
Showalter, M. \& Hamilton, D. 2015, Nature, 522, 45

\bibitem[{Showman \& Malhotra(1997)}]{sho97}
Showman, A.~P. \& Malhotra, R. 1997, Icarus, 127, 93

\bibitem[{Tittemore(1990)}]{tit90}
Tittemore, W.~C. 1990, Science, 250, 263

\bibitem[{Tittemore \& Wisdom(1988)}]{TW1990}
Tittemore, W.~C. \& Wisdom, J. 1988, Icarus, 74, 172

\bibitem[{Yoder(1979)}]{yod79}
Yoder, C.~F. 1979, Nature, 279, 767

\bibitem[{Yoder \& Peale(1981)}]{yod81}
Yoder, C.~F. \& Peale, S.~J. 1981, Icarus, 47, 1

\end{thebibliography}


\newpage

\appendix
\section{Hamiltonian function of the secular and resonant parts of the mutual satellite
interactions}\label{sec:app}

The secular and resonant parts of the Hamiltonian that we used to
model the mutual interaction of the satellites for the 3:2\&3:2
resonance are given by

\begin{equation*}
\label{reshamA1}
\begin{aligned}
H_P^{1,2} &= -{\frac{Gm_1m_2}{a_2}}\big(B_0(\alpha_{1,2})\\
&+{f_A}_1^{1,2}({e_1}^2 + {e_2}^2) \\
&+{f_A}_2^{1,2} e_1 \cos(3\lambda_2-2 \lambda_1-\varpi_1) \\
&+{f_A}_3^{1,2} e_2 \cos(3\lambda_2-2 \lambda_1-\varpi_2) \\
&+ {f_A}_4^{1,2} {e_1}^2 \cos(6\lambda_2 -4\lambda_1 -2\varpi_1)\\ 
&+ {f_A}_5^{1,2} {e_2}^2 \cos( 6\lambda_2 -4\lambda_1 -2 \varpi_2) \\
&+ {f_A}_6^{1,2} e_1 e_2 \cos(6\lambda_2 -4\lambda_1 -\varpi_1-\varpi_2)\\ 
&+ {f_A}_7^{1,2} e_1 e_2 \cos(\varpi_2 - \varpi_1) \\
&+ {f_A}_8^{1,2} {s_1}^2 \cos(6\lambda_2 -4\lambda_1 -2\Omega_1)\\
&+ {f_A}_9^{1,2} {s_2}^2 \cos( 6\lambda_2 -4\lambda_1 -2 \Omega_2) \\
&+ {f_A}_{10}^{1,2} {s_1} {s_2} \cos(6\lambda_2 -4\lambda_1 -\Omega_1 - \Omega_2)\\ 
&+ {f_A}_{11}^{1,2} {s_1} {s_2} \cos(\Omega_1 - \Omega_2) \big) \\[.1cm]
\end{aligned}
\end{equation*}
\begin{equation*}
\label{reshamA2}
\begin{aligned}
H_P^{2,3} &= -{\frac{Gm_2m_3}{a_3}}\big(B_0(\alpha_{2,3})\\
&+{f_A}_1^{2,3}({e_2}^2 + {e_3}^2) \\
&+{f_A}_2^{2,3} e_2 \cos(3\lambda_3-2 \lambda_2-\varpi_2)\\ 
&+{f_A}_3^{2,3} e_3 \cos(3\lambda_3-2 \lambda_2-\varpi_3) \\
&+ {f_A}_4^{2,3} {e_2}^2 \cos(6\lambda_3 -4\lambda_2 -2\varpi_2)\\ 
&+ {f_A}_5^{2,3} {e_3}^2 \cos( 6\lambda_3 -4\lambda_2 -2 \varpi_3) \\
&+ {f_A}_6^{2,3} e_2 e_3 \cos(6\lambda_3 -4\lambda_2 -\varpi_2-\varpi_3)\\  
&+ {f_A}_7^{2,3} e_2 e_3 \cos(\varpi_3 - \varpi_2) \\
&+ {f_A}_8^{2,3} {s_2}^2 \cos(6\lambda_3 -4\lambda_2 -2\Omega_2)\\ 
&+ {f_A}_9^{2,3} {s_3}^2 \cos( 6\lambda_3 -4\lambda_2 -2 \Omega_3) \\
&+ {f_A}_{10}^{2,3} {s_2} {s_3} \cos(6\lambda_3 -4\lambda_2 -\Omega_2 - \Omega_3)\\ 
&+ {f_A}_{11}^{2,3} {s_2} {s_3} \cos(\Omega_2 - \Omega_3) \big) \\[.1cm]
\end{aligned}
\end{equation*}
\begin{equation*}
\label{reshamA3}
\begin{aligned}
H_P^{1,3} &= -{\frac{Gm_1m_3}{a_3}}\big(B_0(\alpha_{1,3})\\ 
&+ {f_A}_1^{1,3}(e_1^2+e_3^2) \\
&+ {f_A}_7^{1,3} e_1 e_3\cos(\varpi_3 - \varpi_1)\\
&+  {f_A}_{11}^{1,3} {s_1} {s_3} \cos(\Omega_1 - \Omega_3) \big)\ . \\[.1cm]
\end{aligned}
\end{equation*}

The resonant part of the Hamiltonian that corresponds to the 2:1\&3:2 resonance is given by
\begin{equation*}
  \label{reshamB1}
\begin{aligned}
H_P^{1,2} &= -{{Gm_1m_2}\over a_2}\big(B_0(\alpha_{1,2})\\
&+{f_B}_1^{1,2}({e_1}^2 + {e_2}^2) \\
&+{f_B}_2^{1,2} e_1 \cos(2\lambda_2- \lambda_1-\varpi_1)\\ 
&+{f_B}_3^{1,2} e_2 \cos(2\lambda_2- \lambda_1-\varpi_2) \\
&+ {f_B}_4^{1,2} {e_1}^2 \cos(4\lambda_2 - 2\lambda_1 -2\varpi_1)\\ 
&+ {f_B}_5^{1,2} {e_2}^2 \cos( 4\lambda_2 - 2\lambda_1 -2 \varpi_2) \\
&+ {f_B}_6^{1,2} e_1 e_2 \cos(4\lambda_2 -2\lambda_1 -\varpi_1-\varpi_2)\\  
&+ {f_B}_7^{1,2} e_1 e_2 \cos(\varpi_2 - \varpi_1) \\
&+ {f_B}_8^{1,2} {s_1}^2 \cos(4\lambda_2 -2\lambda_1 -2\Omega_1)\\ 
&+ {f_B}_9^{1,2} {s_2}^2 \cos( 4\lambda_2 -2\lambda_1 -2 \Omega_2) \\
&+ {f_B}_{10}^{1,2} {s_1} {s_2} \cos(4\lambda_2 -2\lambda_1 -\Omega_1 - \Omega_2)\\ 
&+ {f_B}_{11}^{1,2} {s_1} {s_2} \cos(\Omega_1 - \Omega_2) \big) \\[.1cm]
\end{aligned}
\end{equation*}
\begin{equation*}
  \label{reshamB2}
\begin{aligned}
H_P^{2,3} &= -{{Gm_2m_3}\over a_3}\big(B_0(\alpha_{2,3})\\
&+{f_A}_1^{2,3}({e_2}^2 + {e_3}^2) \\
&+{f_A}_2^{2,3} e_2 \cos(3\lambda_3-2 \lambda_2-\varpi_2)\\ 
&+{f_A}_3^{2,3} e_3 \cos(3\lambda_3-2 \lambda_2-\varpi_3) \\
&+ {f_A}_4^{2,3} {e_2}^2 \cos(6\lambda_3 -4\lambda_2 -2\varpi_2)\\ 
&+ {f_A}_5^{2,3} {e_3}^2 \cos( 6\lambda_3 -4\lambda_2 -2 \varpi_3) \\
&+ {f_A}_6^{2,3} e_2 e_3 \cos(6\lambda_3 -4\lambda_2 -\varpi_2-\varpi_3)\\  
&+ {f_A}_7^{2,3} e_2 e_3 \cos(\varpi_3 - \varpi_2) \\
&+ {f_A}_8^{2,3} {s_2}^2 \cos(6\lambda_3 -4\lambda_2 -2\Omega_2)\\ 
&+ {f_A}_9^{2,3} {s_3}^2 \cos( 6\lambda_3 -4\lambda_2 -2 \Omega_3) \\
&+ {f_A}_{10}^{2,3} {s_2} {s_3} \cos(6\lambda_3 -4\lambda_2 -\Omega_2 - \Omega_3)\\ 
&+ {f_A}_{11}^{2,3} {s_2} {s_3} \cos(\Omega_2 - \Omega_3) \big) \\[.1cm]
\end{aligned}
\end{equation*}
\begin{equation*}
  \label{reshamB3}
\begin{aligned}
H_P^{1,3} &= -{{Gm_1m_3}\over a_3}\big(B_0(\alpha_{1,3})\\ 
&+ {f_A}_1^{1,3}(e_1^2+e_3^2) \\
&+ {f_A}_7^{1,3} e_1 e_3\cos(\varpi_3 - \varpi_1)\\
&+ {f_A}_{11}^{1,3} {s_1} {s_3} \cos(\Omega_1 - \Omega_3) \big)\ .
\end{aligned}
\end{equation*}

The resonant part of the Hamiltonian that corresponds to the 3:1\&3:1 resonance is given by
\begin{equation*}
  \label{reshamC1}
\begin{aligned}
&H_P^{1,2} = -{{Gm_1m_2}\over a_2}\big(B_0(\alpha_{1,2})\\
&+{f_C}_1^{1,2}({e_1}^2 + {e_2}^2) \\
&+{f_C}_2^{1,2} {e_1}^2 \cos(3\lambda_2- \lambda_1-2 \varpi_1)\\ 
&+{f_C}_3^{1,2} e_1 e_2 \cos(3\lambda_2- \lambda_1 -\varpi_1 -\varpi_2) \\
&+ {f_C}_4^{1,2} {e_2}^2 \cos(3\lambda_2 -\lambda_1 -2\varpi_2)\\ 
&+ {f_C}_5^{1,2} e_1 e_2 \cos(\varpi_2 - \varpi_1)  \\
&+{f_C}_6^{1,2} {s_1}^2 \cos(3\lambda_2- \lambda_1-2 \Omega_1)\\ 
&+{f_C}_7^{1,2} s_1 s_2 \cos(3\lambda_2- \lambda_1 -\Omega_1 -\Omega_2) \\
&+ {f_C}_8^{1,2} {s_2}^2 \cos(3\lambda_2 -\lambda_1 -2\Omega_2)\\ 
&+ {f_C}_9^{1,2} s_1 s_2 \cos(\Omega_2 - \Omega_1)\big) \\[.1cm]
\end{aligned}
\end{equation*}
\begin{equation*}
  \label{reshamC2}
\begin{aligned}
H_P^{2,3} &= -{{Gm_2m_3}\over a_3}\big(B_0(\alpha_{2,3})\\
&+{f_C}_1^{2,3}({e_2}^2 + {e_3}^2) \\
&+{f_C}_2^{2,3} {e_2}^2 \cos(3\lambda_3- \lambda_2-2 \varpi_2)\\ 
&+{f_C}_3^{2,1} e_2 e_3 \cos(3\lambda_3- \lambda_2 -\varpi_2 -\varpi_3) \\
&+ {f_C}_4^{2,3} {e_3}^2 \cos(3\lambda_3 -\lambda_2 -2\varpi_3)\\ 
&+ {f_C}_5^{2,3} e_2 e_3 \cos(\varpi_3 - \varpi_2) \\
&+{f_C}_6^{2,3} {s_2}^2 \cos(3\lambda_3- \lambda_2-2 \Omega_2) \\
&+{f_C}_7^{2,1} s_2 s_3 \cos(3\lambda_3- \lambda_2 -\Omega_2 -\Omega_3) \\
&+ {f_C}_8^{2,3} {s_3}^2 \cos(3\lambda_3 -\lambda_2 -2\Omega_3)\\ 
&+ {f_C}_9^{2,3} s_2 s_3 \cos(\Omega_3 - \Omega_2)\big) \\[.1cm]
\end{aligned}
\end{equation*}
\begin{equation*}
  \label{reshamC3}
\begin{aligned}
H_P^{1,3} &= -{{Gm_1m_3}\over a_3}\big(B_0(\alpha_{1,3})\\ 
&+ {f_C}_1^{1,3}(e_1^2+e_3^2) \\
&+ {f_C}_5^{1,3} e_1 e_3\cos(\varpi_3 - \varpi_1)\\  
&+ {f_C}_9^{1,3} s_1 s_3\cos(\Omega_3 - \Omega_1) \big)\ .
\end{aligned}
\end{equation*}

The resonant part of the Hamiltonian that corresponds to the 2:1\&3:1 resonance is given by
\begin{equation*}
  \label{reshamD1}
\begin{aligned}
H_P^{1,2} &= -{{Gm_1m_2}\over a_2}\big(B_0(\alpha_{1,2})\\
&+{f_B}_1^{1,2}({e_1}^2 + {e_2}^2) \\
&+{f_B}_2^{1,2} e_1 \cos(2\lambda_2- \lambda_1-\varpi_1)\\ 
&+{f_B}_3^{1,2} e_2 \cos(2\lambda_2- \lambda_1-\varpi_2) \\
&+ {f_B}_4^{1,2} {e_1}^2 \cos(4\lambda_2 - 2\lambda_1 -2\varpi_1)\\ 
&+ {f_B}_5^{1,2} {e_2}^2 \cos( 4\lambda_2 - 2\lambda_1 -2 \varpi_2) \\
&+ {f_B}_6^{1,2} e_1 e_2 \cos(4\lambda_2 -2\lambda_1 -\varpi_1-\varpi_2)\\  
&+ {f_B}_7^{1,2} e_1 e_2 \cos(\varpi_2 - \varpi_1) \\
&+ {f_B}_8^{1,2} {s_1}^2 \cos(4\lambda_2 -2\lambda_1 -2\Omega_1)\\ 
&+ {f_B}_9^{1,2} {s_2}^2 \cos( 4\lambda_2 -2\lambda_1 -2 \Omega_2) \\
&+ {f_B}_{10}^{1,2} {s_1} {s_2} \cos(4\lambda_2 -2\lambda_1 -\Omega_1 - \Omega_2)\\ 
&+ {f_B}_{11}^{1,2} {s_1} {s_2} \cos(\Omega_1 - \Omega_2) \big)\\[.1cm]
\end{aligned}
\end{equation*}
\begin{equation*}
  \label{reshamD2}
\begin{aligned}
H_P^{2,3} &= -{{Gm_2m_3}\over a_3}\big(B_0(\alpha_{2,3})\\
&+{f_C}_1^{2,3}({e_2}^2 + {e_3}^2) \\
&+{f_C}_2^{2,3} {e_2}^2 \cos(3\lambda_3- \lambda_2-2 \varpi_2)\\ 
&+{f_C}_3^{2,1} e_2 e_3 \cos(3\lambda_3- \lambda_2 -\varpi_2 -\varpi_3) \\
&+ {f_C}_4^{2,3} {e_3}^2 \cos(3\lambda_3 -\lambda_2 -2\varpi_3)\\ 
&+ {f_C}_5^{2,3} e_2 e_3 \cos(\varpi_3 - \varpi_2) \\
&  +{f_C}_6^{2,3} {s_2}^2 \cos(3\lambda_3- \lambda_2-2 \Omega_2)\\ 
&+{f_C}_7^{2,1} s_2 s_3 \cos(3\lambda_3- \lambda_2 -\Omega_2 -\Omega_3) \\
&+ {f_C}_8^{2,3} {s_3}^2 \cos(3\lambda_3 -\lambda_2 -2\Omega_3)\\ 
&+ {f_C}_9^{2,3} s_2 s_3 \cos(\Omega_3 - \Omega_2)\big) \\[.1cm]
\end{aligned}
\end{equation*}
\begin{equation*}
  \label{reshamD3}
\begin{aligned}
H_P^{1,3} &= -{{Gm_1m_3}\over a_3}\big(B_0(\alpha_{1,3})\\ 
&+ {f_C}_1^{1,3}(e_1^2+e_3^2) \\
&+ {f_C}_5^{1,3} e_1 e_3\cos(\varpi_3 - \varpi_1)\\  
&+ {f_C}_9^{1,3} s_1 s_3\cos(\Omega_3 - \Omega_1) \big)\ .
\end{aligned}
\end{equation*}

In these expressions, $\alpha_{ij} = \frac{a_i}{a_j}$ is the
ratio of the semimajor axes of the $i$-th and $j$-th satellite, $B_0(\alpha_{i,j}) = \frac{1}{2} {b_{1/2}}^{(0)}
(\alpha_{i,j}) -1$, and the functions ${f_{Ak}}^{i,j}$,
${f_{Bk}}^{i,j}$, ${f_{Ck}}^{i,j}$ are linear combinations of the
Laplace coefficients ${b_s}^{(n)}$ and their derivatives
(\cite{mur99}, \cite{ell00}, \cite{cel19}), as given below:
\begin{equation*}
{f_{A1}}^{i,j} = \bigg( \frac{1}{4}\alpha_{i,j} \frac{d}{d \alpha_{i,j}} + \frac{1}{8} {\alpha_{i,j}}^2 \frac{d^2}{{d \alpha_{i,j}}^2}  \bigg) {b_{1/2}}^{(0)} (\alpha_{i,j})
\end{equation*}
\begin{equation*}
{f_{A2}}^{i,j} = -3 {b_{1/2}}^{(3)} (\alpha_{i,j}) + \frac{1}{2} \alpha_{i,j} \frac{d}{d \alpha_{i,j}} {b_{1/2}}^{(3)} (\alpha_{i,j})
\end{equation*}
\begin{equation*}
{f_{A3}}^{i,j} = \frac{5}{2} {b_{1/2}}^{(2)} (\alpha_{i,j}) + \frac{1}{2} \alpha_{i,j} \frac{d}{d \alpha_{i,j}} {b_{1/2}}^{(2)} (\alpha_{i,j})
\end{equation*}
\begin{equation*}
\begin{aligned}
{f_{A4}}^{i,j} = & \frac{1}{8}\bigg[ 114 {b_{1/2}}^{(6)} (\alpha_{i,j}) + 22 \alpha_{i,j} \frac{d}{d \alpha_{i,j}} {b_{1/2}}^{(6)} (\alpha_{i,j})\\ 
&+ {\alpha_{i,j}}^2 \frac{d^2}{d\alpha_{i,j}^2} {b_{1/2}}^{(6)} (\alpha_{i,j}) \bigg]
\end{aligned}
\end{equation*}
\begin{equation*}
\begin{aligned}
{f_{A5}}^{i,j} =& \frac{1}{8}\bigg[ 104 {b_{1/2}}^{(4)} (\alpha_{i,j}) + 22 \alpha_{i,j} \frac{d}{d \alpha_{i,j}} {b_{1/2}}^{(4)} (\alpha_{i,j}) \\
&+ {\alpha_{i,j}}^2 \frac{d^2}{d\alpha_{i,j}^2} {b_{1/2}}^{(4)} (\alpha_{i,j}) \bigg]
\end{aligned}
\end{equation*}
\begin{equation*}
\begin{aligned}
{f_{A6}}^{i,j} =& \frac{1}{4}\bigg[ 110 {b_{1/2}}^{(5)} (\alpha_{i,j}) + 22 \alpha_{i,j} \frac{d}{d \alpha_{i,j}} {b_{1/2}}^{(5)} (\alpha_{i,j})\\ 
&+ {\alpha_{i,j}}^2 \frac{d^2}{d\alpha_{i,j}^2} {b_{1/2}}^{(5)} (\alpha_{i,j}) \bigg]
\end{aligned}
\end{equation*}
\begin{equation*}
\begin{aligned}
{f_{A7}}^{i,j} =& \frac{1}{4}\bigg[ 2 {b_{1/2}}^{(1)} (\alpha_{i,j}) - 2 \alpha_{i,j} \frac{d}{d \alpha_{i,j}} {b_{1/2}}^{(1)} (\alpha_{i,j})\\ 
&- {\alpha_{i,j}}^2 \frac{d^2}{d\alpha_{i,j}^2} {b_{1/2}}^{(1)} (\alpha_{i,j}) \bigg]
\end{aligned}
\end{equation*}
\begin{equation*}
{f_{A8}}^{i,j} = \frac{1}{2} \alpha_{i,j} {b_{3/2}}^{(5)} (\alpha_{i,j}) 
\end{equation*}
\begin{equation*}
{f_{A9}}^{i,j} = \frac{1}{2} \alpha_{i,j} {b_{3/2}}^{(5)} (\alpha_{i,j}) 
\end{equation*}
\begin{equation*}
{f_{A10}}^{i,j} = - \alpha_{i,j} {b_{3/2}}^{(5)} (\alpha_{i,j}) 
\end{equation*}
\begin{equation*}
{f_{A11}}^{i,j} = \alpha_{i,j} {b_{3/2}}^{(1)} (\alpha_{i,j})
\end{equation*}

\begin{equation*}
{f_{B1}}^{i,j} = \bigg( \frac{1}{4}\alpha_{i,j} \frac{d}{d \alpha_{i,j}} + \frac{1}{8} {\alpha_{i,j}}^2 \frac{d^2}{{d \alpha_{i,j}}^2}  \bigg) {b_{1/2}}^{(0)} (\alpha_{i,j})
\end{equation*}
\begin{equation*}
{f_{B2}}^{i,j} = -2 {b_{1/2}}^{(2)} (\alpha_{i,j}) + \frac{1}{2} \alpha_{i,j} \frac{d}{d \alpha_{i,j}} {b_{1/2}}^{(2)} (\alpha_{i,j}) 
\end{equation*}
\begin{equation*}
{f_{B3}}^{i,j} = \frac{3}{2} {b_{1/2}}^{(1)} (\alpha_{i,j}) + \frac{1}{2} \alpha_{i,j} \frac{d}{d \alpha_{i,j}} {b_{1/2}}^{(1)} (\alpha_{i,j})  - 2\alpha_{i,j}
\end{equation*}
\begin{equation*}
\begin{aligned}
{f_{B4}}^{i,j} =& \frac{1}{8}\bigg[ 44 {b_{1/2}}^{(4)} (\alpha_{i,j}) + 14 \alpha_{i,j} \frac{d}{d \alpha_{i,j}} {b_{1/2}}^{(4)} (\alpha_{i,j})\\ 
&+ {\alpha_{i,j}}^2 \frac{d^2}{d\alpha_{i,j}^2} {b_{1/2}}^{(4)} (\alpha_{i,j}) \bigg] 
\end{aligned}
\end{equation*}
\begin{equation*}
\begin{aligned}
{f_{B5}}^{i,j} =& \frac{1}{8}\bigg[ 38 {b_{1/2}}^{(2)} (\alpha_{i,j}) + 14 \alpha_{i,j} \frac{d}{d \alpha_{i,j}} {b_{1/2}}^{(2)} (\alpha_{i,j})\\ 
&+ {\alpha_{i,j}}^2 \frac{d^2}{d\alpha_{i,j}^2} {b_{1/2}}^{(2)} (\alpha_{i,j}) \bigg] 
\end{aligned}
\end{equation*}
\begin{equation*}
\begin{aligned}
{f_{B6}}^{i,j} =& - \frac{1}{4}\bigg[ 42 {b_{1/2}}^{(3)} (\alpha_{i,j}) + 14 \alpha_{i,j} \frac{d}{d \alpha_{i,j}} {b_{1/2}}^{(3)} (\alpha_{i,j})\\ 
&+ {\alpha_{i,j}}^2 \frac{d^2}{d\alpha_{i,j}^2} {b_{1/2}}^{(3)} (\alpha_{i,j}) \bigg] 
\end{aligned}
\end{equation*}
\begin{equation*}
\begin{aligned}
{f_{B7}}^{i,j} =& \frac{1}{4}\bigg[ 2 {b_{1/2}}^{(1)} (\alpha_{i,j}) - 2 \alpha_{i,j} \frac{d}{d \alpha_{i,j}} {b_{1/2}}^{(1)} (\alpha_{i,j})\\
&- {\alpha_{i,j}}^2 \frac{d^2}{d\alpha_{i,j}^2} {b_{1/2}}^{(1)} (\alpha_{i,j}) \bigg] 
\end{aligned}
\end{equation*}
\begin{equation*}
{f_{B8}}^{i,j} = \frac{1}{2} \alpha_{i,j} {b_{3/2}}^{(5)} (\alpha_{i,j}) 
\end{equation*}
\begin{equation*}
{f_{B9}}^{i,j} = \frac{1}{2} \alpha_{i,j} {b_{3/2}}^{(5)} (\alpha_{i,j}) 
\end{equation*}
\begin{equation*}
{f_{B10}}^{i,j} = - \alpha_{i,j} {b_{3/2}}^{(5)} (\alpha_{i,j}) 
\end{equation*}
\begin{equation*}
{f_{B11}}^{i,j} = \alpha_{i,j} {b_{3/2}}^{(1)} (\alpha_{i,j})
\end{equation*}

\begin{equation*}
{f_{C1}}^{i,j} = \bigg( \frac{1}{4}\alpha_{i,j} \frac{d}{d \alpha_{i,j}} + \frac{1}{8} {\alpha_{i,j}}^2 \frac{d^2}{{d \alpha_{i,j}}^2}  \bigg) {b_{1/2}}^{(0)} (\alpha_{i,j})
\end{equation*}
\begin{equation*}
\begin{aligned}
{f_{C2}}^{i,j} =& \frac{1}{8}\bigg[ 21 {b_{1/2}}^{(3)} (\alpha_{i,j}) + 10 \alpha_{i,j} \frac{d}{d \alpha_{i,j}} {b_{1/2}}^{(3)} (\alpha_{i,j})\\ 
&+ {\alpha_{i,j}}^2 \frac{d^2}{d\alpha_{i,j}^2} {b_{1/2}}^{(3)} (\alpha_{i,j}) \bigg] 
\end{aligned}
\end{equation*}
\begin{equation*}
\begin{aligned}
{f_{C3}}^{i,j} =& \frac{1}{4}\bigg[ -20 {b_{1/2}}^{(2)} (\alpha_{i,j}) - 10 \alpha_{i,j} \frac{d}{d \alpha_{i,j}} {b_{1/2}}^{(2)} (\alpha_{i,j})\\ 
&- {\alpha_{i,j}}^2 \frac{d^2}{d\alpha_{i,j}^2} {b_{1/2}}^{(2)} (\alpha_{i,j}) \bigg] 
\end{aligned}
\end{equation*}
\begin{equation*}
\begin{aligned}
{f_{C4}}^{i,j} =& \frac{1}{8}\bigg[ 17 {b_{1/2}}^{(1)} (\alpha_{i,j}) + 10 \alpha_{i,j} \frac{d}{d \alpha_{i,j}} {b_{1/2}}^{(1)} (\alpha_{i,j})\\ 
&+ {\alpha_{i,j}}^2 \frac{d^2}{d\alpha_{i,j}^2} {b_{1/2}}^{(1)} (\alpha_{i,j}) \bigg] -\frac{27}{8} \alpha_{i,j} 
\end{aligned}
\end{equation*}
\begin{equation*}
\begin{aligned}
{f_{C5}}^{i,j} =& \frac{1}{4}\bigg[ 2 {b_{1/2}}^{(1)} (\alpha_{i,j}) - 2 \alpha_{i,j} \frac{d}{d \alpha_{i,j}} {b_{1/2}}^{(1)} (\alpha_{i,j})\\
&- {\alpha_{i,j}}^2 \frac{d^2}{d\alpha_{i,j}^2} {b_{1/2}}^{(1)} (\alpha_{i,j}) \bigg] 
\end{aligned}
\end{equation*}
\begin{equation*}
{f_{C6}}^{i,j} = \frac{1}{2} \alpha_{i,j} {b_{3/2}}^{(2)} (\alpha_{i,j}) 
\end{equation*}
\begin{equation*}
{f_{C7}}^{i,j} = \frac{1}{2} \alpha_{i,j} {b_{3/2}}^{(2)} (\alpha_{i,j}) 
\end{equation*}
\begin{equation*}
{f_{C8}}^{i,j} = - \alpha_{i,j} {b_{3/2}}^{(2)} (\alpha_{i,j}) 
\end{equation*}
\begin{equation*}
{f_{C9}}^{i,j} = \alpha_{i,j} {b_{3/2}}^{(1)} (\alpha_{i,j}).
\end{equation*}

\end{document}